\documentclass[12pt]{article}
\usepackage{latexsym}
\usepackage{amssymb}
\usepackage{amsmath}
\usepackage{graphicx}
\usepackage{enumitem}
\usepackage{bbold}
\usepackage{slashed}
\usepackage{hyperref}

\newcommand\mathC{\mkern1mu\raise2.2pt\hbox{$\scriptscriptstyle|$}
        {\mkern-7mu\rm C}}              
\newcommand{\mathR}{{\rm I\! R}}         

\def\be{\begin{equation}}
\def\ee{\end{equation}}
\def\bear{\begin{eqnarray}}
\def\eear{\end{eqnarray}}
\def\nn{\nonumber}

\newcommand\bra[1]{{\langle {#1}|}}
\newcommand\ket[1]{{|{#1}\rangle}}

\def\a{\alpha}
\def\b{\beta}

\def\d{\delta}

\def\th{\theta}

\def\dd{\mbox{d}}

\def\bra{\langle}
\def\ket{\rangle}
\def\a{\alpha}
\def\b{\beta}
\def\d{\delta}
\def\D{\Delta}

\def\f{\phi}
\def\F{\Phi}

\def\l{\lambda}

\def\m{\mu}

\def\th{\theta}

\def\pa{\partial}

\newcommand{\ti}[1]{\tilde{#1}}

\newcommand{\sm}[1]{\mbox{\scriptsize #1}}
\newcommand{\tn}[1]{\mbox{\tiny #1}}
\renewcommand{\@}[1]{\sqrt{#1}}
\renewcommand{\le}[1]{\label{#1}\end{eqnarray}}
\newcommand{\bea}{\begin{eqnarray}}
\newcommand{\eea}{\end{eqnarray}}

\newcommand{\eq}[1]{(\ref{#1})}
\def\nn{\nonumber\\}

\def\ffract#1#2{\raise .35 em\hbox{$\scriptstyle#1$}\kern-.25em/
\kern-.2em\lower .22 em \hbox{$\scriptstyle#2$}}

\def\half{{1\over2}\,}

\begin{document}

\pagestyle{empty}

\centerline{{\Large \bf A Schema for Duality, Illustrated by Bosonization}}
\vskip 1.5truecm

\begin{center}
{\large  
Sebastian De Haro and Jeremy Butterfield}\\

\vskip 1truecm
{\it Trinity College, Cambridge, CB2 1TQ, United Kingdom}\\
{\it Department of History and Philosophy of Science, University of Cambridge\\
Free School Lane, Cambridge CB2 3RH, United Kingdom}\\

\vskip .4truecm
{\tt sd696@cam.ac.uk, jb56@cam.ac.uk}

\vskip .4truecm
Friday 22 Dec 2017\\

\end{center}

\vskip 3truecm

\begin{center}
\textbf{\large \bf Abstract}
\end{center}

In this paper we present a schema for describing dualities between physical theories (Sections \ref{schema} and \ref{symdual}), and illustrate it in detail with the example of bosonization: a boson-fermion duality in two-dimensional quantum field theory (Sections \ref{bfd} and \ref{bfdual}). 

The schema develops proposals in De Haro (2016, 2016a): these proposals include construals of  notions related to duality, like representation, model, 
symmetry and interpretation. The aim of the schema is to give a more precise criterion for duality than has so far been considered.


The  bosonization example, or boson-fermion duality, has the feature of being  {\it simple} yet {\it rich enough} to illustrate the most relevant aspects of our schema, which also apply to more sophisticated dualities. The richness of the example consists, mainly, in its concern with two non-trivial {\it quantum field theories}: including massive Thirring-sine-Gordon duality, and non-abelian bosonization.

This prompts 
two comparisons with the recent philosophical literature on dualities:--- 

(a)~~Unlike the standard cases of duality in quantum field theory and string theory, where only specific simplifying limits of the theories are explicitly known, the boson-fermion duality is known to hold {\it exactly}. This exactness can be exhibited explicitly.


(b)~The bosonization example illustrates  both the cases of isomorphic and {\it non-isomorphic} models: which we believe the literature on dualities has not so far discussed.

\newpage
\pagestyle{plain}

\tableofcontents


\section{Introduction}\label{intro}

In this paper we present a schema for describing dualities between physical theories (Sections \ref{schema} and \ref{symdual}). Then we illustrate it in detail with the example of bosonization: a boson-fermion duality in two-dimensional conformal field theories (Sections \ref{bfd} and \ref{bfdual}). 

Before we introduce these two parts in turn (Sections \ref{introschema}, \ref {introbosonzn}), we briefly set our project in the context of  the legacy of Hilbert's work, a hundred and more years ago, on the foundations of physics and its axiomatisation---work which it is an honour to commemorate. This legacy is of course so broad and deep that we can only touch on it. We will confine ourselves to recalling two Hilbertian ideas about the role of axiomatizing a theory (either mathematical or physical): ideas which obviously relate to our project, and which we will return to in Sections \ref{RandI} and \ref{abstraction}.

The background for both ideas, indeed for all Hilbert's work in axiomatisation (such as his axiomatisation in 1899 of Euclidean geometry, and his choosing as the sixth Problem in his famous 1900 `To Do' list, the axiomatisation of mechanics and geometry) was, of course, the development of formal methods, in particular axiomatic studies, in all of mathematics from about 1850.\footnote{Again, we can only touch on the vast literature.  For Hilbert's Problems of 1900, cf.~e.g.~Gray (2000), Grattan-Guinness (2000). For the sixth Problem, cf.~Corry (1999, 2004, 2006, 2018), and Stachel (2018). For some context for Hilbert's famous `beer-mug' remark, cf.~Kennedy (1972). Finally, we note that Gray (2008) makes an interesting case that this broad development represented a rise of `modernism', in a sense analogous to that in art and literature.} 

First, there is the idea that an axiom system can be realized, i.e.~made true, by very different models.  (Recall Hilbert's famous remark that `one must be able to say at all times---instead of points, straight lines and planes---tables, chairs and beer-mugs'.) We shall see that duality, in the sense nowadays used in physics,  gives illustrations of this idea. Indeed, very {\em vivid} illustrations. For duality, in physicists' current jargon, involves there being two theories that look very different (not just in their formulation and concepts, but also apparently in the topics they are about) that are in some sense equivalent. In particular, there is a `dictionary' that pairs off the concepts in one theory with those in the other. Thus in our example of bosonization, one theory will describe fermions, while the other describes bosons: very different field-contents. So duality illustrates this Hilbertian idea: but on a grand scale! For now, it is entire physical theories that are the very different realizations of some {\em common core} axiom system. Indeed, as explained in Section \ref{introschema}: we shall call the two sides, i.e.~the items that are dual to each other, {\em models} (viz.~models of the single common core), rather than {\em theories}. So this usage echoes the Hilbertian idea.\footnote{The Hilbertian idea has, of course, other important facets: for example, in fostering the idea that an axiom system---or more generally, a doctrine expressed in language---`implicitly defines' its terms. This has been very influential in the foundations of logic and mathematics, beginning with Hilbert's debate with Frege. It has also of course been contested: in the face of non-categoricity, the claim to `define' terms by a body of doctrine containing them is questionable.} 

Second, Hilbert sees the activity of axiomatisation, {\em not} as giving a theory its final form and so best undertaken when (one hopes!) it is fully understood, but as worthwhile even when we recognize that the theory is far from its final form. For it is worthwhile precisely in order to deepen our understanding of the theory. Again this idea has been very influential. In physics, the best known example of its influence is no doubt von Neumann's monumental treatise on quantum mechanics (1932), which over the decades has spawned so many axiomatic studies of quantum mechanics, most directly the quantum logic approach. But also in philosophy, the idea was very influential. Reichenbach and other logical empiricists saw axiomatisation as the way by which philosophers could clarify scientific theories (and in particular, distinguish their factual and conventional contents---a project that, for the logical empiricists, was the distinctive task of philosophy). Thus we think of our own project---to formulate in general, almost formal, terms, the notion of duality (Sections \ref{schema} and \ref{symdual}), and to illustrate this in bosonization (Sections \ref{bfd} and \ref{bfdual})---as an exercise in the tradition of this idea.\footnote{For some `post-Hilbert' history of axiomatisation as `deepening the foundations', cf.~St\"oltzner (2000, 2001). But we should add that we do not endorse the logical empiricist project of distinguishing, once and for all, the factual and conventional parts of a theory: our misgivings are essentially those of Putnam (1962).}

In Section \ref{introschema}, we briefly introduce our notions of theory and model, and of duality as an isomorphism between models. We motivate our usage and compare the notion of duality to the analogous notion of symmetry. In Section \ref{introbosonzn}, we introduce our main example, of bosonization, and compare this example to other examples used in the literature on dualities.

\subsection{The schema}\label{introschema}

The schema develops proposals in De Haro (2016, 2016a). Like other authors, we take duality to be a suitable relation of equivalence between physical theories. The main features of our schema are that:\\
\indent \indent 1): we distinguish uninterpreted theories, which we call {\em bare theories}, from interpreted theories;\\
\indent \indent 2): we emphasize that, wholly independently of issues of interpretation, a bare theory can have many realizations, which we call {\em models};\\
\indent \indent 3): we take duality to be an isomorphism between two models of a single bare theory.

Of these three features, it is 2) and 3) that are the distinctive ones. For several authors also define duality in terms of uninterpreted theories.
This has the advantage of making verdicts of duality not beholden to semantic issues, and so less vague or even controversial. And it allows cases of duality without any sort of physical or semantic equivalence---which certainly occur, e.g.~Kramers-Wannier duality between the high and low temperature regimes of the statistical mechanics of a lattice. But features 2) and 3) make duality an equivalence (formally: an isomorphism) between items that are not only uninterpreted, but also {\em more specific} than an uninterpreted theory: viz.~realizations---which we will call `models'---of a (single) bare theory. A prototypical example is: taking a bare theory to be an abstract algebra of quantities (maybe also equipped with a dynamics, viz.~as a 1-parameter group of automorphisms, and a set of abstract states, i.e.~rules for evaluating (i.e.~assinging values to) the quantities): a model or realization is given by a representation (in the mathematical sense) of the algebra, together with a realization of the rules for evaluating the quantities, for the representation in question, i.e.~a set of maps to the relevant field, of complex or real numbers.

We shall say `model' rather than `realization', not least for brevity. But we should disavow, here at the outset, some misleading connotations of the word `model'. Indeed, there are three misleading connotations. The word `model', as contrasted with `theory', often connotes: \\
\indent \indent  (i): a specific solution (at a single time: or for all times, i.e.~a possible history) for the physical system concerned, whereas the `theory' encompasses all solutions---and in many cases,  for a whole class of systems; \\
\indent \indent  (ii): an approximation, in particular an approximate solution, whereas the `theory' deals with exact solutions; \\
\indent \indent (iii): being part of the physical world (in particular, being empirical, and-or observable) that gives the interpretation, whereas the `theory' is of course not part of the physical world, and so stands in need of interpretation.\\
So we stress that our use of `model' rejects all three connotations. As we just said: for us, a model is a specific realization---one might say `version'---of a theory. But it is a version of a bare, i.e.~uninterpreted, theory, and the version is itself bare, i.e.~uninterpreted. So a model adds details---we shall say: `specific structure'---to its bare theory. But these details are {\em not} a matter of specifying: (i) a solution or history of the system; or (ii) approximation(s); or (iii) interpretation(s). Rather, the extra details are extra mathematical structure: just like a representation of a group or an algebra has extra details or structure, beyond that of the group or  algebra of which it is a representation.\footnote{{\label{modelusage}}{We agreed that for our notion, the word `model' has disadvantages. But note that other words also have disadvantages. For example:  `formulation'  connotes that any two formulations of a theory are `notational variants', i.e.~fully equivalent: they say exactly the same thing about the world. But that is far from true for our notion (and this matches the connotations of `model'): for us, two models of a bare theory are in general not isomorphic, and not in any sense equivalent; and so typically, it is surprising to find  two isomorphic models, i.e.~to find a duality. Other examples:  `realization', `instance' and `instantiation' connote being part of the physical world, as in `the mechanism/hardware which realizes some specific function/software', or `the object is an instance/instantiation of the predicate'---which is the misleading connotation  (iii) above. 

Notice that in theoretical physics, the use of model is, roughly, between: (a) our use, and (b) (ii) and (iii) above: e.g.~the `massive Thirring model' or the `sine-Gordon model'.}} 

Indeed,  for clarity later on, we should distinguish two broad kinds of extra detail or structure that a model adds. Again, group representations provide obvious---and countless---examples.\\
\indent \indent  (A): The `concreteness' of a specific mathematical object: such as $\mbox{GL}(n,\mathC)$, the general linear group over $\mathC^n$, or any subgroup of it---any of which is a `concrete', not abstract, group. (Agreed, the concrete vs.~abstract contrast is flexible; but this will not matter for anything that follows.)  \\
\indent \indent   (B): The fact that the mathematical notion of representation requires homomorphism, not isomorphism: i.e.~it allows non-injectivity and non-surjectivity. Thus two representations of an abstract group $G$ can be non-isomorphic as groups---i.e.~different, even as described in only abstract group-theoretic terms---to $G$; and of course also, non-isomorphic to each other.\\
Of course, these kinds (A) and (B) of `extra detail' usually occur together: just think of how every abstract group can be represented by the trivial one element subgroup of $\mbox{GL}(n,\mathC)$, the $n \times n$ identity matrix. 

Our picture is therefore of a bare theory, that can be realized (we will say: modelled) in various ways: like the different representations of an abstract group or algebra. And these models are in general {\em not} isomorphic, since they differ from one another in their specific structure: like inequivalent representations of a group. But we say: {\em when the models are isomorphic, we have a duality}. 

In Sections \ref{schema} and \ref{symdual}, we will develop this view of duality (with Section \ref{schema} dealing with theories, models and interpretations, and Section \ref{symdual} with symmetries). We end this Subsection with two further remarks about our schema. The first motivates our usage of `theory' and `model'; the second compares duality with that more familiar topic, symmetry. \\

(1): {\em Motivating our usage}:--- Dualities in physics give a rationale for our usage of `theory' and `model', as introduced above.  (This rationale does not depend on the contrast between interpreted and uninterpreted (bare) theories; and so we temporarily set that aside.)  Recall that in both physics and philosophy of physics, `theory' is usually taken as something like a set of differential equations, and `model' is usually taken as something like a solution to such a set. But a duality often shows us that what we first considered as distinct theories can, or should, be seen as the same theory, in two guises. Agreed, that is very rough speaking: which will of course be clarified in what follows. But for now, we only need the point that this kind of surprising discovery prompts us to move our usage of  `theory' ``one level up''. After all: if two sets of differential equations somehow express the same theory, then a theory cannot be identified with such a set. Besides: if we thus move our usage of  `theory' one level up, we can still keep the usual intuitive idea of how `theory' and `model' are related---viz.~that a model is a realization, or instance, of a theory---by correspondingly moving our usage of `model' one level up. And this is what we have proposed. 

To sum up: the broad and widely-agreed idea, that in physics a duality often suggests that the two theories concerned, though they look different, are in fact `the same', motivates our proposed usage of `theory' and `model'. \\

(2): {\em Analogy with symmetry}:--- The analogy is (as is often remarked) that `a duality is like a symmetry, but at the level of a theory'. Here, and for the rest of this Subsection, we will {\em temporarily} set aside our jargon just announced, of `theory' vs.~`model'.  We will temporarily join the literature's usual jargon of taking a theory to be interpreted, and a model to be---not a `version' of the theory with some specific structure of its own---but a solution (or representative of a solution) of the theory.
   
That is, the analogy is: while a symmetry carries a state to another state that is `the same' or `matches it', a duality carries a theory to another theory that is `the same' or `matches it'. We will endorse this analogy. So the interesting questions, for both sides of the analogy, will concern the different ways  to make precise `the same' or `matches'. We give details (respecting our proposed `theory' vs.~`model' usage!) in Sections \ref{Symm} and \ref{DlyDef}. But the questions about making precise `the same'/`matches' can be introduced as follows.  

 A symmetry $a$ (we write $a$ for `automorphism')  carries a state $s$ in a state space ${\cal S}$ 
 to another state $a(s)$: thanks to $a$ being a symmetry, the two states $s$ and $a(s)$ assign the same values to all the quantities (i.e.~magnitudes) in some salient, usually large, set of quantities. The question then arises: do $s$ and $a(s)$ represent the very same physical state of affairs, or scenario---or in philosophers' jargon: the same possible world? 
 
 The answer, in full generality, is of course: `No'. That is: not always. But for a large enough set of quantities being preserved; and in particular for a theory that is a `toy cosmology' (i.e.~a theory whose system of interest is a cosmos, with no external environment, so that there are no relational quantities whose values are {\em not} preserved by $a$): there is a tradition of answering `Yes'.  
 
 Debate then ensues about:\\
 \indent \indent (i)~~what are the general conditions for the `Yes' answer being correct? and \\
 \indent \indent (ii)~what does the `Yes' answer imply about the propriety of---perhaps even the requirement of---moving to a reduced formalism, i.e.~one in which states are taken as the orbits, in the given formalism, of the action of the symmetry $a$?\footnote{A bit more precisely: states would be taken as the union of the orbits for all the  symmetries for which the `Yes' answer is true. For recent work on the debate about (i) and (ii), cf.~Caulton (2015), Dewar (2016) and Weatherall (2015).}

So, turning to our topic of dualities: we endorse the analogy. We will say, roughly speaking,  that: a theory $T$ is mapped by duality $d$ to a theory $d(T)$ which is `the same' as $T$. This will be made precise in various ways. But it is worth stressing now, in line with the three features 1), 2) and 3) we listed at the start of this Subsection, that:\\
 \indent \indent (a): We take theories to be initially uninterpreted: so it will not follow from the existence of a duality map $d$ that $T$ and $d(T)$ are wholly equivalent (`state the very same propositions').\\
 \indent \indent (b): We make explicit the interpretation of a theory's formalism: so there will be
interpretation maps $I$ acting on both the theory $T$ and its dual $d(T)$.\\
 \indent \indent (c): For a given theory, we distinguish different realizations of it, which we call `models'. Duality is an isomorphism between such models: an isomorphism that is often surprising since the models, despite their common core, ``look different''.
 
\subsection{Bosonization and other dualities}\label{introbosonzn}

In this Subsection, we motivate our choice of bosonization as the illustration of our schema. We first sketch a spectrum of examples of dualities (Section \ref{exdlies}). Then we describe how bosonization strikes a balance between mathematical rigour and physical interest, and introduce its main features (Section \ref{bosointro}).

\subsubsection{Examples of dualities}\label{exdlies}

Recent philosophical literature on duality and theoretical equivalence has dealt with three main kinds of examples:---\\
 \indent (a):~~equivalence between models (in our sense!) formulated in {\it first-order} (maybe {\em many-sorted}) {\em logic}: e.g.~definitional equivalence, Morita equivalence, and-or categorical equivalence (e.g.~Barrett and Halvorson (2016, 2016a)); \\
 \indent (b):~~~categorical equivalence of models (in our sense) of {\it classical theories} (e.g.~Weatherall (2015), Teh (2016)); \\
 \indent (c):~~dualities between models (in our sense) of {\it quantum theories} whose classical descriptions are very different (e.g.~Dieks et al.~(2015), De Haro (2017), De Haro et al.~(2017), Fraser (2017), Huggett (2017), Rickles (2017), Matsubara (2013)). \\
The classification (a)-(c) is arranged in increasing order of physical (not mathematical!) sophistication. Consequently, there is also decreasing mathematical rigour, as one moves from kind (a) to  kind (c):\\
\\
Kind (a):~These examples have the advantage of being very simple, in their reliance on first-order logic only: and so, the notions of equivalence in question can be defined rigorously. But in their simplicity, the notions developed, and the examples given, generally do not seem to have sufficient structure that they could describe in detail the sorts of examples that physicists would be interested in. (At any rate, the authors cited do not describe how such logical models can illustrate even the simplest physical models of, say, classical, source-free Maxwell theory: which is, of course, not to claim that this is impossible!).\\
\\
Kind (b):~These examples include some important models of {\it classical theories}, such as Newtonian gravitation, general relativity, and Yang-Mills theory models. But these examples also have some limitations. (1): To physicists, the example is, typically, not  surprising (e.g.~Newtonian gravitation being equivalent to {\it geometrized} Newtonian gravitation). (2): When it {\it is} surprising (e.g.~Teh (2016)), it is not a case of equivalence, but rather of analogy. Furthermore, (3): categorical equivalence has been criticised by Barrett and Halvorson (2016a) for being `too liberal'. In our view, the element of `surprise' (see Section \ref{orient}) seems to come  with models of {\it quantum theories}, i.e.~examples of (c): \\
\\
Kind (c):~Typical examples of this kind are dualities between very different-looking models of {\it quantum field theories}, or of {\it string theories} (cf.~Witten (2018)). 

This explains the recent interest, shown by both physicists and philosophers of physics, in such dualities. Physicists tend to view dualities as powerful epistemic statements: the epistemic gain being both mathematical and physical. As the mathematical aspect: mirror symmetry  is the prime example.\footnote{Despite the name of `symmetry', mirror symmetry falls under what we here call a duality. That mirror symmetry is a case of duality---of two different models, rather than a single model, being related---is uncontroversial, and reflected in the literature.} Michael Atiyah has characterised the discovery of mirror symmetry as `spectacular': since it established a new link between complex geometry and symplectic geometry, later proven (in one of its simplified versions) by mathematicians (2009, p.~83). As to the physical aspect: gauge-gravity duality is an important example, which has led to both new theoretical developments in quantum gravity, and to new experimental results and ideas (like the explanation of the shear viscosity-to-entropy ratio in a quark-gluon plasma, and recent applications to cosmology: cf.~e.g.~Ammon and Erdmenger (2015), ~De Haro et al.~(2016)). Other examples are T-duality (related to mirror symmetry) and S-duality: which falls under the same class of dualities as our bosonization example, viz.~exchanging Noether charges and topological charges (Castellani (2017)).  An important idea of these dualities is that it is the models of the {\it quantum theories} which are equivalent, while their classical limits are very disparate (differing in the number or the size of the dimensions, the matter content, etc.). These two aspects---physical and mathematical---will be developed in Section \ref{orient}'s discussion of the {\em scientific importance} of dualities. 

But there is a second reason these dualities interest philosophers of physics: which the recent literature has emphasised. Namely, these dualities obviously bear on philosophical questions such as the distinction between theoretical and physical equivalence,  emergence (of spacetime, and-or other entities), and  realism vs.~structuralism. We will return to these questions in another paper.

Agreed: examples of kind (c) also have limitations, as follows. (1): The models (in our sense)  are mathematically very difficult; and typically, no exact formulation of the models that are dual is yet known. So the duality, e.g.~in the case of gauge-gravity duality, is still---despite all the favourable evidence, in various limits etc.---a conjecture. (2): The physics involved is  not yet established, since the models involved either deal with quantum gravity situations (a regime of energies about which experiment has so far given no direct clues: cf.~Smolin (2018)), or involve simplifying assumptions about the world, typically a high degree of symmetry (e.g.~supersymmetric quantum field theory models). 

\subsubsection{Bosonization introduced}\label{bosointro}

It is clear, in the light of Section \ref{exdlies}, that to illustrate our schema, we should choose an example that judiciously balances the desiderata: on the one hand, (i): mathematical precision and established physics, as in kinds (a) and (b); on the other, (ii):  scientific importance, as in kind (c). As we will see in detail in Sections \ref{bfd} and \ref{bfdual}: bosonization or, more precisely, boson-fermion duality in two dimensions, is just such an example. 

As to (i): Boson-fermion duality allows a treatment of the quantum theory model that does not need to rely on techniques of approximation such as perturbation theory. For this reason, boson-fermion duality explicitly illustrates our schema: the common core {\em theory} can be formulated according to our construal in Section \ref{bare;was22A}, and the two sides of the duality are {\em models} in the sense of Section \ref{models;was22B}. The physics of these models is not speculative, and these 1+1-dimensional models describe systems that can be realised in the lab, e.g.~as one-dimensional spin chains (Giamarchi (2003:~Chapter 2), Altland and Simons (2010:~Sections 4.3 and 9.4.4)). 

As to (ii): The scientific importance of the duality is witnessed by three facts. (1): Bosonization involves rich models of {\it quantum field theories}, and not just classical theories; (2) it is an active area of physics (see e.g.~Gogolin et al.~(2004), Kopietz (2008)); and (3) it illustrates the surprise that we discuss in Section \ref{orient}, viz.~by relating a model of bosons and a model of fermions.

Bosonization was discovered in two papers by Coleman (1975) and Mandelstam (1975), which built on previous work on the sine-Gordon and Thirring models (Dell'Antonio et al.~(1972)). Coleman discovered that the sine-Gordon model (a scalar field whose interaction potential is the cosine of the field) in 1+1 dimensions was equivalent to the {\it charge-zero sector} of the massive Thirring model (a massive Dirac fermion field with quartic interaction) in 1+1 dimensions.\footnote{See Section \ref{TSG}; for the simple, free case, see Section \ref{bfd}. Here of course we adopt the usual theoretical physics usage of `model': cf.~the end of footnote \ref{modelusage}.} `Charge-zero sector' here refers to the restriction of the physical quantities of the model to pairs of fermionic fields. Thus Coleman wrote: \\
\indent`... under the assumption that one can only make particle-antiparticle pairs out of the vacuum, not single particles ... For massless particles in two dimensions, it is quite possible to make a pair that never separates. Such a pair consists of two particles moving in the same direction. The wave functions do not spread; they just move on steadily at the speed of light, and the particles never get away from each other [since there is no other direction in which they could turn]. If the particles had a mass, or if the world were of greater than two dimensions, this would not be possible.' (p.~2094).\\

While Coleman's analysis was perturbative, Mandelstam constructed a map which was exact, and went both ways.  Not only could a boson be mapped to a pair of fermions; but also the map could be inverted, so as to map a single fermion to a coherent state of bosons. The construction was non-perturbative, i.e.~it did not use perturbation theory. This is related to the fact, already recognised by Coleman (1975: p.~2088), that all divergences that occur in perturbation theory can be removed by normal-ordering the Hamiltonian. Mandelstam also did a canonical treatment of the model, working out canonical commutation relations between the fields and the currents constructed from them, using regularisation and renormalisation. Therefore, boson-fermion duality was proven to be {\it exact}.\\

There are three significant features of this duality: features that both (a) justify our claim, two paragraphs above, that boson-fermion duality balances the desiderata (i) and (ii), and (b) bear out the conceptual relevance of the example.

(A): {\it The duality is exact}. That is: it is valid for all physically interesting values of the parameters, and it does not require the use of perturbation theory. In this respect, boson-fermion duality is closer to kinds (a) and (b) than kind (c). Yet the models  related as duals are non-trivial (because massive, or massless, and interacting!: see Section \ref{mgd}) quantum field theory models---and in {\it that} sense they are in kind (c)!

(B) {\it The duality goes both ways}. It relates boson operators to fermion operators, and vice versa. This is, of course, surprising, since these two kinds of operators have very different properties: both mathematically (e.g.~different statistics) and physically (they describe particles with distinct properties). We will explain, in Section \ref{bfd}, how two models with such disparate formulations can nevertheless be isomorphic to each other.

(C) {\it The duality maps the weak-coupling regime of one model to the strong-coupling regime of the other,} and vice versa (Coleman (1975:~p.~3027)): 
\bea\label{couplings}
{g\over \pi}={4\pi\over\b^2}-1~.
\eea
Here, $\b$ is the coupling constant of the bosonic model (the sine-Gordon model) and $g$ is the coupling constant of the fermionic model (the Thirring model: for more details, see Section \ref{TSG}). Clearly, when $\b\rightarrow0$, $g\rightarrow\infty$. This attests to the physical richness and, indeed, the non-trivial character of the duality. This weak coupling/strong coupling correspondence has later been found to be a feature of most dualities of kind (c), i.e.~dualities in models of quantum field theory and string theory: especially S-duality and gauge-gravity duality.\footnote{In T-duality and mirror symmetry, it is not the size of the {\it couplings} that is inverted by the duality map but, roughly speaking, the sizes of the {\it spaces}.} This richness is the main reason why physicists are interested in dualities: since they can learn about the strong-coupling regime of one model (where perturbation theory cannot be used effectively nor reliably, so that the model is in general much harder to deal with) from the weak-coupling regime of the other model (where perturbation theory is usually a good guide). For more discussion of how Eq.~\eq{couplings} contributes to scientific importance, see Section \ref{orient}-(2).\\

Features (A) and (B) are needed in order that the example illustrate our schema with  mathematical precision. We will spell this out in Section \ref{bfdual}. Indeed, we believe this is the first conceptual and technical exposition in the philosophical literature of a duality combining the physical interest of kind (c), with features (A)-(B).

Feature (C) relates to another important topic relating to dualities, viz.~that of {\it emergence.} Indeed, a recent theme in the philosophy of physics literature has been the close connection between duality and emergence.\footnote{On the connection between duality and emergence, see: Dieks et al.~(2015), Rickles (2013), Teh (2013).} A framework for understanding the connection between dualities and emergence was developed in De Haro (2016): it was argued that the two notions (duality as isomorphism, and emergence as novel and robust behaviour relative to a comparison class), while close to each other, also exclude one another. But we must leave the topic of emergence for another occasion. 

Feature (B) also prompts the question of {\it fundamentality}. Coleman himself wrote: \\
\indent`I am led to conjecture a form of duality, or nuclear democracy in the sense of Chew, for this two-dimensional theory. A single theory has two equally valid descriptions in terms of Lagrangian field theory: the massive Thirring model and the quantum sine-Gordon equation. The particles which are fundamental in one description are composite in the other: In the Thirring model, the fermion is fundamental and the boson a fermion-antifermion bound state; in the sine-Gordon equation, the boson is fundamental and the fermion a coherent bound state' (1975:~p.~2096). 

The issue of fundamentality in boson-fermion duality, and in electric-magnetic duality, has been addressed by Castellani (2017). Her account is, in philosophers' jargon, deflationary. That is: she argues that all the manifestations of the fields (as bosonic or as fermionic; as electric or magnetic) are ontologically equally fundamental: `What the duality specifically implies here, concerns, not mutual composition of the particles, but rather their different modes of appearance when considering the different classical limits of the quantum theory, i.e.~the dual perspectives' (2017:~Section 3.3).

Our construal of duality as an isomorphism, in Sections \ref{schema} and \ref{symdual}, is in agreement with such a deflationary account. For the content of the theory will be taken to be based on the {\em common core} of the models: and this common core includes {\it both} bosons and fermions, on an equal footing. We will discuss some of these issues in Section \ref{DlyInterpn}, but we will not emphasise this point: (for it  was worked out in detail for a general duality, and illustrated for gauge-gravity duality,  in De Haro (2016a:~Section 1), under the heading of `physical equivalence').

\section{The Schema Introduced: Theories and Models }\label{schema}

In this Section and the next, we develop the treatments of theory, model, interpretation, symmetry and duality, given in De Haro (2016:~Section 1, 2016a:~Section 1) (and foreshadowed in De Haro, Teh, and Butterfield (2016)). This Section deals with theory, model and interpretation; Section \ref{symdual} will deal with symmetry and duality itself. 

We begin with the scientific importance of dualities, and the comparison of duality with gauge (Section \ref{orient}). Then we further specify our notions of  theory and model (Section \ref{TandM}). Then we discuss: interpretations (Section \ref{interp}), representations and isomorphisms (Section \ref{RandI}).

\subsection{Duality's scientific importance}\label{orient}

Recall from Section \ref{intro}, our overall proposal. A bare theory can be realized (we will say: modelled) in various ways, like the different representations of an abstract algebra. These models are in general {\em not} isomorphic, since they differ from one another in their specific structure. But when they are isomorphic, we have a duality.

To develop this proposal, we begin with four clarifying remarks. Each remark leads in to the next. The first three  defend our taking duality as a notion that is both logically weak and independent of a theory's interpretation. The first is, in effect, just the point that `duality' is a term of art; so one can choose how to use it: and our choice of a logically weak definition makes for a {\it strong} physical notion! But the second and third are substantive---about the scientific importance of dualities. The fourth remark is a contrast with the notion of {\em gauge}.\\

(1): {\em A logically weak but physically strong definition}:--- We agree that at first sight, it looks profligate to say that there is duality whenever two models are isomorphic. For it means there are countless dualities. For example: if a group or an algebra, endowed with a set of rules for evaluating quantities, can be a bare theory, any two isomorphic representations will yield a duality, as long as the isomorphism preserves the values of the quantities. Accordingly, the notion of duality is sometimes narrowed  by adding physical conditions, not just on `bare theory', but also on the isomorphism between models, e.g.~by requiring the isomorphism to link the weak and strong coupling regimes of the two models (see Eq.~\eq{couplings}). But we will maintain in (2) and (3) below that it is best to leave `duality' broadly defined, as we have done: with such extra conditions being articulated in individual cases as the need arises. As we will see in Section \ref{logweak}, the strengthening will be given by the kind of physical degrees of freedom that one wishes to describe. And so, our notion of duality will be physically strong. In particular, it cannot be argued that two given models which share some structure are dual, unless the common structure is exactly equal to what the models regard as physical. In short:  this apparently profligate verdict can be accepted. \\

(2):  {\em Duality as surprising}:---  So far we have spoken mainly of logico-semantic issues, and ignored epistemological ones: we have said what a duality is, but not how surprising and fruitful it can be. Our case-study, in Sections 4 et seq., will of course bring out these issues. It is surprising indeed to learn that a theory we thought of as having as its quantum particles fermions also contains bosons---and even more surprising to learn that  conversely the theory can be presented in the first place as having bosons, and then shown to contain the fermions with which we first began. For the moment, we note three clarifying comments---which are suggested by phrases like `a theory we thought of', and `the theory can be presented'. Each comment leads in to the next.\\

\indent \indent (i): We usually discover a duality in the context, not of a bare theory, but of an interpreted theory; for of course  we work with interpreted theories.\footnote{Agreed, pure mathematicians sometimes work with uninterpreted theories; and duality is a grand theme in mathematics, just as it is in physics. But although comparing duality in mathematics and in physics would be a very worthwhile project, we set it aside. Cf.~Corfield~(2017).} \\
\indent \indent (ii): Indeed, we usually work with what we have called `a model of the theory', indeed an interpreted model. That is: usually, before the duality is discovered, we  have two interpreted models (usually called `physical theories'!) which we do not believe to be isomorphic in any relevant sense.\\
\indent \indent (iii): Usually,  we do not initially believe the two models are models of any single relevant theory (even of a bare one: i.e.~even if we let ourselves completely suspend our antecedent interpretation of the models). The surprise is to discover that they are such models---indeed are isomorphic ones.\\

The word `relevant' in (ii) and (iii) signals the fact that of course `isomorphism', `model' and `theory' are very flexible words. For example: almost any two items can be considered isomorphic, i.e.~as having a common structure, under a weak enough construal of `structure'. Thus physicists might well in some specific context notice that the two models in question are both groups, or both algebras. But they rightly do not announce this as discovering a duality: not even if they also notice that the two groups (or algebras) are isomorphic. They set it aside as irrelevant, since the abstract notion of group or of algebra is so general that having it identified as a bare theory in common between the models is scientifically useless.  

On the contrary, what is surprising, and scientifically valuable, is to find very specific, {\em not} general, structures in common between different models: especially when \\
\indent \indent  (a) the models as presented (so: as interpreted) are very disparate, and-or \\
\indent \indent  (b) the common structure is not only detailed (like `10-dimensional semisimple Lie group', as against `group') but amounts to an isomorphism of that detailed structure (like `isomorphic as 10-dimensional semisimple Lie groups'). \\
As noted above, what will give physical theories their specificity, thus making duality a more powerful tool than its logically weak definition might make it seem, is the fact that physical theories, even bare ones, come with sets of maps from groups and algebras to appropriate fields (in the mathematical, not physical, sense!), i.e.~maps that assign values to the physical quantities. These maps are defined at the level of the abstract structure, but must also be instantiated in each of the models (according to the relevant sense of instantiation, as either `representation' or `realization': cf.~Section \ref{models;was22B}). And this set of maps is usually so rich, that it often suffices to reconstruct a model. And so, the fact that duality preserves these maps can be very non-trivial, and surprising, especially when combined with (a)-(b) above. 

This discussion of (a)-(b) returns us to (1) above.  We doubt that there can be a general characterization of when the models as presented are disparate enough, and-or the discovered isomorphism is detailed enough, for scientific importance. Instead, one can only articulate in any specific case how the disparity and-or the details are enough: e.g.~because the isomorphism links the weak and strong coupling regimes of the two models. So it is not worth trying to tighten the {\em definition} of `duality' with conditions beyond the logically weak ones we advocate. One just needs to use one's judgment about which cases count as scientifically important enough to analyse. \\

(3):  {\em Examples}:--- The conclusion of (2) is supported by some famous examples of duality in physics. Apart from boson-fermion duality, which we already introduced in Section \ref{introbosonzn}, it is worth illustrating this with two other examples.\\
\indent \indent (A): Gauge-gravity duality. In this case, the models differ in the dimensions they assign to spacetime ($d$ in the gravity model, $d-1$ in the gauge model), in their field content and classical equations of motion (Einstein's equations coupled to matter in the gravity model, the Yang-Mills equations in the gauge model), and in much more. In this case, the common core consists only in a class of asymptotic operators and a conformal class of $(d-1)$-dimensional metrics. Of course, it is very surprising to learn that a gauge theory model in $d-1$ dimensions, and a model of quantum gravity in $d$ dimensions, despite their very disparate guises, nevertheless have the same common core, and represent the same theory. See De Haro~(2016, 2016a) for a discussion in the context of our schema.\\
\indent \indent  (B): Electric-magnetic, or S-duality. This relates two models by mapping the electric charges of one model to the magnetic charges of the other. Furthermore, it does so by mapping a small electric charge to a large magnetic charge, analogously to \eq{couplings} (since the charges play the role of couplings, in gauge models). Nevertheless, the common structure is the same in the two models, i.e.~the quantum theory is invariant under the replacement of one gauge group by its dual. \\

(4):  {\em A contrast with `gauge'}:--- This discussion of dualities' scientific importance brings out a contrast between our treatment of duality, and the notion of gauge.  Physicists sometimes make remarks like: `two dual theories are like different gauge formulations of a single theory'. We agree that this remark is {\em analogous} to our view: indeed, in two ways.\\
\indent (i): A gauge formulation of a theory has specific structure (viz.~the gauge variables) going beyond that mandated by the ideas (gauge-invariant ideas!) of the theory; just like for us, a model has specific structure going beyond that mandated by the bare theory.\\
\indent (ii): The idea of gauge as `descriptive redundancy' means that two gauge formulations of a single theory must `say the same thing'; just like we say that in a duality, two models are isomorphic, and so (if interpreted: could) `say the same thing'.  

But we submit that this is {\em only} an analogy. There are two differences. First, we want to allow for cases where the two duals are not physically equivalent (as in Kramers-Wannier duality, mentioned above): {\em pace} the suggestion in (ii). Second (and more importantly), the extra structure in a model is usually {\em not} gauge, i.e.~descriptively redundant: think of how the extra structure in a representation of a group usually carries physical information (e.g.~a representation of the Poincare group carrying mass and spin information). Again, as stressed in (2) above: the surprising and scientifically important discovery is that in two  models, with apparently very disparate structures, there is in fact an exact correspondence of structures. We shall return to these two differences in Section \ref{DlySymmies}, comment (3). 

\subsection{Theories and Models}\label{TandM}

In this subsection, we add details about the ideas of a  {\it bare theory} and its {\em models}, already introduced in Section \ref{intro}. 

\subsubsection{Bare theories}\label{bare;was22A}

Following De Haro (2016, 2016a), we take a {\it bare theory} to be a triple $T:=\bra \cal S,Q,D\ket$ comprising a structured set of states, a structured set of quantities, and a dynamics: together of course with the rules for evaluating the physical quantities on the states. (We will later discuss symmetries, which we will take as automorphisms $a:\cal S\rightarrow\cal S$ of the set of states; or as the dual maps on the set $\cal Q$ of quantities.)

Two immediate points of clarification:---\\
\indent \indent (1): We stress that, despite the physical connotations of the words `states', `quantities', and `dynamics', a bare theory is {\em not} interpreted physically. Interpretation comes later (cf.~Section \ref{interp}). Thus it will help to think of a bare theory as given by an entirely abstract algebra of quantities, together with a similarly abstract state-space and dynamics. For example, the quantities might be (the self-adjoint elements of) an abstract C*-algebra $A$, the state-space might be determined by $A$, viz.~as the positive linear functionals on $A$, and the dynamics might be an arbitrary automorphism of $A$.    \\
\indent \indent (2): Indeed, it will help to think of a bare theory yet more generally. The reason is that most of what we need to say throughout Section \ref{schema}, about theories and accordingly about their models, is independent of taking a bare theory as a triple---even an uninterpreted one {\em a la} (1). It depends only on a bare theory having two features:\\
\indent \indent \indent (a) being uninterpreted, yet ready to be interpreted as a physical theory (hence the idea of the abstract set of states, quantities etc. `standing ready' for interpretation);\\
\indent \indent \indent  (b) being augmentable, i.e.~able to be supplemented with extra (again: uninterpreted) structure, in various ways, yielding different realizations, which we will call `models': (to which we will  turn directly, in Section \ref{models;was22B}).\\
Clearly, a theory does not need to be a triple  $\bra \cal S,Q,D\ket$ in order to have features (a) and (b). It could, for example, be a theory in logicians' traditional sense of a deductively closed set of formulas in a formal language: such a theory is uninterpreted, and can be augmented in many ways, for example just by adding an extra set of formulas and then closing under deduction. (Here we again connect with the Hilbertian and logical empiricist tradition of formulating physical theories axiomatically, mentioned at the start of Section \ref{intro}. 
But we must postpone discussion till Section \ref{RandI}.)

\subsubsection{Models}\label{models;was22B}

So we define a {\it model} $M$ of a bare theory $T$ to be a realization  of  $T$. The word `realization' can be taken in two senses:---\\
\indent \indent (i): In the broad sense of a mathematical instantiation: i.e.~a mathematical entity having the structure of the theory,  and usually of course some specific structure of its own (cf.~(2)(b) in Section \ref{bare;was22A}). Thus if  $T$ is a theory in the logicians' sense of a deductively closed set of formulas, a realization is an entity that in some sense `satisfies' all the formulas of $T$, and usually of course some formulas of its own. So any deductively closed superset of $T$ would count as a realization of $T$; but so of course would any model of $T$ in logicians' usual sense of `model'.  \\
\indent \indent (ii): In the mathematical sense of `representation', as in representation theory. This requires $T$ to have some structure, so that a representation is a homomorphism (of that structure) from $T$ to some given, structured object. Since homomorphism need not be isomorphism, the homomorphism's range---the  structured object that represents---may have only a `coarse-grained' version of $T$'s structure. But like in (i): since the representing object is `given', it will also have specific structure of its own. (Recall the two kinds, (A) and (B), of `extra detail' in Section \ref{introschema}.) Again, the obvious examples  are when $T$ is an abstract group or algebra, endowed with a set of rules for evaluating quantities, and a model is a group/algebra representation: for example a subgroup of the general linear group on a complex vector space, ~$\mbox{GL}(n,\mathC)$, endowed with a set of maps to the complex numbers, invariant under similarity transformations, e.g.~the group characters.

In our specific example of duality, bosonization (Section 4 et seq.), we will use the second more specific notion, i.e.~representation, (ii). But again: much of what we say in this Section needs only (i): the first, more general, sense of realization. And we believe that this notion applies more generally, to the dualities in quantum field theory and string theory: gauge-gravity duality (De Haro (2016a)), mirror symmetry, T-duality, and S-duality; cf.~Sections \ref{RandI} and \ref{abstraction}.

\subsubsection{Notations for models; model roots and model triples}\label{notation;was22C}

It is helpful to have a schematic notation for models that exhibits how they augment the structure of a theory with specific structure of their own. This will also introduce some jargon which will be important for us.

One's first thought is to write the model as the ordered pair of the theory and its specific structure, $\bar M$ say: $ M = \bra T, \bar M\ket$. But we need to reflect the fact that (in almost all cases) the way that a model incorporates the theory's structure is not by simply containing a `copy' of the theory `beside' its specific structure $\bar M$: but instead, by using $\bar M$ to build a realization---in one or other of Section \ref{models;was22B}'s two senses---of the theory's structure. Again, the obvious example of group representations illustrates. We should not think of a matrix representation of, say, the symmetric group $S_N$ as containing a copy of $S_N$ `beside' its specific structure of a vector space $V$ and $N!$   linear maps on $V$; (or maybe less than $N!$ maps---recall that a representation need only be a homomorphism). Rather, $V$ and the (upto!) $N!$ linear maps realize, give a `concrete copy' of, $S_N$. 

Similarly for  examples of dualities in physics, including our  example of bosonization. In  a typical physics example, the specific structure $\bar M$  consists of a set of fields, endowed with a set of symmetries, a dynamics for the fields, and a set of states of the fields. (So here, fields play the role of quantities in our conception of theories as triples, though of course not all fields are observable: in quantum theories, self-adjoint.) These fields etc. are used to build a `concrete copy' of the bare theory's structure (or maybe a `coarse-grained' copy). In our own example: a concrete copy of the enveloping affine Lie algebra or Kac-Moody algebra (see Eq.~\eq{ALA} in the Appendix), which is the algebra of which both the bosons and the fermions are representations.

  So a better notation that reflects how $\bar M$ is used to build a realization of $T$ is to write: $ M = \bra T_M, \bar M\ket$. The occurrence of $T$ in the notation encodes that the model $M$ is indeed a model of $T$.  But the subscript $M$ on $T$ reflects that the specific structure $\bar M$ is used to realize $T$. In short, $T_M$ is not `given before' $M$ itself: rather, $T_M$ realizes $T$---in one or other of Section \ref{models;was22B}'s two senses---by making use of $\bar M$.
  
    So one should not think of the model as an ordered pair made from two already-given items $T_M$ and $\bar M$. Rather, the decomposition $ M = \bra T_M, \bar M\ket$ is conceptual.\footnote{Besides, in physics examples the actual mathematical structure of a model is often very rich, e.g.~in gauge-gravity dualities the metric of a $(d+1)$-dimensional spacetime will belong to the specific structure $\bar M$, but this metric will be a {\it fibration} over the metric of a $d$-dimensional spacetime, which belongs to $T_M$: see De Haro (2016:~Section 2.1, 2016a:~Section 2.2).}
  
It will be convenient to have a word for $T_M$, the `part' of $M$ that realizes $T$. We call it the {\bf model root}. It will also be convenient to have a notation for the model root that does not mention $T$, just for simplicity in discussions where it is obvious that one theory  $T$ is in question. We use $m$. Thus a theory $T$ can have various models and model roots, $M_i$ and $m_i$, where $i$ is in some index set $I$. This notation $m$, i.e.~without mention of $T$, will be useful in Section \ref{RandI}.

This discussion carries over intact to the more detailed conception of a theory as a triple, $T = \bra \cal S,Q,D\ket$. We write a model, not as a pair, but as a quadruple:
\bea\label{modelq}
M=\bra{\cal S}_M,{\cal Q}_M,{\cal D}_M,\bar M\ket=:\bra m,\bar M\ket~,
\eea
where $m:= T_M := \bra {\cal S}_M, {\cal Q}_M, {\cal D}_M\ket$ will be dubbed the {\bf model triple}, as well as {\it model root}. As before, $\bar M$ is the specific structure that distinguishes one model, and so model root/triple, from another; and it is $\bar M$ that is used to build the model triple. 

This jargon of model root, and model triple, will be important in relation to our proposed schema for duality, that a duality is an isomorphism of models. For this isomorphism is of course isomorphism as regards the bare theory's structure, rather than any other structure: in particular, this isomorphism sets aside the specific structure (even though the model's realizing the bare theory is built from its specific structure). Accordingly, we will often talk of duality as an isomorphism of model roots; and also, when a theory is conceived as a triple, as an isomorphism of model triples.

Finally, it will also be convenient (especially in Section \ref{Symm}) to have notation for a model considered in itself, not by comparison with the bare theory of which it is a model. A model is of course {\em itself} also a triple of a set of states, quantities and a dynamics: i.e.~its own states etc., not that of the bare theory. And we will again use the overbar to indicate what is specific to the model. So we write:  $M = \bra {\cal {\bar S}}, {\cal {\bar Q}}, {\cal {\bar D}} \ket$. 

This prompts the question: what is the relation between the unbarred items  ${\cal S}, {\cal Q}, {\cal D}$ that make up a bare theory $T$, and the barred items ${\cal {\bar S}}, {\cal {\bar Q}}, {\cal {\bar D}}$? We will not attempt a general answer to this: we will not need one, and indeed we doubt that there is one. But bearing in mind the prototypical cases of representations of groups or algebras (e.g.~representations of the Poincare or conformal algebras which add a specific field content), the tempting, broad answer is that the barred items ``are bigger''/``have more structure''. (This is another aspect of the analogy between duality and gauge, despite the contrasts we stressed in remark (4) at the end of Section \ref{orient}:  both for two dual models and for a gauge formulation of a theory, there is the intuition of ``being bigger''/``having more structure''.)

In the example of boson-fermion duality, as we will see in Sections \ref{fmB} and \ref{fdirac}, the theory consists of a specific enveloping algebra of the affine algebra (see Eq.~\eq{ALA} in the Appendix), together with a representation space for this algebra and appropriate functionals to the real numbers, which represent the quantities that (in the interpreted theory) become the physical quantities. But the models contain a complex variable $z$, which represents two-dimensional spacetime. Using this variable, one can construct, from the operators of the algebra, the bosonic (Klein-Gordon) or fermionic (Dirac fermion) field of the bosonic and the fermionic model, respectively. We shall argue in Section \ref{abstraction}:(c) that, at the level of the bare theory, the specific dependence of the fields on the spacetime variable $z$ has no particular significance, and is not needed: so that this suggests the formulation of a more general theory, with less structure.

\subsection{Interpretations of theories and models}\label{interp}

So far, so abstract! Both theories and models have so far been bare, i.e.~uninterpreted. We now sketch how we envisage their physical interpretations.  We will adapt the elementary ideas of the Frege-Carnap-Lewis framework for semantics (Frege 1892, Carnap 1947, Lewis 1970). This will be easy work: two interpretation maps,
$I_{\tn{Int}}$ and $I_{\tn{Ext}}$, will map from our theories and models, to `meanings' and to `the world' respectively. These maps will later be useful for discussing symmetries (in Section \ref{symminterp:was25B}).  \\

We recall that according to the Frege-Carnap-Lewis framework:---\\
\indent \indent (i) A word gets assigned: first, an intension (Carnap's word: Frege's is `Sinn', or in English, `sense': roughly, `linguistic meaning'); and second, an extension (Carnap's word: Frege's is `Bedeutung', or in English, `reference/referent': roughly, `the object or worldly item mentioned'). Roughly speaking: our interpretation maps, $I_{\tn{Int}}$ and $I_{\tn{Ext}}$, will assign intensions and extensions, respectively. \\
\indent \indent (ii) A word's intension is assigned to it, once and for all (making the simplifying idealization that all words are univocal, and their linguistic meanings do not change). But  a word's extension is assigned to it, relative to a possible world and to other features of the context of use that together determine the reference. For example, the reference of `the tallest Swede alive today' depends not just on the possible world, but also on the day of use. And in general, the set of features that together determine the reference is large and open-ended (cf.~Lewis 1980). \\
\indent \indent (iii) A singular term (such as a proper name `Aristotle' or a definite description `the capital of Denmark') has as its extension, its bearer (in these examples: the man, the city Copenhagen); while a one-place predicate (such as `...walks') has as its extension its set of instances (the set of walkers at the possible world and time in question); while similarly, a two-place predicate (such as `... loves ...') has as its extension {\em its} set of instances, i.e.~the set of ordered pairs where the first loves the second (at the possible world and time in question), e.g.~$\bra {\mbox{Romeo, Juliet}} \ket$; and so on for  predicates with three or more places.    \\
\indent \indent (iv)  Compositional rules describe how to assign intensions and extensions to grammatical phrases and thus to complete sentences, in terms of the intensions and extensions of the component words. For example, `John walks'  is assigned, at a world $W$ and time $t$, the extension True (as against False) iff the reference of `John' for $(W,t)$ is in the set of walkers for $(W,t)$.\\   

To adapt these ideas, (i) to (iv), to our theories and models, there are two points to bear in mind. They seem to be stumbling blocks, or sources of confusion. But they are easily surmounted and dispelled.\\
\indent \indent (a): Notice how the ideas above express {\em contingency} and {\em transience}: by postulating a background set of possible worlds and times, they secure that a sentence's truth-value can be contingent (vary across the worlds) and transient (vary across the times). In philosophical discussion of physics, contingency and  transience are often expressed in a corresponding way: the theory has many solutions, and typically a solution changes with time. This is often expressed using the word {\em model}: a state at a time, or a temporal sequence of states (a trajectory through the state-space) is called `a model of the theory'. Thus recall that this was connotation (i) at the start of Section \ref{introschema}. The stumbling block is of course that since Section \ref{intro}, we have reserved `model' for a very different use: for what many would call `specific theory', i.e.~for a notion that encompasses many solutions throughout time. But we take it that one can surmount this stumbling block, and avoid confusion, just by recognising our stipulated usage.\\
\indent \indent (b): The ideas, (i) to (iv), were of course developed to give semantics for language about ordinary objects, such as people and towns, like Aristotle, Romeo and Copenhagen. But when one considers one of Section \ref{TandM}'s theories or models, one is hard pressed to find mention of objects: at least, of objects in the plural. For undoubtedly, `most of the talk' in the theory or model is about the various states and quantities, about which so many details are given. But these are surely not objects, but rather properties. Namely, properties of the one object---the physical system itself---being theorized about. Agreed: if the system is composite, one naturally regards its component systems as objects in their own right.\footnote{Also agreed: it is common, and  mathematically natural, to consider the set $\cal S$ of all states, with quantities as extra structure on $\cal S$: e.g.~in classical mechanics, as real-valued functions on the phase space $\cal S$, and in quantum mechanics, as linear operators on the Hilbert space $\cal S$.  But this does {\em not} make it compulsory to treat states as the basic objects in a semantics of a physical theory. For it is equally legitimate, though less common in textbooks, to start with the set  $\cal Q$ of quantities, and take states as extra structure on $\cal Q$. And the legitimacy of both these approaches shows that {\em au fond}, a state is an assignment of numerical values to all quantities; and {\em mutatis mutandis} a quantity is an assignment of numerical values to all states. For now, the point is just that if one is asked to classify states and quantities in either of the philosophical categories of `object' and ` property', undoubtedly one should classify both states and quantities as properties.} But the main point remains: most of the talk in a theory or model is about (numerically quantifiable) properties, and their intricate (quantitative) relationships, not about objects. But this second stumbling block is, like the first, minor. For nothing prohibits interpretations using just one object, viz.~the system; or just the system and its component subsystems. (Besides, though we will not go into details: nothing prohibits interpretations treating as objects what are in fact properties; cf.~e.g.~Lewis (1970a: 429).)\\

Bearing in mind (a) and (b), we can now spell out how interpretation maps, $I_{\tn{Int}}$ and $I_{\tn{Ext}}$, assign intensions and extensions, respectively---once we are given a bare theory $T$, or a bare model  $M$. We will discuss the assignment to an element of the set of quantities: i.e.~for $T$, an element of $\cal Q$ in the triple $T = \bra {\cal S}, {\cal Q}, {\cal D} \ket$; and for $M$, an element of ${\cal {\bar Q}}$ in the triple $M = \bra {\cal {\bar S}}, {\cal {\bar Q}}, {\cal {\bar D}} \ket$. (Recall from the end of Section \ref{notation;was22C} that ${\cal {\bar Q}}$ is all the quantities in the model, and is intuitively `larger' than ${\cal Q}_M$, which is the realization of $T$'s ${\cal Q}$ in the  quadruple $M = \bra {\cal S}_M, {\cal Q}_M, {\cal D}_M, {\bar M} \ket$.) But to stress that this element is uninterpreted/abstract, we will call it $a$, not $Q$. It will be obvious how the corresponding assignments get made for an element interpreted as a state, or as a dynamics.  

For $I_{\tn{Int}}$, the idea is: $I_{\tn{Int}}$ assigns to an element $a$ of $T$ or $M$, a physical quantity understood in general terms. For example, $a$ could be an element of an abstract C*-algebra, abelian for classical mechanics and non-abelian for quantum mechanics. And $I_{\tn{Int}}(a)$ could be the quantity, position: which in philosophical terms, is a property with numerically measurable degrees. Or perhaps $I_{\tn{Int}}(a)$ is, more specifically, position in the $x$ direction, using such-and-such point as spatial origin, and using axes and length-unit thus and so. 

Two comments are in order, here. First: We thus envisage a `Platonic realm' of numerically measurable properties as the codomain of the function $I_{\tn{Int}}$. But our `realism' about quantities is milder than it might appear; and anyway, nothing in what follows will depend on it. In particular, (i): nothing will depend on our taking intensions to encode conventional choices such as spatial origin, axis-direction and length-unit. Besides, (ii): we do not need `trans-theoretic identity' for quantities. That is, a quantity like position need not be `the very same quantity' in different theories: especially if they are radically different, e.g.~classical mechanics and quantum mechanics. Despite the single word `position', $I_{\tn{Int}}(a)$ can be different  quantities, according as $a$ is in an abelian, or non-abelian, C*-algebra.

Second: Note that we do not need to be precise about the exact domain of definition of $I_{\tn{Int}}$. For we are only sketching how we envisage interpretation proceeding; and there is of course a great deal of convention about how to formalize both $T$ and $M$, and thus about what is the exact domain of definition of $I_{\tn{Int}}$. 
 
For $I_{\tn{Ext}}$, the idea is similar, except that we need to allow for the fact that extensions are assigned relative to   a possible world and to all the other features of the context of use that together determine reference. Interpreting any theory or model $T$ or $M$ means taking it to be used to describe some empirical phenomena: i.e.~taking there to be, in an appropriate possible world, a context of use with a rich enough set of features to determine reference for the elements of $T$ or $M$.\footnote{We say `appropriate' so as to signal that of course, for any $T$ or $M$, not every possible world has a context rich enough to determine reference for all $T$'s or $M$'s elements. Indeed: for many worlds, all their contexts will determine reference for none of $T$'s or $M$'s elements. For example, take  $T$ or $M$ to be supersymmetric theories, and a world with no supersymmetric physics. So whatever our precise definition of the domain (i.e.~set of arguments) of the map $I_{\tn{Ext}}$, the map will surely be partial, i.e.~undefined on some, maybe the majority, of its arguments. But that is no problem. Formal semantics and philosophy of language in the Frege-Carnap-Lewis framework have long had various proposals for how to treat words and phrases that lack extensions (called `bearerless terms'); and these proposals can be adapted to our  $T$ or $M$.\label{partialfunctions}} 

 For example, let $a$  be again an element of an abstract C*-algebra: say, an abelian one, because $T$ is a bare theory that is to be interpreted as classical mechanics. Suppose that the possible world $W$ contains two classical point particles; and we take $T$ to be used in a context sufficiently rich that $a$ successfully refers to the position of the more massive particle; or perhaps more specifically, its position in the $x$ direction, using such-and-such place in $W$ as spatial origin, and using axes and length-unit thus and so.  Then relative to $W$ and this assumed context of use, $I_{\tn{Ext}}(a)$ is defined to be: the heavier particle's position. 

\subsection{Isomorphisms: defining theories by abstraction from models}\label{RandI}

So far, we have taken a theory $T$ as given, and then considered its models $M$. In this Subsection, we note that one can argue in the opposite direction: i.e.~one can approach defining a theory starting from a class of models. The idea is a widespread one: to define a notion (here, a theory) as those features in common between a suitably varied class of examples (here, models).\footnote{This line of thought is not only widespread, but has a long tradition: for thousands of years in philosophical accounts of abstraction; and for a hundred and fifty years in mathematics, with e.g.~Frege's proposal to define notions as equivalence classes (e.g.~a direction as an equivalence class of parallel lines).} This is what earlier we called the `common core' of the models (in Sections \ref{introschema}-(c), and \ref{bosointro}, \ref{orient}-(3)). We will begin within our previous perspective, i.e.~with a theory as given, to introduce notation; then we will sketch how to define a theory, `arguing in the opposite direction'.

So recall from Sections \ref{models;was22B} and \ref{notation;was22C} that a bare theory $T$ can have various models and model roots, $M_i$ and $m_i$, where $i$ is in some index set $I$. These `realize' the theory, in the senses (`instantiate' or `represent') discussed in Section \ref{models;was22B}. But they are in general  not isomorphic to each other, nor to the theory: we in general do not have $m_i\cong m_j\cong T$. And even if there {\it is} an isomorphism: it is not an {\em identity}, because each $m_i$ is a realization of $T$ built using the model's specific structure $\bar M_i$: it is not a `pure copy' of $T$.

Supposing there is an isomorphism between a model root $m_i$ and the theory $T$, we denote it by $f_i:m_i\rightarrow T$, $i\in I$. Of course, $m_i$ is in general not a single set, but an $n$-tuple or family of sets: often endowing some base set in the family with structure, e.g.~the structure of an algebra.  In particular, with a theory taken as a triple, $T = \bra \cal S,Q,D\ket$, each  $m_i$ is a triple $m_i = \bra {\cal S}_{M_i}, {\cal Q}_{M_i}, {\cal D}_{M_i} \ket$. So the domain and codomain of $f_i$ are triples, so that $f_i$ is actually a triple of maps that map respectively:  states to  states, quantities to  quantities, and dynamics to dynamics.\footnote{{\label{triplenote}}{The first two maps will be isomorphisms, the last an equivariance condition: we will say more about this in Section \ref{DlyIsom}.}} We will, however, not need to indicate this in our notation. 

Likewise, let the maps $f_{ij}:m_i\rightarrow m_j$ ($i,j\in I$) denote isomorphisms---when they exist---between two model roots. But again, our notation will not need to indicate the precise domains and codomains of these maps. 


Now in the opposite direction. Suppose we are given, not a theory $T$, but a set of models indexed by $I$: we write this set as $\{M_i : i \in I \}$. Similarly, we write the set of model roots  indexed by $I$ as $\{m_i : i \in I \} $. Thus we are not assuming that these models, or these model roots, are given as realizing a theory. But we do assume that the model roots (and the models) have some kind of structure, so that it makes sense to say that some pair of model roots is isomorphic with respect to that kind of structure. The idea now is to define a theory as `what is in common' among a suitable set of model roots; where `what is in common' will be expressed as an equivalence class of an appropriate equivalence relation. 

The most obvious implementation of this idea is to define equivalence as just the given notion of isomorphism that we assumed applies to the model roots. Thus we might define two model roots $m_i$, $m_j$ ($i,j\in I$) to be equivalent, $m_i \sim m_j$, just in case there is an isomorphism $f_{ij}:m_i\rightarrow m_j$ between them. 
Then the proposal would be: a theory is an equivalence class under this equivalence relation. Recall for example the way in which Frege (1884: Sections 64-67), defined a {\em direction} as an equivalence class of straight lines under the relation of being parallel.

But if we apply this proposal to model roots, it has the trivial consequence that all model roots of a theory thus defined {\em are} isomorphic. And as we have discussed above, we must allow  a theory to have non-isomorphic model roots.\footnote{Similarly, if we apply this proposal to models: all models of a theory thus defined {\em are} isomorphic. Of course: we expect that since a model $M$ has specific structure $\bar M$ going beyond its model root $m$, isomorphism for models will in general be stronger---i.e.~lead to smaller equivalence classes---than does isomorphism of model roots. But in this Section we will not need  to linger on this model vs.~model root contrast. 
For our main concern is defining a theory using isomorphism of model roots. As we will argue below (at the end of this Subsection), there is a natural constraint that model roots must be {\em sufficiently varied}. For mistaking the presence of accidental similarities between the model roots one happens to have at hand for necessary similarities between {\it all} the model roots of the theory one is trying to define, leads to unnecessarily, or perhaps undesirably, restrictive theories. This is {\it a fortiori} true of the models $M_i$ of the theory: since the specific structure $\bar M_i$ is {\it specific} to $M_i$, and so in general not shared with the another model.}

So the obvious implementation of the initial idea stumbles. If we want to define a theory as `what is in common' among model roots, and express `what is in common' as an equivalence class, then we need a more judicious---no doubt, a weaker---choice of equivalence relation than isomorphism. We will not go into details about how to make this choice. We doubt that there are general rules. But we note that it will be guided by two  desiderata:\\
\indent \indent (i):  our previous understanding (maybe partial understanding) of the theory we are trying to define: the  choice is meant to pick out just the structure of the theory we intuitively intend; \\
\indent \indent (ii): our expectation that  the model roots will end up being representations, in the mathematical sense, of the theory thus defined (and so, in general, non-isomorphic to each other).\\

We will close by mentioning another constraint. It is partly independent of the question of choosing an equivalence relation. For it is about the membership of the class of models or model roots one begins with, rather than the judicious choice one must make of an equivalence relation over it. 

Namely:  if this whole approach to defining theories is to work---i.e.~is to define theories of  the kind  we intended in our original conception---then the model roots must be sufficiently varied that there are no `accidental commonalities' between them, which would then be inadvertently encoded in the theory defined as an equivalence  class---thereby limiting the theory's possibilities of representation. The point is familiar, e.g.~from Frege's example. If one imagines the lines can be coloured, then Frege's approach to defining direction needs the lines to vary in colour sufficiently. For if all the lines in a given parallelism equivalence class were the same colour, Frege's definition of the corresponding direction, viz.~as that class, would inadvertently be ambiguous between (i) the direction, as we originally intended it, and (ii) the unintended common colour. Hence our mentioning, in the opening paragraph of this Subsection, `a suitably varied class of examples'.

We will not go into details about this constraint. But we stress that obviously, it is substantive. For recall how on the original `theory-first' approach in  Section \ref{TandM}, we stressed that a model root $m = T_M$ is not a `pure copy' of the theory $T$, but is built from the model $M$'s specific structure $\bar M$. So on the present `reverse', or `models-first', approach: the danger is that  if a class of model roots is not sufficiently varied, they may have considerable specific structure in common (like colour for Frege's lines)---which will therefore be inadvertently encoded in theories defined as equivalence classes of model roots. We will come back to this point in Section \ref{abstraction}.\footnote{Agreed: one does not always get to `choose' one's model roots (or models), and so this constraint cannot always be implemented. Thus there is judgment involved in this process of abstraction, viz.~of (i) how many, and how varied, the model roots should be, to provide representations of one's theory, and (ii) how to make the distinction, for a given model, between model root and specific structure (since part of the specific structure of a model could be mistaken for e.g.~additional information about the theory). We therefore maintain that this reverse approach, from model roots to theory, is not deductive but inductive---which brings us back to our Hilbertian theme from Section \ref{intro}. It only stops when one is happy with the theory---based on whatever independent criteria one uses to judge one's theory and models.}

\section{Duality and Symmetry}\label{symdual}

In this Section, we first develop our treatment of symmetry (Section \ref{Symm}). This, together with our discussion in Section \ref{schema},  sets us up to present (at last!) our schema for duality as isomorphism of models (Section \ref{DlyDef}).

\subsection{Symmetries of theories and their models}\label{Symm}

We mentioned symmetry in Section \ref{introschema}'s closing analogy between a symmetry mapping a state to a state that is `the same', and a duality mapping a theory to a theory that is `the same'. But we now can say more about symmetries, using: (A) our distinction between theories and models (from Section \ref{TandM}); and (B) our interpretation maps (from Section \ref{interp}).  We take up these topics in Sections \ref{symmthies:was25A} and \ref{symminterp:was25B}, respectively. 

About (A), our main point will be that the symmetries of a bare (or indeed, an interpreted) theory, and the symmetries of a model of it, are in general overlapping, but distinct, sets. In particular, the symmetries of a theory  can be a proper subset of the symmetries of its model: and beware---this inclusion is in the opposite direction from that for the other, more common, use of `model', viz.~as a solution, or representative of a solution, of a theory. 

 About (B), our main point will be that a symmetry must `commute' in an appropriate sense with interpretation. Both these points, and our other ones, will be uncontroversial. 

\subsubsection{Symmetries of uninterpreted theories and models}\label{symmthies:was25A}

Recall the usual conception of symmetry as a map $a$ on states that preserves the values of a salient, usually large, set of quantities: the state $s$ and the image-state $a(s)$ have the same values for quantities. This prompts three immediate comments. \\

\indent \indent  (i): Agreed, it is also usual to think of a symmetry as a map on quantities that preserves values on a salient, usually large, set of states: i.e.~for a given state, the value of the argument-quantity equals the value of the image-quantity. But there is no conflict here: the two conceptions are related by duality---in the mathematical, not physical, sense! That is: one map is the (mathematical) dual of the other.  

In more detail: given any map $a: {\cal S} \rightarrow {\cal S}$, we can define its {\it dual map} (not to be confused with a `duality map!') on quantities, $a^*: {\cal Q} \rightarrow {\cal Q}$, by requiring that for any $s \in {\cal S}$ and $Q \in {\cal Q}$: $\bra a^*(Q), s \ket := \bra Q, a(s) \ket$. And similarly, starting with quantities: given any map $a: {\cal Q} \rightarrow {\cal Q}$, we say that its dual map on states, $a^*: {\cal S} \rightarrow {\cal S}$, is defined by requiring for all arguments: $\bra Q, a^*(s) \ket := \bra a(Q), s \ket$. \\
\indent \indent (ii): Recall the question at the end of Section \ref{introschema}: do a state $s$, and its image $a(s)$ under a symmetry, represent the same physical state of affairs (`possible world')? Our answer there was, roughly: `In general, No: but the `Yes' cases are a natural focus of interest'. We will return to this when discussing interpretation, in Section \ref{symminterp:was25B}.\footnote{For the moment, we just note that it is also common to think that a symmetry as a map on states is `active', i.e.~the image-state must be a different  physical state of affairs (so the question's answer is `No'), while a symmetry as a map on quantities is `passive', i.e.~the image-quantity and the argument-quantity (each with their common value) describe the single given physical state (so that now the question's answer is `Yes'). 

We will {\em deny} this. There is no universal association of symmetry as a map on states as `active', and symmetry as a map on quantities as `passive'. The reason lies, essentially, in the distinction between a mathematical state and a physical state: (in the jargon of `gauge', the latter is a gauge-equivalence class of the former). That is: we of course concede that a symmetry as a map on states is `active', in the sense that it changes the states. That is a tautology: (except for the degenerate case where the symmetry is given as being the identity map!). But this concession does not imply that  a symmetry as a map on states must change the physical state of affairs represented: for the states in question could yet be  `merely' mathematical. That is: one still needs a further argument why a difference of these states must imply a difference of physical state (and thus why the question's answer is `No'). This distinction, between a mathematical and a physical symmetry, was labelled, in De Haro et al.~(2017:~Section 2), with the label (Redundant); and in De Haro (2016a:~Section 1.1.2.b), as (Physical)-(Redundant). It also roughly corresponds to the distinction, in Caulton (2015), between an `analytic' and a `synthetic' symmetry.}     \\
\indent \indent (iii): We have discussed symmetries as preserving values. But it is common to also require that a symmetry  `preserves the dynamics'. Taking a symmetry as a map on states, this means, roughly: if a sequence of states is possible under the dynamics, so is the sequence of image-states. That is: if a possible time-evolution is represented by a temporal sequence of states, with the state at time $t$ being $s(t)$ (a `Schr\"{o}dinger picture' of time-evolution), then the sequence $a(s(t))$ of states is also possible.\footnote{If the dynamics is deterministic, we can write $s(t) = D_{t, t_0}(s(t_0))$ where $D_{t, t_0}$ represents the deterministic dynamics; and then `preserving the dynamics' is equivalent to the commutation i.e.~equivariance condition, $a(s(t)) \equiv a(D_{t, t_0}(s(t_0))) = D_{t, t_0}(a(s(t_0)))$.\label{dynsequivariance}} A corresponding definition can be given for when we take a symmetry as a map on quantities, and use a `Heisenberg picture' of time-evolution as given by a sequence of quantities, i.e.~by the sequence of their values on a single `fixed' state.\\

These comments (i)-(iii) bring out two points. The second is longer: it addresses symmetries of {\em models}, pointing out that  these are in general overlapping but distinct (as abstract groups) from symmetries of theories; and that in the special case where a model triple is isomorphic to the bare theory, the symmetries of the theory can be a proper subset of the symmetries of its model.\\ 

 First: it is clear that discussing symmetries returns us to Section \ref{TandM}'s more detailed conception of a theory, even a bare one, as a triple comprising a set of states, a set of quantities and a dynamics: $T:=\bra \cal S,Q,D\ket$, together with a set of rules for evaluating quantities. \\

Second: it is clear that comments (i)-(iii) carry over exactly to {\em models} in our sense, viz.~a realization (`a more detailed version') of a theory $T$, with specific structure of its own. Recall that Section \ref{notation;was22C} introduced two notations for models in this sense. Both notations will be useful in what follows: the first notation immediately, the second notation in the next Subsection. (The second notation was simpler: since a model is  itself a triple of its own sets of states, quantities and a dynamics, we wrote: $M = \bra {\cal {\bar S}}, {\cal {\bar Q}}, {\cal {\bar D}} \ket$. It means we can define and discuss `symmetry of a model' just as we did symmetries of theories, e.g.~as an automorphism of the state-space ${\cal {\bar S}}$ preserving values of a large salient subset of ${\cal {\bar Q}}$.)

Section \ref{notation;was22C}'s first notation distinguishes the realization of the theory's triple from the specific structure $\bar M$, and gives a subscript $M$ to the former to signal that it is built out of the latter: $M=\bra{\cal S}_M,{\cal Q}_M,{\cal D}_M,\bar M\ket$ (Eq.~\eq{modelq}). We also wrote this as $M =:\bra m,\bar M\ket~,$ where $m:= T_M := \bra {\cal S}_M, {\cal Q}_M, {\cal D}_M\ket$ is the {\em model triple}. This  notation brings out that for any theory $T$ and any of its models $M$, there is a natural condition for a symmetry $a$ of $T$ to be itself realized in $M$: for it to have, so to speak, a `shadow' in the model $M$, i.e.~in the model triple. This condition is that a diagram should commute, and is not automatic. \\
\indent To state this condition, however, we need a bit more notation about realization. We will write it as a map  $\theta$. And we will suppose that in $T$ we treat symmetries as maps on states, so that $a: {\cal S} \rightarrow {\cal S}$ preserves the value of all quantities in a salient subset, say ${\cal Q}^0$, of the set of all quantities ${\cal Q}$. Then in the usual case where `realization' means `representation', we can take $\theta$ as an appropriate structure-preserving map: from ${\cal S}$ in the theory $T$ itself, to ${\cal S}_M$ in the representing model triple $m = T_M = \bra {\cal S}_M, {\cal Q}_M, {\cal D}_M\ket$. Then the condition in question---that the symmetry $a$ is itself realized in $M$---is that there should be a map $a_M: {\cal S}_M \rightarrow {\cal S}_M$, such that the diagram in Figure \ref{commuteformodeltripleautomm} commutes.
\begin{figure}
\begin{center}
\bea
\begin{array}{ccc}{\cal S}&\xrightarrow{\makebox[.6cm]{$\sm{$a$}$}}&{\cal S}\\
~~\Big\downarrow {\sm{$\th$}}&&~~\Big\downarrow {\sm{$\th$}}\\
{\cal S}_M&\xrightarrow{\makebox[.6cm]{$\sm{$a_M$}$}}&{\cal S}_M\nonumber
\end{array}\nonumber
\eea
\caption{Commutativity diagram of the symmetry $a$ with the representation map $\th$.}
\label{commuteformodeltripleautomm}
\end{center}
\end{figure}

To convey this idea less abstractly, think of the simplest case. Let the bare theory be just a group $G_1$,\footnote{Together with a set of maps to the real numbers, to express evaluation of the quantities. But for simplicity we ignore these maps for the moment.} with an automorphism $a_1: G_1 \rightarrow G_1$; and suppose a group $G_2$ represents $G_1$ thanks to the existence of a homomorphism $\theta: G_1 \rightarrow G_2$. So  $G_2 \cong G_1/\mbox{ker}\,\theta$. For there to be a homomorphism of $G_2$, $h: G_2 \rightarrow G_2$ (even homomorphism: let alone automorphism), that realizes $a_1$ (counts as $a_1$'s `shadow'  in $G_2$) requires commutation: i.e.~for all $g_1 \in G_1, \theta(a_1(g_1)) = h(\theta(g_1))$. Or as a diagram, see: Figure \ref{commuteforelmtarygroups1}.
\begin{figure}
\begin{center}
\bea
\begin{array}{ccc}G_1&\xrightarrow{\makebox[.6cm]{$\sm{$a_1$}$}}&G_1\\
~~\Big\downarrow {\sm{$\th$}}&&~~\Big\downarrow {\sm{$\th$}}\\
G_2&\xrightarrow{\makebox[.6cm]{$\sm{$h$}$}}&G_2
\end{array}\nonumber
\eea
\caption{Commutativity of group automorphism $a_1$ with group homomorphism $\th$.}
\label{commuteforelmtarygroups1}
\end{center}
\end{figure}

In the special case where the model triple is isomorphic to the bare theory, this discussion of course simplifies. Then the map $\theta$ will be an isomorphism, and the map $a_M$ (or $h$ in the toy example of groups) will trivially exist, and both the above diagrams will trivially commute.
In this case, a symmetry of the theory has a `duplicate', or `replica', in the symmetries of the model---so that in effect, the symmetries of the theory are a {\em subset} of the symmetries of the model. We say `in effect' just because of the different domains of definition: ${\cal S}$ vs.~${\cal S}_M$. Apart from this `in effect', there are two comments to make about this special, simplified, case.  

(1): First, we note that on the other use of `model' as an individual solution of a theory, a model is  in general {\em less} symmetric than its theory---as is often remarked, with the buzz-word `symmetry-breaking'. A solution of a dynamics with a spherically symmetric Hamiltonian need not be spherically symmetric; a cubical crystal lattice with one particular placing of its lattice points, and one particular orientation of its edges, can be a solution of a dynamics that is translation-invariant and isotropic; and so on. So: the subset-inclusion in our case above is in the opposite direction from that holding for the other use of `model'.

(2): Besides, `subset' here will usually mean {\em proper subset}. That is: a model's specific structure $\bar M$---its `content' that goes beyond its being a model/realization of $T$---will mean the model has symmetries  additional to those that are `duplicates' of the symmetries of $T$. And we expect that if these additional symmetries are well-defined on the model triple, or if they naturally induce a symmetry there, that symmetry is trivial, i.e.~just the identity map on the model triple. Our prototypical cases of representations of a group or algebra give examples. Perhaps the simplest is as follows. Let $T$ be the real numbers $\mathR$; and let $M$ be the complex numbers $\mathC$ which of course represents $\mathR$ as the real axis, i.e.~the complex numbers with zero imaginary part,  $\{ z \in \mathC \, | \, z = x + i0, \, x \in \mathR \}$. So this latter set, the real axis, is like the model triple. Then $M$ has the symmetry of complex conjugation $z \mapsto {\bar z}$ which is indeed well-defined on the real axis: but there, it is just the identity map. 

And there are examples in interesting cases of dualities. In gauge-gravity dualities, De Haro (2017a) showed that a certain subgroup of the diffeomorphism group of the gravity model of the theory (roughly, the diffeomorphisms which preserve the asymptotic boundary conditions) was `invisible' to the gauge model of the theory, in the sense of not representing any difference on that model: and so these diffeomorphisms are not in the common core between the two models, and they are trivially represented on the theory. The same verdict was made in De Haro (2016a:~\S2.2.3) for the `gauge symmetries' of the gauge side of the duality. These are not visible on the gravity side: they are symmetries of the formulation of the gauge model of the theory, and are trivially represented on the theory. \\

   To sum up this discussion of the symmetries of a bare theory, and those of its models, and of its model-triples: there are really three points here:\\
\indent \indent (i):    A bare theory $T$ is realized---typically: represented in the mathematical sense--- by one of its model triples, $m$. The model $M$ then consists of  $m$ and some specific structure $\bar M$; (cf.~Section \ref{notation;was22C}). And representation requires only a homomorphism, not an isomorphism. Hence our articulating in this Section the condition---in terms of a commuting diagram---for a symmetry of $T$ to be itself realized in $m$. \\
\indent \indent (ii): And even if in some given case, the representation is an isomorphism, i.e.~the representing model triple is isomorphic to the theory $T$, so that any symmetry $a$ of $T$ will indeed have a `duplicate' or `replica' symmetry in the model triple: still, we must expect that the {\em model} (as against the model triple) has its own specific structure $\bar M$. And this specific structure may have symmetries that $m$, and the theory $T$, `knows nothing of': (cf.~comment (2) just above).\\
\indent \indent (iii): Furthermore, different models, and therefore model triples, of a bare theory are in general {\em not isomorphic} (as we also discussed in Section \ref{RandI}). However, our example of boson-fermion duality, in Section \ref{bfd}, will not illustrate this in detail: i.e.~all the model triples {\em will} be isomorphic. (But see the comments in Section \ref{abstraction}.) 

\subsubsection{Interpretations respect symmetries}\label{symminterp:was25B}

Recall from Section \ref{interp} that once we are given a bare theory $T$, or a bare model  $M$, the interpretation maps $I_{\tn{Int}}$ and $I_{\tn{Ext}}$ assign intensions and extensions (respectively) to, for example, an element $a$ of the set of quantities: i.e.~for $T$, an element of $\cal Q$ in the triple $T = \bra {\cal S}, {\cal Q}, {\cal D} \ket$; and for $M$, an element of ${\cal {\bar Q}}$ in the triple $M = \bra {\cal {\bar S}}, {\cal {\bar Q}}, {\cal {\bar D}} \ket$. (And similarly for assigning intensions and extensions to states and dynamics.) Thus $I_{\tn{Int}}$ assigns to an element $a$ of $T$ or $M$, a physical quantity, e.g.~position (or more specifically, position in the $x$ direction, using such-and-such spatial origin), understood in general terms. Similarly, $I_{\tn{Ext}}$ assigns $a$ an extension relative to a possible world $W$ and the other features of the context of use that together determine reference. For example, $W$ and the  context of use may determine that $a$ is assigned the position of the more massive of two classical point particles that are in $W$. 

We now propose that these interpretation maps should satisfy appropriate meshing conditions whereby they form commuting diagrams with symmetry maps. The reason is simply that this reflects what one usually means by `interpretation' of the formalism of a physical theory. And this is so whether `formalism of a physical theory' is (in our senses) a bare theory, or a bare model; and whether `interpretation' refers to intension or extension.  

Thus to take a very simple example: suppose that a state describes three classical point particles forming a scalene triangle, stationary in an absolute Newtonian space, and  (for more simplicity) that the particles do not interact; and suppose the bare theory, or model,  at issue has spatial translation and rotation as symmetries. Taking symmetries as maps on states (as usual): these suppositions mean that a spatially translated and-or rotated state has the same values as the given state, for a large and salient set of quantities. So far, these suppositions are, officially, at the level of a bare theory or model---though of course words like `point particles',`scalene triangle' and `Newtonian space' suggest interpretation. And indeed: reading these suppositions, one tends to read them as interpreted, i.e.~to unconsciously apply the interpretation maps $I$---where $I$ is short for both $I_{\tn{Int}}$ and $I_{\tn{Ext}}$. In any case: applying these maps to the two bare states, say $s$ and $T(s)$ (`T' for `transform' or `translate and-or rotate'), one concludes that for the act of interpretation to respect the given symmetries,  the interpretations $I(s)$ and $I(T(s))$ (where again: $I$ is short for both $I_{\tn{Int}}$ and $I_{\tn{Ext}}$) must have the same values for a large and salient set of quantities---namely, of course, for the interpretations of the bare quantities. This is exactly the condition that symmetry and interpretation form a commuting diagram. But to write such diagrams down, we will need a bit more notation.\footnote{Agreed, to impose this commutation condition for every symmetry and every interpretation is contentious. It seems best justified when we envisage that the bare theory or model describes the whole universe; so that for the example of  three point particles, there are no other material bodies in the universe. But in this paper, we do not need to assess exactly when the commutation condition is justified. See De Haro (2016a:~\S1.3-\S1.4), especially the condition called `unextendability'.}

The reasons we need more notation are as follows. So far:\\
\indent \indent (1): We did not spell out that the domain of each of $I_{\tn{Int}}$ and $I_{\tn{Ext}}$ would include states and quantities and dynamics; so that each of $I_{\tn{Int}}$ and $I_{\tn{Ext}}$ is really a triple of three maps from states to states, from quantities to quantities, and from dynamics to dynamics. (This is just like an isomorphism $f_i$ in Section \ref{RandI} being really a triple of such maps; cf.~footnote \ref{triplenote}.)\\
\indent \indent (2): We have not introduced notation for the codomains of the interpretation maps: for what one might call the `realm of intension', or `meanings', for $I_{\tn{Int}}$, and for what one might call  the `realm of extension', or the `world', for $I_{\tn{Ext}}$.\\
\indent \indent (3): Nor did we introduce a notation for symmetry maps defined on the `realm of intension' or on the `realm of extension'. Indeed, we did not yet do this: neither (a) on the states therein (generally understood, in the realm of intension, and specific to a particular physical system, in the realm of extension), analogous to the symmetry maps $a: {\cal S} \rightarrow {\cal S}$ on the states of a bare theory; nor (b) on the quantities therein (generally understood, in the realm of intension, and specific to a particular physical system, in the realm of extension), analogous to the symmetry maps $a^*: {\cal Q} \rightarrow {\cal Q}$ on the quantities of a bare theory (which are mathematical duals of the maps $a: {\cal S} \rightarrow {\cal S}$).\\

To avoid a lot of extra notation, we shall only act on (2) and (3) above. That is: \\
\indent \indent (2'): We now denote the codomains of the interpretation maps $I_{\tn{Int}}$ and $I_{\tn{Ext}}$ by, respectively: `Sinn' (in honour of Frege's German word for the realm of intension) and `Bed' (short for `Bedeutung', which was Frege's  word for referent, such as the bearer of a name). \\
\indent \indent (3'): We confine ourselves to treating symmetries as maps on states (treating them as maps on quantities would be parallel). So we now denote a symmetry on the states in `Sinn' as $a^{\tn{Int}}$, where the superscript ${\tn{Int}}$ corresponds to the subscript in $I_{\tn{Int}}$. And we now denote a symmetry on the states in `Bed' as $a^{\tn{Ext}}$, where the superscript ${\tn{Ext}}$ corresponds to the subscript in $I_{\tn{Ext}}$.\\
Putting (2') and (3') together, we write a state-space in the realm of intension as ${\cal S}_{\tn{Sinn}}$, and a state-space in the realm of extension as ${\cal S}_{\tn{Bed}}$. So we write: $a^{\tn{Int}}: {\cal S}_{\tn{Sinn}} \rightarrow {\cal S}_{\tn{Sinn}}$; and we write $a^{\tn{Ext}}: {\cal S}_{\tn{Bed}} \rightarrow {\cal S}_{\tn{Bed}}$. 

But we shall not act on (1) above: the  notation would be cumbersome, and without compensating advantages. In short: acting only on (2) and (3) above---i.e.~introducing `Sinn' and `Bed', with symmetry maps $a^{\tn{Int}}$ and $a^{\tn{Ext}}$ respectively---is enough to enable us to draw the required commuting diagrams. \\

We spell these out: first for the realm of intension, then for the realm of extension. In each case, we first draw the diagram for a bare theory $T$ taken as a triple, $T = \bra {\cal S}, {\cal Q}, {\cal D} \ket$; and then draw the diagram for a bare model $M$ taken as a triple $M = \bra {\cal {\bar S}}, {\cal {\bar Q}}, {\cal {\bar D}} \ket$. (Note that these diagrams do not reflect Section \ref{symmthies:was25A}'s discussion about the overlap but distinctness (in general) of symmetries of a theory $T$, and symmetries of its model $M$.)

Thus for the realm of intension, we have: for a bare theory $T$ with state-space $\cal S$, the diagram in Figure \ref{interpsymmdiag1}. For a bare model $M$ with state-space ${\cal {\bar S}}$, we have the diagram in Figure \ref{interpsymmdiag2}.
\begin{figure}
\begin{center}
\bea
\begin{array}{ccc}{\cal S}~~&\xrightarrow{\makebox[.6cm]{$\sm{$a$}$}}&{\cal S}~~\\
~~\Big\downarrow {\sm{$I_{\tn{Int}}$}}&&~~\Big\downarrow {\sm{$I_{\tn{Int}}$}}\\
{\cal S}_{\tn{Sinn}}&\xrightarrow{\makebox[.6cm]{$\sm{$a^{\tn{Int}}$}$}}&{\cal S}_{\tn{Sinn}}
\end{array}\nonumber
\eea
\caption{Commutativity of the symmetry $a$ with the interpretation map $I_{\tn{Int}}$ for $T$.}
\label{interpsymmdiag1}
\end{center}
\end{figure}
\begin{figure}
\begin{center}
\bea
\begin{array}{ccc}{\cal\bar S}~~&\xrightarrow{\makebox[.6cm]{$\sm{$a$}$}}&{\cal\bar S}~~\\
~~\Big\downarrow {\sm{$I_{\tn{Int}}$}}&&~~\Big\downarrow {\sm{$I_{\tn{Int}}$}}\\
{\cal\bar S}_{\tn{Sinn}}&\xrightarrow{\makebox[.6cm]{$\sm{$a^{\tn{Int}}$}$}}&{\cal\bar S}_{\tn{Sinn}}
\end{array}\nonumber
\eea
\caption{Commutativity of the symmetry $a$ with the interpretation map $I_{\tn{Int}}$ for $M$.}
\label{interpsymmdiag2}
\end{center}
\end{figure}

Strictly speaking, we should in diagrams in Figure \ref{interpsymmdiag2} and Figure \ref{interpsymmdiag1} distinguish two interpretation maps---both of which we have in fact  written as $I_{\tn{Int}}$---according as the domain of definition is the state-space of a bare theory or of a bare model. But like in comment (1) above: the precision is not worth the burden of extra notation. And similarly of course, for the next two diagrams about the realm of extension.

For the realm of extension, we have, at some given possible world $W$ and context of use sufficiently rich to determine references (i.e.~to avoid the interpretation maps being undefined on the given arguments): for a bare theory $T$ with state-space $\cal S$, the diagram in Figure \ref{interpsymmdiag3}.
\begin{figure}
\begin{center}
\bea
\begin{array}{ccc}{\cal S}~~&\xrightarrow{\makebox[.6cm]{$\sm{$a$}$}}&{\cal S}~~\\
~~\Big\downarrow {\sm{$I_{\tn{Ext}}$}}&&~~\Big\downarrow {\sm{$I_{\tn{Ext}}$}}\\
{\cal S}_{\tn{Bed}}&\xrightarrow{\makebox[.6cm]{$\sm{$a^{\tn{Ext}}$}$}}&{\cal S}_{\tn{Bed}}
\end{array}\nonumber
\eea
\caption{Commutativity of the symmetry $a$ with the interpretation map $I_{\tn{Ext}}$ for $T$.}
\label{interpsymmdiag3}
\end{center}
\end{figure}

For a bare model $M$ with state-space ${\cal {\bar S}}$, we have the diagram in Figure \ref{interpsymmdiag4}.
\begin{figure}
\begin{center}
\bea
\begin{array}{ccc}{\cal\bar S}~~&\xrightarrow{\makebox[.6cm]{$\sm{$a$}$}}&{\cal\bar S}~~\\
~~\Big\downarrow {\sm{$I_{\tn{Ext}}$}}&&~~\Big\downarrow {\sm{$I_{\tn{Ext}}$}}\\
{\cal\bar S}_{\tn{Bed}}&\xrightarrow{\makebox[.6cm]{$\sm{$a^{\tn{Ext}}$}$}}&{\cal\bar S}_{\tn{Bed}}
\end{array}\nonumber
\eea
\caption{Commutativity of the symmetry $a$ with the interpretation map $I_{\tn{Ext}}$ for $M$.}
\label{interpsymmdiag4}
\end{center}
\end{figure}

\subsection{Duality as isomorphism of models}\label{DlyDef}

We turn in this last Subsection to our proposal 
that a duality is an isomorphism of models of a bare theory. To be precise: it is {\em an isomorphism of   model triples} of a bare theory.  Indeed, after all the stage-setting of the previous Subsections (!), the proposal is straightforward. We first give its details, using the notations we have established (Section \ref{DlyIsom}). In Section \ref{logweak}, we argue that our notion of duality is logically weak but physically strong. Then in Sections \ref{DlyInterpn} and \ref{DlySymmies}, we turn to how duality relates to the topics of Sections \ref{interp} and \ref{Symm}: interpretations and symmetries.

\subsubsection{Duality as isomorphism}\label{DlyIsom}

Our basic idea is that a duality is an isomorphism between two triples, each comprising a state-space  endowed with appropriate structure, a set of quantities endowed with appropriate structure, and a dynamics, consistent with that structure.\footnote{As we mentioned in Section \ref{bare;was22A}, `appropriate structure' here refers to: (i) the structure of the sets of spaces and quantities, (ii) the rules for evaluating quantities, (iii) the structure which the dynamics satisfies, (iv) the set of symmetries of the theory. We can now be more specific about these, for the examples of quantum theories, which will illustrate our schema: (ia) the set of states will be a separable Hilbert space; (ib) the quantities will be elements (normally the self-adjoint, renormalisable elements) of an algebra; (ii) the rules for evaluating quantities are maps to the appropriate field: for most quantum theories, the inner product on the Hilbert space, and the usual rules for evaluating matrix elements; (iii) dynamical evolution will usually be a (unitary) map, satisfying appropriate commuting diagrams with the other maps in the theory; (iv) the group of symmetries will comprise the automorphisms of the algebra: and possibly additional symmetries, on the states and on the quantities. For classical theories, these comments get modified in familiar ways: e.g.~(ia) would say that the set of states is a manifold, with structure appropriate to e.g.~Lagrangian or Hamiltonian mechanics.  \label{struct}} 
`Isomorphism between triples' is of course short for a triple of maps: an isomorphism between the two state-spaces, and isomorphism between the sets (almost always: algebras, cf.~footnote \ref{struct}) of quantities, and an equivariance condition on the dynamics.\footnote{Our proposal does not depend on the formulation of models as triples. A model root can be presented in many different forms, and the isomorphism should then preserve the corresponding structure. Even for triples, one can envisage isomorphisms which do not respect the triple structure, though they map the model roots isomorphically. Compare Section \ref{logweak}. But it will suffice for our purposes to restrict to model roots defined as triples, whose structure is preserved by the duality.\label{isomod}}
In addition, the isomorphism must commute with the symmetries of the theory, as sketched in Section \ref{Symm}.

More important is the question of {\em which} kinds of triples are related by duality. Recalling our distinction between bare theories and their more specific models, the answer is clear: {\em a duality relates two model triples of a single bare theory}. 

The crucial point here is that the model triple is separated from the model's own specific structure, and expresses only the model's realizing (typically: representing in the mathematical sense) the bare theory. Recall the notation from Eq.~\eq{modelq} in Section \ref{notation;was22C}: $M=\bra{\cal S}_M,{\cal Q}_M,{\cal D}_M,\bar M\ket  =:\bra m,\bar M\ket~,$ where $m:= T_M := \bra {\cal S}_M, {\cal Q}_M, {\cal D}_M\ket$ is the model triple. We emphasised already  in Sections \ref{models;was22B} and \ref{notation;was22C} that two model triples are in general {\em not} isomorphic to each other, nor to the bare theory. So the assertion of duality is substantive: it asserts that two model triples are in fact isomorphic.

 But this is {\em not} to say that the two {\em models}, each `considered in their entirety', are isomorphic. They each have their own  specific structure, and are (in almost all cases) {\em not} isomorphic. Recall our other notation from Section \ref{notation;was22C} for models `considered in their entirety': $M = \bra {\cal {\bar S}}, {\cal {\bar Q}}, {\cal {\bar D}} \ket$. Indeed, their being non-isomorphic is usually part of what makes the duality surprising and (if Nature is kind to us!) empirically fruitful, i.e.~of scientific importance. And `the more non-isomorphic'---i.e.~the more disparate the two models, considered in their entirety, are---the more surprising, and (one hopes) empirically fruitful, is the duality (cf.~(2) and (3) in Section \ref{orient}).\footnote{We should put this last point more precisely, since our notion of bare theory is logically weak, with even a group or an algebra, together with a set of maps to the real numbers, counting as a legitimate bare theory: (cf.~(1) in Section \ref{orient}). And it is in general not surprising, nor likely to be empirically fruitful, to learn that two very disparate models are both groups, or both algebras: (unless the maps to the real numbers are so disparate that the existence of an isomorphism is not easy to guess). Thus the point here, more precisely, is that, for a given degree of detail or logical strength in the bare theory (and the more, the better!): the more disparate its models (considered in their entirety), the more surprising, and one hopes fruitful, is their both realizing the bare theory. That is: the more surprising is the duality.} \\

We now introduce some notation for dualities as isomorphisms between model triples. This will require first giving:\\
\indent \indent (1) some new notation for the value of a quantity on a state, and \\
\indent \indent (2) a more detailed discussion of dynamics (in both the `Schr\"{o}dinger' and `Heisenberg' pictures). \\
Both (1) and (2) can be given wholly independently of our distinctions (i) between theories and their models, and (ii) between interpreted and uninterpreted theories. So for the moment, please consider a generic triple of a state-space, a set of quantities, and a dynamics:  $\bra {\cal {S}}, {\cal {Q}}, {\cal {D}} \ket$.\footnote{This simpler idea of a triple was used in our earlier---cruder!---discussion of duality: cf.~De Haro et al.~(2017:~Section 3.2). There, the simplicity engendered no errors, since our general description of duality was but a preamble to a specialist topic: an assessment of gauge symmetries in gauge-gravity duality.}\\

(1): Suppose we are given a set of states ${\cal S}$, a set of quantities ${\cal Q}$ and a dynamics ${\cal D}$: $\bra {\cal {S}}, {\cal {Q}}, {\cal {D}} \ket$. We will write $\langle Q , s \rangle$ for the value of quantity $Q$ in state $s$. This prompts two further general points. 

First: it is common to think of a state $s \in {{\cal S}}$ as a maximal  specification of the instantaneous properties of the system in question; and a quantity  $Q \in {{\cal Q}}$ as a numerically measurable property of it. In effect, this makes states and quantities nothing but assignments of values to each other. Second: for classical physics, one naturally takes quantities as real-valued functions on states, so that $\langle Q , s \rangle := Q(s) \in \mathR$ is the system's possessed or intrinsic value of the quantity; and for quantum physics, one naturally takes quantities as linear operators on a Hilbert space of states, so that  $\langle Q , s \rangle := \langle s |{\hat Q} | s \rangle \in \mathR$ is the system's Born-rule expectation value of the quantity.  But  for quantum physics it is often important to consider the non-diagonal matrix elements of a given quantity/operator $\hat Q$, without requiring this to be adequately encoded in the Born-rule expectation values of various other quantities. So for a quantum theory---as in the bosonization example of Section 4 et seq.!---we should understand a value written schematically as $\langle Q , s \rangle$ to also represent all the matrix elements $\langle s_1 |{\hat Q} | s_2 \rangle$. Thus $\bra Q,s\ket$ is a short-hand for an expression like  $\bra Q; s_1,s_2\ket := \langle s_1 |{\hat Q} | s_2 \rangle$,\footnote{Therefore duality will imply unitary equivalence.} i.e.~$Q$ is regarded as a map: ${\cal S}\times{\cal S}\rightarrow\mathC$. 

(2)  We turn to the dynamics $\cal D$, i.e.~a specification of how the values of quantities change over time. We will keep the discussion very simple. First, we assume the dynamics is deterministic: also in quantum theories, despite the threat of Schr\"{o}dinger's cat. Then it can be presented in two ways, for which we adopt the quantum terminology, viz.~the `Schr\"{o}dinger' and `Heisenberg' pictures; (though the ideas occur equally in classical mechanics: for example the remark, frequent in the textbooks, that in Hamiltonian mechanics time-evolution can be regarded as a sequence of canonical  transformations, is in effect a statement of the Heisenberg picture). But we shall not need to distinguish otherwise between the different detailed formalisms for dynamics, such as Hamiltonian vs.~Lagrangian, and the path-integral. Besides,  we will adopt for simplicity the Schr\"{o}dinger picture.

So we say: $D_S$ is an action of the real line $\mathR$ representing time on ${\cal S}$. There is an equivalent Heisenberg picture of dynamics with $D_H$,  an action of  $\mathR$ representing time on ${\cal Q}$. The pictures are related by, in an obvious notation:
\be
D_S:   \mathR \times {{\cal S}} \ni (t,s) \mapsto D_S(t,s) =: s(t) \in {{\cal S}} \;  {\mbox{iff}} \; 
D_H:   \mathR \times {{\cal Q}} \ni (t,Q) \mapsto D_H(t,Q) =: Q(t) \in {{\cal S}} 
\ee 
where for all $s \in {{\cal S}}$ considered as the initial state, and all quantities $Q \in {{\cal Q}}$, the values of physical quantities at the later time $t$ agree in the two pictures:
\be
\langle Q, s(t) \rangle = \langle Q(t), s \rangle \; .
\ee \\

With the notations and notions of remarks (1) and (2) in hand, we can now present the notation for dualities as isomorphisms between model triples. Let $M_1, M_2$ be two models, with model triples $m_1 =  \bra {\cal S}_{M_1}, {\cal Q}_{M_1}, {\cal D}_{M_1}\ket$ and $m_2 =  \bra {\cal S}_{M_2}, {\cal Q}_{M_2}, {\cal D}_{M_2}\ket$. We can suppose that  $M_1, M_2$ are both models of a bare theory $T$. Or we can proceed in the `opposite direction' discussed in Section \ref{RandI}: that is, we can suppose that  $M_1, M_2$ are given independently of a bare theory $T$, but their model triples (model roots in the more general language of Section \ref{RandI}) are isomorphic. Either way, the notation for dualities is as follows.\\

To say that the model triples $m_1, m_2$ are isomorphic is to say, in short, that: there are isomorphisms between their respective state-spaces and sets of quantities, that (i) make values match, and (ii) are equivariant for the two triples' dynamics (in the Schr\"{o}dinger and Heisenberg pictures, respectively). We now spell this out. Though retaining the $M$s in the subscripts is cumbersome, we will do so, in order to emphasise our main conceptual point: that duality is a relation between model triples in our sense---it is {\em not} between theories, or between generic triples $\bra {\cal {S}}, {\cal {Q}}, {\cal {D}} \ket$ as in remarks  (1) and (2).

Thus we say:--- A duality between $m_1 =  \bra {\cal S}_{M_1}, {\cal Q}_{M_1}, {\cal D}_{M_1}\ket$ and $m_2 =  \bra {\cal S}_{M_2}, {\cal Q}_{M_2}, {\cal D}_{M_2}\ket$ requires:\footnote{See footnote \ref{isomod} and Section \ref{logweak} for a brief discussion of more general cases.} \\
\indent \indent an isomorphism between the state-spaces (almost always: Hilbert spaces, or for classical theories, manifolds): 
\be
d_s: {{\cal S}_{M_1}} \rightarrow {{\cal S}_{M_2}} \;\;  {\mbox{using $d$ for  `duality'}} \; ;
\ee 
\indent \indent and an isomorphism between the sets  (almost always: algebras) of quantities\\
\be
d_q: {{\cal Q}_{M_1}} \rightarrow {{\cal Q}_{M_2}} \;\;  {\mbox{using $d$ for `duality'}} \; ;
\ee 
such that: (i) the values of quantities match: 
\be\label{obv1}
\langle Q_1, s_1 \rangle_1 = \langle d_q(Q_1), d_s(s_1) \rangle_2 \; , \;\; \forall Q_1 \in {{\cal Q}_{M_1}}, s_1 \in {{\cal S}_{M_1}}. 
\ee
and: (ii) $d_s$ is equivariant for the two triples' dynamics, $D_{S:1}, D_{S:2}$, in the Schr\"{o}dinger  picture; and 
$d_q$ is equivariant for the two triples' dynamics, $D_{H:1}, D_{H:2}$, in the  Heisenberg picture: see Figure \ref{obv2}.
\begin{figure}
\begin{center}
\bea
\begin{array}{ccc}{\cal S}_{M_1}&\xrightarrow{\makebox[.6cm]{$\sm{$d_s$}$}}&{\cal S}_{M_2}\\
~~\Big\downarrow {\sm{$D_{S:1}$}}&&~~\Big\downarrow {\sm{$D_{S:2}$}}\\
{\cal S}_{M_1}&\xrightarrow{\makebox[.6cm]{\sm{$d_s$}}}&{\cal S}_{M_2}
\end{array}~~~~~~~~~~~~
\begin{array}{ccc}{\cal Q}_{M_1}&\xrightarrow{\makebox[.6cm]{$\sm{$d_q$}$}}&{\cal Q}_{M_2}\\
~~\Big\downarrow {\sm{$D_{H:1}$}}&&~~\Big\downarrow {\sm{$D_{H:2}$}}\\
{\cal Q}_{M_1}&\xrightarrow{\makebox[.6cm]{\sm{$d_q$}}}&{\cal Q}_{M_2}
\end{array}\nonumber
\eea
\caption{Equivariance of duality and dynamics, for states and quantities.}
\label{obv2}
\end{center}
\end{figure}

Eq.~\eq{obv1}  appears to favour $m_1$ over $m_2$; but in fact does not, thanks to the maps $d$ being bijections.


It is already clear that a duality reduces to a symmetry, in the case where there is just one model, and one model triple, at issue, i.e.~$M_1 = M_2$ and $m_1 = m_2$. We shall return to this topic in Section \ref{DlySymmies}. First, we turn to the questions (i) whether our notion of duality is too weak (Section \ref{logweak}) and (ii) how it relates to Section \ref{interp} topic of interpretation (Section \ref{DlyInterpn}). 

\subsubsection{A logically weak but physically strong notion of duality}\label{logweak}

In Section \ref{orient}, we admitted that our definition of duality is logically weak because there is a duality whenever two models are isomorphic. Thus one might worry that, whenever two given models share some common structure smaller than the model triples, they are dual with respect to the substructures they share.
In this Section, we argue that this worry is unfounded, for models that purport to describe physical systems---which is our concern in this paper. Thus the notion of duality is physically stronger than it would at first seem. The point will be to distinguish between a purely formal model vs.~a physical (although uninterpreted) model. Our schema is intended for the latter: and it is only the model triples, not their specific structure, that is physically significant. 

We will illustrate this in an example. Consider, for simplicity, the following model, based on the $\mbox{su}(2)$ algebra: 
\bea
M_0=\bra{\cal S}_0,{\cal Q}_0,{\cal D}_0\ket:=\bra{\cal S}_J,U(\mbox{su}(2)),C_2(J)\ket~.
\eea
${\cal S}_J$ is here the Hilbert space of irreducible representations of su(2) with total quantum number $J$.  $U(\mbox{su}(2))$ is the universal enveloping algebra of su(2), i.e.~roughly, all powers of the algebra elements, quotiented by the algebra relations. The dynamics is the Hamiltonian of the model, which we take to be $C_2(J)$, the second Casimir.

Let us compare $M_0$ with a model $M$ with which it shares a common structure, and which is based on the $\mbox{su}(2)\otimes\mbox{su}(2)$ algebra. The state space is: ${\cal S}={\cal S}_J\otimes{\cal S}_K$, where ${\cal S}_J$ and ${\cal S}_K$ are the state spaces of the first and the second su(2), respectively. $J$ is the total quantum number of the first su(2), and $K$ is the total quantum number of the second su(2). 
We take the dynamics to be given by the sum of the Casimirs of the two su(2)'s, ${\cal D}=C_2(J)+C_2(K)$. This model is written as:
\bea\label{M}
M=\bra{\cal S},{\cal Q},{\cal D}\ket=\Big\bra{\cal S}_J\otimes{\cal S}_K,U\left(\mbox{su}(2)\otimes\mbox{su}(2)\right),C_2(J)+C_2(K)\Big\ket,
\eea
where $U$ again indicates the universal enveloping algebra.

$M_0$ and $M$ are of course both representations of $M_0$, i.e.~$M$ is a representation of su(2) with as `extra structure' the second su(2). 
In fact, $M_0$ is isomorphic to $M$ for the trivial representation with $K=0$, i.e.~$M_0\cong M|_{K=0}$.


But $M$ and $M_0$ are {\it not} dual for arbitrary values of $K$ (as in Eq.~\eq{M}). To see this, we rewrite $M$ in a way which makes explicit the common structure they share, i.e.~the first su(2). So, define:
\bea
M':=\bra m,\bar M\ket~,
\eea
where $m$ contains the first su(2), and the specific structure $\bar M$ contains the second su(2). Explicitly, $m=\bra{\cal S}_J,U(\mbox{su}(2)),C_2(J)\ket$ and $\bar M=\bra{\cal S}_K,U(\mbox{su}(2)),C_2(K)\ket$. Thus, $m$ and $\bar M$ are both triples, and they are both isomorphic to $M_0$, in particular $m\cong M_0$.

We can summarise the above definitions introducing the following short notation: 
\bea
M_0\,\cong~m&=&{\bf J}\nn
M&=&{\bf J}\,\otimes\,{\bf K}\nn
M'&=&\bra{\bf J},{\bf K}\ket~,
\eea
where ${\bf J}$ is short for the first factor of the tensor product and ${\bf K}$ for the second. 

To reconstruct $M={\bf J}\,\otimes\,{\bf K}$ from $M'=\bra{\bf J},{\bf K}\ket$, one takes the tensor products of the states in ${\bf J}$ and ${\bf K}$, takes all the products of the quantities, and adds up the dynamics, to reproduce Eq.~\eq{M}. 

If we were to say that $M={\bf J}\,\otimes{\bf K}$ and its model quadruple counterpart, $M'=\bra{\bf J},{\bf K}\ket$, are `the same', then since it is true that $M'=\bra{\bf J},{\bf K}\ket$'s model triple and $M_0\cong{\bf J}$ are isomorphic: it would follow that $M={\bf J}\,\otimes{\bf K}$ and $M_0$ would be dual in the relevant sense. 

But notice that $M'$ is {\it not} the same as $M$, nor is it isomorphic to it, because they differ in what they regard as physically significant (cf.~De Haro (2017:~p.5)). Only the first model triple, ${\bf J}$, is physically significant in $M'$, whereas both model triples are physically significant in $M$. Thus there can be no isomorphism between $M$ and $M'$ as candidate models of physics, for they differ in their physical content. 

In other words, the difference is in a tensor product model presented as such (i.e.~$M={\bf J}\,\otimes{\bf K}$) vs.~a model quadruple that has as model triple the first factor ${\bf J}$, and as its specific structure the second factor ${\bf K}$. 

This argument reinforces the point that it is not necessary, nor desirable, to define {\it bare theories} as equivalence classes of models. This means that the condition to have a {\it duality} is only that two model triples be isomorphic: but the model triples need not be isomorphic to the bare theory, only homomorphic to it. And since, as we have just seen, the isomorphism between dual models is essentially unique (i.e.~it is not possible to weaken the isomorphism to get dual structures, without changing their physical content), there is no gain in requiring that duality must also involve the theory. If one starts with a bare theory which is weaker than two isomorphic model triples that represent it, it may be possible to strengthen it so as to match the two model triples: but there is no gain in this. So, it is best to keep the notion of {\it bare theory} physically weak, and the notion of {\it duality} physically strong.

We end with a contrast of the notions of duality in physics and mathematics. In mathematics, just as in physics, `duality' does not have a fixed meaning; however, all the examples of duality involve just {\it two} theories. More precisely, the duality operation generates the two-element group $\mathbb{Z}_2$. This is not so in the physics literature, where duality can involve more than two models, and the duality group can be rich. For example, the S-duality group of electric-magnetic duality is $\mbox{SL}(2,\mathbb{Z})$, and string theories realize so-called U-duality groups, which involve orthogonal and exceptional groups. Our schema allows for dualities among many models, and so it is closer to the notion in physics. This also strengthens the analogy between duality and symmetry, mentioned in Section \ref{introschema}-(2). 

\subsubsection{Duality and interpretation}\label{DlyInterpn}

So far, our discussion of interpretation has concerned a {\em single} theory or model.  Thus recall that Section \ref{interp} introduced interpretation maps  $I_{\tn{Int}}$ and $I_{\tn{Ext}}$ in a rather informal way, as mapping from a bare i.e.~uninterpreted theory or a bare model, to the realm of intension (`Sinn'), or to the realm of extension (`Bed'), respectively. Then Section \ref{symminterp:was25B} laid out how $I_{\tn{Int}}$ and $I_{\tn{Ext}}$ are to mesh with symmetry maps. This amounted to  a commutation condition, i.e.~$I_{\tn{Int}}$ and $I_{\tn{Ext}}$ forming a commuting diagram with symmetry maps, which for simplicity we only considered as defined on state-spaces: either on a bare state-space, or on a state-space in the realm of intension (`Sinn'), or on a state-space in the realm of extension (`Bed'). (Cf.~the diagrams in Figures \ref{interpsymmdiag1} to \ref{interpsymmdiag4}.) But again, everything in Section \ref{symminterp:was25B} concerned a {\em single} theory or model.

Since duality is about relations between theories/models, there is, at first sight, little to say about duality and interpretation. That is: interpretation should simply proceed independently on the two sides of the duality---for example, we just require the interpretation-symmetry commuting diagram on both sides of the duality. Indeed: we said already at the start of Section \ref{introschema} that in some cases of duality, the two sides were clearly not---nor intended to be---physically or semantically equivalent: e.g.~the high and low temperature regimes in Kramers-Wannier duality. And our definition of duality as formal (viz.~an isomorphism of model triples) certainly allows this idea of `distinct but isomorphic sectors of reality'---namely as the codomains of the interpretation maps on the two sides of the duality.
   
This verdict---`there is little to say'---is true, so far as it goes. And of course, it does not forbid the other sort of case: where the two sides of the duality {\em are} physically/semantically equivalent, i.e.~do describe `the same sector of reality'. In our schema, this would be modelled by the interpretation maps on the two sides having the same images/values in their codomain---so as to give a {\em triangular}, rather than {\em square}, commuting diagram. We shall spell this out as regards the interpretation of (bare) quantities: similar diagrams could of course be drawn for states. 

\begin{figure}
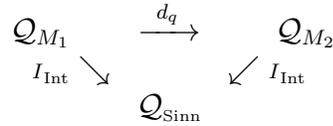

\begin{center}
\bea
\begin{array}{ccc}{\cal Q}_{M_1}\!\!&\xrightarrow{\makebox[.6cm]{$\sm{$d_q$}$}}&\!{\cal Q}_{M_2}\\
~~~~~~~{\sm{$I$}}_{\tn{Int}}\searrow&&\swarrow \sm{$I$}_{\tn{Int}}~~~~~~~~\\
&{\cal Q}_{\tn{Sinn}}&\end{array}\nonumber
\eea
\caption{The two sides of the duality describe `the same sector of reality', in the realm of intension.}
\label{interpdlydiag1}
\end{center}
\end{figure}

For (bare) quantities being mapped by $I_{\tn{Int}}$ into the realm of intension `Sinn', the two sides of a duality describing `the same sector of reality' amounts to the diagram in Figure \ref{interpdlydiag1}.

\begin{figure}
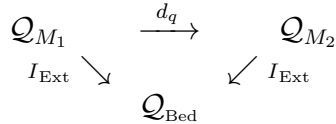

\begin{center}
\bea
\begin{array}{ccc}{\cal Q}_{M_1}\!\!&\xrightarrow{\makebox[.6cm]{$\sm{$d_q$}$}}&\!{\cal Q}_{M_2}\\
~~~~~~~{\sm{$I$}}_{\tn{Ext}}\searrow&&\swarrow \sm{$I$}_{\tn{Ext}}~~~~~~~~\\
&{\cal Q}_{\tn{Bed}}&\end{array}\nonumber
\eea
\caption{The two sides of the duality describe `the same sector of reality', in the realm of extension.}
\label{interpdlydiag2}
\end{center}
\end{figure}

Similarly:  for (bare) quantities being mapped by $I_{\tn{Ext}}$ into the realm of extension `Bed'---relative to some given possible world $W$ with a context rich enough to determine references, of course---the two sides of a duality describing `the same sector of reality' amounts to Figure \ref{interpdlydiag2}.

So far, so straightforward. But the above verdict is a bit quick: there are two further points to make.\\

(1): {\em What determines equivalence?}:--- First, there is the question what determines whether the two sides of a duality are physically/semantically equivalent, i.e.~describe the same `sector of reality'. De Haro (2016) and Dieks et al.~(2015)) have argued that the choice between these two options should depend on whether the models in question are given what was called an  `internal' or an `external' interpretation. The idea is that for the ranges of the interpretation maps to be distinct, there must be other facts, external to the triples themselves and our use of them, that determine the distinct ranges. Typically, these facts will be other pieces of physics to which the system described by each model triple is coupled---with different pieces of physics on the two sides of the duality. This coupling `breaks the symmetry' between the two sides, and secures that the two model triples are about distinct, albeit isomorphic, subject matters (`sectors of reality'). In the proposed jargon: the coupling provides an `external interpretation' of the model triple. On the other hand: sometimes we propose a physical theory as a putative theory of the whole universe, i.e.~as a putative cosmology, so that according to the theory there are no physical facts beyond those about the system (viz.~universe) it describes.  If in such a case, there is a duality---which in our framework, means there are two isomorphic model triples, each putatively describing the whole universe---then there can be no such coupling to other pieces of physics. (Gauge-gravity duality provides, of course, a putative example of such a duality between theories of the universe.)  An interpretation of each triple must therefore be what was labelled an `internal interpretation'; and this prompts the conclusion that the two triples describe the very same `sector of reality'. That is: the interpretation maps have the same range; and there is a triangular diagram, as in Figures \ref{interpdlydiag1} and \ref{interpdlydiag2}. \\

(2): {\em Interpreting the specific structure}:--- Second, there is more to say about the interpretation of a model's specific structure, especially in the latter sort of case, i.e.~two sides of a duality describing the same `sector of reality'.

Recall that a model $M$ is more than the model triple $m$, by which it realizes a bare theory, and which relates to another model triple in a duality. $M$ also has a specific structure $\bar M$: as we stressed at the start of Section \ref{DlyIsom}, this structure is {\em not} related by the duality to `the other side'. But the specific structure $\bar M$ {\em does} get interpreted---it supplies arguments for the interpretation maps  $I_{\tn{Int}}$ and $I_{\tn{Ext}}$---just as much as the model triple $m$ gets interpreted.\footnote{At least, this is what we would in general expect. Agreed, one might interpret a model without interpreting {\em all} of the specific structure $\bar M$: recall footnote \ref{partialfunctions} on the need to allow the interpretation maps to be partial, i.e.~to deliver no value for certain arguments. For more details, see Section 1.1.2.a of De Haro (2016a).}
This was emphasised by the other notation for models introduced at the end of Section \ref{notation;was22C}: viz.~a model is itself, like a bare theory, a triple. $M$'s states and quantities, being specific to $M$, are in general `bigger'/`more structured' than the states and quantities of the bare theory that $M$ models/realizes. We wrote them with a `bar': thus $M = \bra {\cal {\bar S}}, {\cal {\bar Q}}, {\cal {\bar D}} \ket$. And recall that then Section \ref{interp} took elements of these `bigger' sets ${\cal {\bar S}}, {\cal {\bar Q}}$ as arguments for the interpretation maps  $I_{\tn{Int}}$ and $I_{\tn{Ext}}$. 

So the first point to make is: our discussion of duality has so far ignored the specific structures ${\bar M}_1$ and ${\bar M}_2$ on the two sides, even though they do get interpreted. This silence is presumably no problem in a case where we agree that the two sides are not physically or semantically equivalent. In such a case, the interpretation of ${\bar M}_1$ and ${\bar M}_2$ just means there are physical facts on each of the two sides, additional to the facts that are isomorphic with (a subset of) facts on the other side.\footnote{In Section \ref{logweak}, we emphasised the fact that only the model triples, and not the specific structure, are physically significant. When we now consider external interpretations that do give a physical meaning to the specific structure, we have to say that these interpretations change the physical content of the model (its physical degrees of freedom). This is correct, because external interpretations do not need to preserve the structure of the model as a quadruple.} Thus for the case of Kramers-Wannier duality, the obvious examples of such non-matching physical facts would be facts about the value of the temperature: high on one side, and low on the other. In short: these additional facts (in the realms of intension and extension, respectively) are: on one side, in the ranges $I_{\tn{Int}}({\bar M}_1)$ and $I_{\tn{Ext}}({\bar M}_1)$;  and on the other side, in the ranges $I_{\tn{Int}}({\bar M}_2)$ and $I_{\tn{Ext}}({\bar M}_2)$.  

But what about the other sort of case: where the two sides {\em do} describe `the same sector of reality'? Is it really satisfactory to say that there are physical facts that:\\
\indent \indent (a) are additional to those facts described by the isomorphic model triples, i.e.~those `caught' by the duality/the common bare theory; yet also \\
\indent \indent (b) fall into two such disparate subsets: one subset expressed by ${\bar M}_1$ and the other subset expressed by ${\bar M}_2$?\\
In short: this world-picture, combining (i) a set of facts expressed by the two sides in the same way (though this sameness may be not obvious---the duality can be surprising), and (ii) two other sets of facts expressed in very different  ways by the two sides, is surprising: and maybe it is odd, or unsatisfactory ...

We saw examples of this in comment (3)  of Section \ref{orient}, for example in gauge-gravity duality. Here, the set of facts that are the common core, {\em a la} (i),  consists only in a class of asymptotic operators and a conformal class of $(d-1)$-dimensional metrics. And the sets of facts  {\em a la} (ii) include, on the bulk side, gravity (such as: Einstein's equations coupled to matter) in $d$ dimensions expressed by ${\bar M}_1$, and on the boundary side,  a conformal field theory (such as:  the Yang-Mills equations)  in $d-1$ dimensions expressed by ${\bar M}_2$. See DeHaro (2016:~Section 2.1, 2016a:~Section 2.2) for a discussion in the context of our schema.

For our example of bosonization, we will see in Section \ref{interpn524} how external interpretations map to different (sets of) worlds, while the internal interpretation maps to the same (set of) world(s). For the latter interpretation to be possible, we will see that the worlds must contain both bosonic and fermionic facts. So, the internal interpretation does not efface the distinction between bosons and fermions, but distinguishes them and identifies them in the world.

\subsubsection{Combining duality and symmetries}\label{DlySymmies}

In this Subsection, we turn to the relations between dualities and symmetries. There are three comments to make. They are not controversial. Indeed, they simply gather some threads from discussions in previous Sections. The first makes the obvious comparison between dualities and symmetries, and notes the conditions for a duality to reduce to being a symmetry. The second is about a duality preserving a symmetry of its model-triples; and so returns us to the contrast between the symmetries of a bare theory, and those of its model-triples. The third returns us to the contrast between duality and gauge, discussed in comment (4) at end of Section \ref{orient}.\\
 
(1): {\em Making the comparison precise}:---\\
 Earlier (at the end of Section \ref{introschema}) we announced that we would endorse a basic analogy between duality and symmetry: `a duality is like a symmetry, but at the level of theory', so that while a symmetry carries e.g.~a state into a `matching' state, a duality carries a theory into a `matching' theory. 

Indeed, we endorse this analogy---allowing of course for the shift of words from `theory' to `model-triple'. This endorsement is clear from:\\
\indent \indent (a): our discussion of symmetries of theories, taken as triples, and symmetries of their models, and their model-triples (Section \ref{symmthies:was25A}); and \\
\indent  \indent (b):  our definition of duality as an isomorphism of model-triples that (i) makes the values of quantities match, and (ii) is equivariant for the two triples' dynamics (cf.~Eq.~\ref{obv1} and Figure \ref{obv2} at the end of Section \ref{DlyIsom}).
  
In particular (as mentioned at the end of Section \ref{DlyIsom}): a duality reduces to a symmetry, in the case where there is just one model, and one model triple, at issue, i.e.~$M_1 = M_2$ and $m_1 = m_2$. Spelling this out will use the notion of a dual map (in the pure mathematical sense!), introduced in (i) at the start of Section \ref{symmthies:was25A}. Recall that this notion is defined by the pairing whereby states $s \in {\cal S}$ and quantities $Q \in {\cal Q}$ assign each other a value $\bra Q, s \ket$. Namely: given any map $a: {\cal S} \rightarrow {\cal S}$, we said that its dual map on quantities, $a^*: {\cal Q} \rightarrow {\cal Q}$ is defined by requiring that for any $s \in {\cal S}$ and $Q \in {\cal Q}$: $\bra a^*(Q), s \ket := \bra Q, a(s) \ket$. And similarly, starting with quantities: given any map $a: {\cal Q} \rightarrow {\cal Q}$, we said that its dual map on states, $a^*: {\cal S} \rightarrow {\cal S}$ is defined by requiring for all arguments: $\bra Q, a^*(s) \ket := \bra a(Q), s \ket$. \\

Thus suppose there is just one model triple at issue. Then $d_s$ is an automorphism of ${\cal S}_{M_1} \equiv {\cal S}_{M_2}$, i.e.~of the state-space in the one model triple; and similarly, for  $d_q$ on ${\cal Q}_{M_1} \equiv {\cal Q}_{M_2}$. So duality's condition (i), that the values of quantities match (Eq.~\eq{obv1}), becomes the condition
\be\label{obv1reducestosymmy}
\langle Q_1, s_1 \rangle_1 = \langle d_q(Q_1), d_s(s_1) \rangle_1 \; , \;\; \forall Q_1 \in {{\cal Q}_{M_1}}, s_1 \in {{\cal S}_{M_1}}. 
\ee
But $d_q$ induces a dual map $d_q^*$ on states, such that: $\langle d_q(Q_1), d_s(s_1) \rangle_1 = \langle Q_1, d_q^*(d_s(s_1)) \rangle_1$. So we conclude that $(d_q^* \circ d_s): {\cal S}_{M_1} \rightarrow {\cal S}_{M_1}$ is a symmetry (written, as usual for us, as a map on states rather than quantities). For we have:  
\be\label{obv1reducestosymmysecondeqn}
\langle Q_1, s_1 \rangle_1 = \langle d_q(Q_1), d_s(s_1) \rangle_1 = \langle Q_1, d_q^*(d_s(s_1)) \rangle_1 \; .
\ee

Finally, the same verdict---that for a single theory, duality reduces to symmetry---applies to dynamics, i.e.~to dynamical symmetries. That is: if a duality concerns just one model triple, then Section \ref{DlyIsom}'s  condition (ii) for duality---that the duality map is equivariant for the two triples' dynamics (i.e.~$d_s$ is equivariant for Schr\"{o}dinger dynamics, and $d_q$ is equivariant for Heisenberg dynamics)---reduces to the condition that the duality is also a dynamical symmetry: for example, that $d_s$ is a dynamical symmetry represented as a map on states. \\

(2): {\em On duality preserving a symmetry}:---\\
It is straightforward to confirm that on Section \ref{DlyIsom}'s definition of duality, a duality preserves any symmetry of its model triples. There are two points here. First: there is a commuting square diagram of isomorphisms. Second: there is the issue of the values of a quantity being equal on a given state, and on its transform under a symmetry. The first point will lead in to the second. 

First: The duality maps $d_s, d_q$ are not only bijections, but isomorphisms: $d_s: {\cal S}_{M_1} \rightarrow {\cal S}_{M_2}$, and $d_q: {\cal Q}_{M_1} \rightarrow {\cal Q}_{M_2}$. And although we did not have to spell out the exact structures of ${\cal S}_{M_i}, {\cal Q}_{M_i}$ that these isomorphisms are to preserve (but cf.~footnote \ref{struct}),
it is obvious from the fact that `is isomorphic to' is both a symmetric and a transitive relation, that the following diagram, with $a$ understood to be any automorphism of  ${\cal S}_{M_1}$, commutes (cf.~Figure \ref{dlykeepsymmystates}).
\begin{figure}
\begin{center}
\bea
\begin{array}{ccc}{\cal S}_{M_1}&\xrightarrow{\makebox[.6cm]{$\sm{$a$}$}}&{\cal S}_{M_1}\\
~~\Big\downarrow {\sm{$d_s$}}&&~~\Big\downarrow {\sm{$d_s$}}\\
{\cal S}_{M_2}&\xrightarrow{\makebox[.6cm]{}}&{\cal S}_{M_2}
\end{array}\nonumber
\eea\caption{Commutativity of duality and symmetry for states.}
\label{dlykeepsymmystates}
\end{center}
\end{figure}

And  of course, this diagram of isomorphisms is just what we mean by saying a duality $d$ preserves an automorphism of the state-space ${\cal S}_{M_1}$ in its domain model triple, and preserves ${\cal S}_{M_1}$'s structure. Namely, $d$ carries the automorphism---a map $a$  on ${\cal S}_{M_1}$---to a corresponding automorphism of states in the codomain (indeed: range) model triple. The diagram defines this corresponding automorphism, i.e.~the map forming the fourth side of the square: $d_s \circ a \circ (d_s)^{-1}: {\cal S}_{M_2} \rightarrow {\cal S}_{M_2}$. 

There is obviously a corresponding point about quantities, as against states. Since $d_q$ is required to be an isomorphism of quantities, the following diagram, with $a$ now understood to be any automorphism of  ${\cal Q}_{M_1}$, must commute, cf.~Figure \ref{dlykeepsymmyqties}.
\begin{figure}
\begin{center}
\bea
\begin{array}{ccc}{\cal Q}_{M_1}&\xrightarrow{\makebox[.6cm]{$\sm{$a$}$}}&{\cal Q}_{M_1}\\
~~\Big\downarrow {\sm{$d_q$}}&&~~\Big\downarrow {\sm{$d_q$}}\\
{\cal Q}_{M_2}&\xrightarrow{\makebox[.6cm]{}}&{\cal Q}_{M_2}
\end{array}\nonumber
\eea\caption{Commutativity of duality and symmetry for quantities.}
\label{dlykeepsymmyqties}
\end{center}
\end{figure}

And again, this diagram is just what we mean by saying a duality $d$ preserves an automorphism of the quantities in its domain model triple, and preserves ${\cal Q}_{M_1}$'s structure. Namely, $d$ carries the automorphism---a map $a$  on  ${\cal Q}_{M_1}$---to a corresponding automorphism of quantities in the codomain (indeed: range) model triple. The diagram defines this corresponding automorphism: $d_q~ \circ~ a ~\circ ~(d_q)^{-1}: {\cal Q}_{M_2} \rightarrow {\cal Q}_{M_2}$. \\

Second: But in physics, the notion of symmetry involves more than the notions of automorphism of the state-space, and of the set (usually algebra) of quantities. It involves the pairing whereby states $s$ and quantities $Q$ assign each other a value: $\bra Q, s \ket$. For these values (for a large and salient set of quantities, though usually not {\em all} quantities)  must be preserved under the symmetry. 

But satisfying this is automatic, for a duality as defined at the end of Section \ref{DlyIsom}. That is: For a duality to respect this aspect of symmetry was already built in to our definition of duality: namely in condition (i), that the values are equal between states and quantities that correspond by the duality. Recall Eq.~\eq{obv1}), which we here repeat:
\be\label{obv1repeat}
\langle Q_1, s_1 \rangle_1 = \langle d_q(Q_1), d_s(s_1) \rangle_2 \; , \;\; \forall Q_1 \in {{\cal Q}_{M_1}}, s_1 \in {{\cal S}_{M_1}}. 
\ee
   
Finally (and just like at the end of (1) above): the same verdict---that a duality preserves any symmetry of its model triples---applies to dynamics, i.e.~to dynamical symmetries. Recall from footnote \ref{dynsequivariance} (in Section \ref{symmthies:was25A}) that a dynamical symmetry is a commutation i.e.~equivariance condition. So for the Schr\"{o}dinger picture of dynamics, the diagram for the `first'  side of a duality, i.e.~$m_1 = \bra  {\cal S}_{M_1}, {\cal Q}_{M_1}, {\cal D}_{M_1} \ket$, is, with $a$ the dynamical symmetry, as in Figure \ref{dynslsymmyfordltysection}.
\begin{figure}
\begin{center}
\bea
\begin{array}{ccc}{\cal S}_{M_1}&\xrightarrow{\makebox[.6cm]{$\sm{$a$}$}}&{\cal S}_{M_1}\\
~~\Big\downarrow {\sm{$D_{t,t_0}$}}&&~~\Big\downarrow {\sm{$D_{t,t_0}$}}\\
{\cal S}_{M_1}&\xrightarrow{\makebox[.6cm]{$\sm{$a$}$}}&{\cal S}_{M_1}
\end{array}\nonumber
\eea
\caption{Commutativity of symmetry and dynamics.}
\label{dynslsymmyfordltysection}
\end{center}
\end{figure}

So we now compose this diagram with Figure \ref{dlykeepsymmystates}, which represents that a duality preserves a symmetry. But since in Figure \ref{dynslsymmyfordltysection}, the `first'  side, `1', of the duality occurs twice, on both top and bottom rows, we now need to compose  Figure \ref{dynslsymmyfordltysection} with Figure \ref{dlykeepsymmystates} twice: both on its bottom row; and also  on its top row (with the duality arrow in Figure \ref{dlykeepsymmystates} reversed). The resulting diagram (Figure \ref{threebox}) shows that the duality isomorphism on state-spaces $d_s$ carries the dynamical symmetry $a$ on the `1'  side of the duality, to a dynamical symmetry on the `2'  side: namely, the symmetry $d_s \circ a \circ d^{-1}_s$ (cf. either the top or bottom square). The Schr\"{o}dinger picture dynamics on ${\cal S}_{M_2}$ is (reading down the columns in the Figure): $d_s \circ D_{t,t_0} \circ d_s^{-1}$.
\begin{figure}
\begin{center}
\bea
\begin{array}{ccc}{\cal S}_{M_2}&\longrightarrow&{\cal S}_{M_2}\\
~~\Big\downarrow {\sm{$d_s^{-1}~~$}}&&~~\Big\downarrow {\sm{$d_s^{-1}~~$}}\\
{\cal S}_{M_1}&\xrightarrow{\makebox[.6cm]{$\sm{$a$}$}}&{\cal S}_{M_1}\\
~~\Big\downarrow {\sm{$D_{t,t_0}$}}&&~~\Big\downarrow {\sm{$D_{t,t_0}$}}\\
{\cal S}_{M_1}&\xrightarrow{\makebox[.6cm]{$\sm{$a$}$}}&{\cal S}_{M_1}\\
~~\Big\downarrow {\sm{$d_s~~~~$}}&&~~\Big\downarrow {\sm{$d_s~~~~$}}\\
{\cal S}_{M_2}&\longrightarrow&{\cal S}_{M_2}
\end{array}\nonumber
\eea
\caption{Commutativity of duality, symmetry, and dynamics.}
\label{threebox}
\end{center}
\end{figure}

So much by way of showing that a duality always preserves a symmetry of its model triples. In conclusion, we should emphasise again the three summarising comments, (i) to (iii), at the end of Section \ref{symmthies:was25A} about the contrasts between the symmetries of a bare theory, and those of its models, and of its model-triples.\\

(3): {\em The contrast between duality and gauge}:---\\
Finally, we should briefly return to our  comment (4) at the end of Section \ref{orient}. We said there that, although there is some truth in the common remark that two dual theories are like gauge formulations of a single theory, there are two important differences. Sometimes the two duals are agreed to {\em not} be physically equivalent (as in Kramers-Wannier duality). And anyway, the specific structure in a model is usually not gauge, in the sense of descriptively redundant. 

Our discussion since Section \ref{orient} reinforces this comment. For we have seen in more detail the idea of specific structure in a model----starting with our notation, $\bar M$, from Section \ref{notation;was22C}. And by relating duality as isomorphism of model triples to our interpretation maps, we saw that duality allows, but does not entail, physical equivalence (cf.~Section \ref{DlyInterpn}). 

Besides, we have also seen a more specific contrast between duality and gauge, that was not foreshadowed in comment (4) at the end of Section \ref{orient}. Namely, we noted in (2) at the end of Section \ref{symmthies:was25A} that if a symmetry of a model's specific structure---a symmetry of $\bar M$---is well-defined on the model triple, we expect it to be trivial, i.e.~the identity map, there. This point  implies that we would in general expect gauge (i.e.~descriptively redundant) structure to {\em not} be carried across intact by a duality. And indeed: this has been illustrated in detail in gauge-gravity dualities.  De Haro (2017a) has shown that a certain subgroup of the diffeomorphism group of the gravity model of the theory (roughly, the diffeomorphisms which preserve the asymptotic boundary conditions) is `invisible' to the gauge model of the theory, in the sense of not representing any difference on that model, and so being trivially represented on the theory (the common core).  Similarly for the `gauge symmetries' of the gauge side of the duality. These are not visible on the gravity side: they are symmetries of the specific structure of the gauge model, and are trivially represented on the theory (the common core); De Haro (2016a:~ Section 2.2.3), De Haro et al.~(2017:~Section 5.2). 

\section{The Basic Boson-Fermion Duality}\label{bfd}

Boson-fermion duality will be our main example of the schema developed in Sections \ref{schema} and \ref{symdual}: for, as we discussed in Section \ref{introbosonzn}, it lies in the middle of the spectrum between (i) mathematical precision and established physics, (ii) scientific importance. 

Boson-fermion duality is a vast field, still an active area of research today, especially in its three- and four-dimensional versions.\footnote{Cf.~e.g.~Gogolin (2004), Kopietz (2008), Kouneiher (2018).} But even just in two dimensions, there is a large number of examples, which we will discuss in the next Section. In this Section, we will start with the basic case: that is, the equivalence between the free, massless scalar field, and the free, massless Dirac fermion, in two Euclidean dimensions.\footnote{The results in Minkowski signature are readily obtained by a Wick rotation, as we discuss in Section \ref{fmB}.} This case already exhibits all of the interesting and non-trivial features of the more involved dualities, and so we will analyse it in some detail. 

Our exposition will be necessarily brief, and will focus on those aspects that best illustrate the schema. Thus we will downplay many other important physical and mathematical aspects of this duality, such as: the existence of classical and quantum soliton solutions in these theories, the integrability of the equations, the notions of Noether and topological charges, the connection with QCD and monopoles. Neglecting these important topics is the price we pay for focussing on illustrating a conceptual schema.

This Section is introductory. We here collect the technical results (especially, about the symmetries, their associated Noether currents, and the algebras the currents generate) that will allow us to illustrate our schema. Section \ref{fmB} introduces the free, massless boson. Section \ref{fdirac} introduces the free, massless Dirac fermion. (In Section \ref{bfdual}, we will present the basic boson-fermion duality, and show how it exemplifies our duality schema from Sections \ref{schema} and \ref{symdual}.)

Our exposition will mainly follow Ginsparg (1990) and Frishman and Sonnenschein~(2010).

\subsection{The free, massless boson}\label{fmB}

In this subsection, we study the free, massless boson. We analyse its symmetries, write down the associated Noether currents, and give relevant details about its quantization, in particular we give the algebra that the Noether currents satisfy. This algebra will be the starting point of the comparison, in Section \ref{bfdual}, with the fermionic model (where we will also justify how the bosonic and fermionic models are `models', in our sense of Sections \ref{models;was22B} and \ref{notation;was22C}).

In two-dimensional quantum field theory, it is very useful to work in complex coordinates. We will use the coordinates $z=x^0+ix^1$, $\bar z=x^0-ix^1$ parametrising $\mathbb{C}\cong\mathbb{R}^2$, the complex and Euclidean planes, respectively. 

We first discuss the classical field and its symmetries, in two points, (i)-(ii), below. A free, massless scalar field $\Phi$ satisfies the massless Klein-Gordon equation, which in the complex coordinates chosen in the previous paragraph takes the form: $\pa\bar\pa\Phi=0$ (here, and elsewhere, we use the short-hand $\pa=\pa/\pa z$, $\bar\pa=\pa/\pa\bar z$). The general solution to this equation allows for much more general functions than it does in higher dimensions, and this is the root of the richness of two-dimensional quantum field theory. The general solution is the sum of a holomorphic and an anti-holomorphic function:\footnote{Classically, we may indeed require the solutions to be holomorphic and anti-holomorphic functions. Quantum mechanically, there are singularities which are both inevitable and the source of interesting physics, as we will see. Thus we will allow $\f(z)$ and $\bar\f(\bar z)$ to have isolated singularities, hence we will allow them to be meromorphic and anti-meromorphic functions (operators, in the quantum version of the model), respectively.\label{mero}}
\bea\label{Fzz}
\F(z,\bar z)=\f(z)+\bar\f(\bar z)~.
\eea
The holomorphic and anti-holomorphic functions $\f(z)$ and $\bar\f(\bar z)$ are often called the left-, respectively right-moving parts of $\F$. This is because a holomorphic function depends only on $z=x^0+ix^1$: and after Wick rotation $x^1\rightarrow -ix$ (with $x^0=t$), the holomorphic part of $\F$ induces a function of $t+x$. For any fixed value of $t+x$, this indeed gives motion to the left (the speed is always negative at fixed $t+x$); whereas $\bar z\rightarrow t-x$, and thus $\bar f(\bar z)$ induces a function of $t-x$, which is right-moving (the speed is always positive at any fixed $t+x$).

The equations of motion (equivalently, the classical action) have two sets of symmetries which will be the starting point of our set-up (once they are generalised to symmetries of the quantum version of the model):\\
\\
(i)~~{\bf Conformal transformations.} In two dimensions, the action of a massless scalar field is invariant under a large group of coordinate transformations, namely conformal transformations: these are scale transformations with a variable scale factor, such that angles are preserved. In complex variables $z,\bar z$, the conformal transformations are parametrised by arbitrary holomorphic and anti-holomorphic functions:
\bea\label{ct}
z&\rightarrow&z'=f(z)~,~~~~\bar z\rightarrow\bar z'=\bar f(\bar z)~.
\eea
This is the two-dimensional version of the conformal transformations. Unlike the conformal group in higher dimensions, which has an finite number of generators (the generators of the Poincar\'e group plus additional generators of conformal transformations), the above transformations form an group whose corresponding algebra has an infinite number of generators. After taking into account quantum effects, this will be the celebrated Virasoro algebra.

It is easy to see that the above transformations contain, in terms of the Euclidean coordinates $x^\m=(x^0,x^1)$, in particular: (a) constant translations, $x^\m\rightarrow x^\m+a^\m$, (b) $\mbox{SO}(2)$ rotations (in Minkowski signature, these are $\mbox{SO}(1,1)$ Lorentz transformations), which in complex coordinates induce a U(1) action, (c) dilations (scale transformations) $z'=\l\,z$, $\bar z'=\bar\l\,\bar z$ ($\l\in\mathbb{R}$).\\
\\
(ii)~{\bf Affine current algebra transformations.} These are translations of the field by holomorphic or anti-holomorphic functions,
\bea\label{aca}
\Phi(z,\bar z)&\rightarrow&\Phi(z,\bar z)+\varphi(z)~,~~~~\Phi(z,\bar z)\rightarrow\Phi(z,\bar z)+\bar\varphi(\bar z)~.
\eea
Again, these transformations generalise the invariance of the action under constant shifts $\Phi\rightarrow\Phi+\varphi_0$, and are specific to two dimensions. 

The conserved currents associated with these two sets of symmetries are obtained through the Noether procedure. The currents for the affine current algebra transformations \eq{aca} are, up to an overall constant:
\bea\label{Jcurrents}
J(z):=\pa \phi(z)~, ~~~\bar J(\bar z):=\bar\pa\bar\phi(\bar z)~,
\eea
and they are anti-holomorphically, respectively holomorphically conserved in virtue of their (anti-) holomorphicity. These currents are called `affine currents' because, in the quantum version of the model, they generate an affine Lie algebra or Kac-Moody algebra.\\

The conserved currents associated with the conformal transformations \eq{ct} are the (holomorphic and anti-holomorphic) components of the stress-energy tensor:
\bea\label{freeT}
T(z)=-\half\,\pa\phi\,\pa\phi=-\half J^2(z)~,~~~\bar T(\bar z)=-\half\bar\pa\phi\,\bar\pa\bar\phi=-\half \bar J^2(\bar z)~.
\eea
The fact that the components of the stress-energy tensor can be written as squares of the affine algebra currents will be important upon quantisation, since it will link together the Virasoro and the Kac-Moody algebras. In the quantum case, the right-hand side of \eq{freeT} will contain the normally ordered product, and the relation is then called the {\it Sugawara construction}.\\

Our next task is to quantise the model. There are two well-known ways to quantise this model (which we briefly discuss in what follows), but we will settle for a third one, which is the more `modern approach', called `radial quantisation'. It is well suited to our perspective because it exploits the conformal symmetry group, and delivers the conformal algebra in the way we want it for Section \ref{bfdual}.

One can adopt conventional canonical quantization, where one writes down canonical commutation relations for the fields and their canonically conjugate momenta. One can realize these commutation relations by choosing a Fock-space representation of the fields, and writing down the algebra of the creation and annihilation operators, which
(in the present case of two dimensions) depend on one-dimensional momenta. One can then write down a Hamiltonian in terms of the creation and annihilation operators, and set up the physical states in the Fock space. See e.g.~Frishman and Sonnenschein (2010:~Section 1.6). 

The model can also be readily quantised using path integral quantisation. See e.g.~Frishman and Sonnenschein (2010:~Section 1.9). Here, we will adopt the third method, {\em radial quantisation}, as follows (ibid.~Section 1.7).

This approach  is based on the complex coordinates $z,\bar z$. First, we will use the conformal symmetry group to choose more convenient coordinates: mapping $\xi:=x^0+ix^1\mapsto z:=e^\xi=e^{x^0+ix^1}$ (for more details, see the Appendix). `Radial' then refers to the fact that, after this conformal transformation, the equal-time slices $x^0=\mbox{constant}$, used for canonical quantisation, become circles of constant radius. The line integrals over space which appear in physical quantities, such as charges, then turn into contour integrals on the complex plane. We will not spell out the details of this procedure, of which there are many good reviews (see e.g.~Section 2.2 of Ginsparg (1990) or L\"ust et al.~(1989:~Section 4.1)), but the key technique we will mention is the use of radial ordering to define the order of operators which are integrated over a contour in the complex plane. 

Remember that our aim for the quantum model of the boson is to obtain its algebra of operators. To this end, we will use the short-distance behaviour of a distinguished set of fields (so-called `primary fields': cf.~next paragraph). The algebra is indeed encoded in the short-distance behaviour of the products of the primary fields among each other and with the stress-energy tensor. This short-distance behaviour of products is called the `operator product expansion'. To this we now turn.

So let us consider a {\it primary field} $\Psi(z,\bar z)$. Primary fields are defined by their transformation properties under conformal transformations (see Eq.~\eq{prim} in the Appendix).\footnote{They are called `primary' because all other fields, which are called `descendants', can be obtained from them, through successive application of derivatives. See De Haro et al.~(2016:~Section 3).} 
It can be readily shown that the short-distance behaviour of the product of the stress-energy tensor with any primary field is:
\bea\label{ope}
T(z)\,\Psi(w,\bar w)&=&{h\over(z-w)^2}\,\Psi(w,\bar w)+{1\over z-w}\,\pa_w\Psi(w,\bar w)+\mbox{finite terms},~(z\rightarrow w)\\
\bar T(\bar z)\,\Psi(w,\bar w)&=&{\bar h\over(\bar z-\bar w)^2}\,\Psi(w,\bar w)+{1\over\bar z-\bar w}\,\pa_{\bar w}\Psi(w,\bar w)+\mbox{finite terms},~(\bar z\rightarrow\bar w)\nonumber
\eea
where $h,\bar h$ are the conformal weights of the primary field $\Psi$, i.e.~the powers with which a field scales under meromorphic, respectively anti-meromorphic transformations \eq{ct}, as in Eq.~\eq{prim} of the Appendix.\footnote{As remarked in footnote \ref{mero}, in the quantum case we allow for (anti-) {\it mero}morphic, rather than (anti-) holomorphic, operators.}  
The conformal weights are usually denoted as $(h,\bar h)$, and they encode how a field transforms under dilatations.\footnote{In more detail: $h+\bar h$ is the eigenvalue of the dilatation operator, and $h-\bar h$ is the eigenvalue of the (Euclidean) rotation operator. Hence, the conformal weights contain information about the mass and the Euclidean spin of a field.} For example, taking $\Psi=\pa\f(w)$, i.e.~the (derivative of the) massless left-moving part of $\f$, we have $h=1$, since this field is a vector, hence the conformal weight is $(1,0)$. Obviously, this is our primary field of interest in this model. As we will see below, after Eq.~\eq{TTope}, the stress-energy tensor is {\it not} a primary field. On the other hand (cf.~Section \ref{fdirac}), a Weyl-Majorana fermion $\Psi=\chi(w)$ has conformal weight $({1\over2},0)$. The anti-holomorphic  bosonic field $\bar\f$ has conformal weight $(0,1)$; and the anti-holomorphic Weyl-Majorana fermion has $(0,{1\over2})$. 

The expansion \eq{ope} is called an {\it operator product expansion}. Operator product expansions are important because they tell us almost everything we need to know about a field: in particular, we will derive from them {\it the algebra of operators} of the model. The form of \eq{ope} is a direct consequence of the quantum nature of the operators, together with the assumption of $\Psi$'s being a primary field, i.e.~that it satisfies \eq{prim}. It can be shown that the operator product expansion is equivalent to the canonical commutation relations of the modes of the fields. Eq.~\eq{ope} thus encodes the short-distance behaviour of $\Psi$, and is often taken to be an alternative {\it definition} of a primary field $\Psi$. 

In the classical model, the stress-energy tensor was given, in Eq.~\eq{freeT}, by the squares of operators evaluated at the same point. In the quantum version of the model, this gives rise to divergences which need to be (and can be) renormalised. For the two-dimensional quantum field theories which we consider in this paper, the divergences are renormalised, to all orders, by the addition of a {\it single} counterterm to the action; alternatively, it suffices to define expressions such as \eq{freeT} by {\it normal ordering}:
\bea\label{normalT}
T(z)=-\half :J(z)\,J(z):~,~~~~\bar T(\bar z)=-\half:\bar J(\bar z)\,\bar J(\bar z) : ~,
\eea
where the affine currents are still given, in the free bosonic scalar case, by \eq{Jcurrents}, now as operator equations. The normal ordering is denoted by the colons. For the details of the normal ordering procedure, see e.g.~Ginsparg (1990:~Section 2.3). 

Similarly to \eq{ope}, the operator product expansion of the stress-energy tensor with itself can be worked out:
\bea\label{TTope}
T(z)\,T(w)={c/2\over(z-w)^4}+{2\over(z-w)^2}~T(w)+{1\over z-w}~\pa T(w)~.
\eea
Here, $c=1$ for the free scalar \eq{freeT}, and it is called the {\it central charge}. Comparing the second term with \eq{ope}, we see that $h=2$, i.e.~the stress-energy tensor $T$ is a conformal field of weight $(2,0)$, as expected from dimensional analysis. However, compared to \eq{ope}, this expansion has, in addition, the first term. So $T(z)$ is not a primary field in the sense of Eq.~\eq{ope}, nor does it transform as in Eq.~\eq{prim}. The term proportional to the central charge $c=1$ is obtained making use of the normalized expression \eq{normalT}. For another way to calculate the central charge, see Frishman and Sonnenschein (2010:~Section 1.10).

In the same way, the operator product expansion of two affine currents \eq{Jcurrents}, as well as that between $T$ and $J$, can be calculated, with the following result:
\bea\label{JJope}
J(z)\,J(w)&=&{1\over(z-w)^2}+\mbox{finite terms},~~~(z\rightarrow w)\nn
T(z)\,J(w)&=&{1\over(z-w)^2}\,J(w)+{1\over z-w}~\pa J(w)~.
\eea
The second equation between $T$ and $J$ is simply \eq{ope} applied to the special case $h=1$, owing to the fact that $J=\pa\f$ is indeed a primary field of weight $(1,0)$. \\

Like in the canonical formalism, where a classical expansion of the field translates, quantum mechanically, into Fock modes: also in radial quantization it is convenient to introduce modes, which will give rise to creation and annihilation operators of the fields upon quantisation. That this is possible is ensured by \eq{Fzz}, which says that the field can be decomposed into meromorphic and anti-meromorphic parts. In particular, we can do a Laurent expansion of the meromorphic and anti-meromorphic parts of the field. Thus, in virtue of \eq{Jcurrents} and \eq{normalT}, the affine currents and the stress-energy tensor will have their own Laurent expansions: 
\bea\label{Laurent}
J(z)&=&\sum_{n\in\mathbb{Z}}{J_n\over z^{n+1}}~,~~~~\bar J(\bar z)=\sum_{n\in\mathbb{Z}}{\bar J_n\over z^{n+1}}\nn
T(z)&=&\sum_{n\in\mathbb{Z}}{L_n\over z^{n+2}}~,~~~~\bar T(\bar z)=\sum_{n\in\mathbb{Z}}{\bar L_n\over\bar z^{n+2}}~.
\eea
Overall factors of $1/z$ and $1/z^2$ can be extracted from $J$, respectively $T$: a fact which will be useful because these currents have $h=1$ and $h=2$, respectively, and so $J_0$ and $L_0$, as defined by \eq{Laurent}, will have special physical significance. The summation range is infinite and therefore this normalization can always be reached, by a simple translation of $n$. 

Because the currents satisfy \eq{Jcurrents} and \eq{normalT}, it is clear that $J_n$ are linear in the creation and annihilation operators of the meromorphic field $\f$, and the $L_n$ are quadratic in (an infinite sum of) the $J_n$'s. This fact will be built into our considerations in what follows, though we do not work it out explicitly.

Finally, we find the algebras satisfied by $J_n$ and $L_n$. Again these can be found either by canonical quantization or, in line with the methods we have used so far, they can be obtained directly from the operator product expansions \eq{JJope} and \eq{TTope}, combined with the radial ordering prescription for the operators which contain circle integrals, mentioned above. Writing $\d_{m+n}$ as short for $\d_{(m+n)0}$ (defined as usual to be 1 or 0 according as $m+n=0$ or $m+n \neq 0$): the result is:
\bea\label{ALAk1}
{}[L_m,L_n]&=&(m-n)\,L_{m+n}+{c\over12}\,n(n^2-1)\,\d_{m+n}\nn
{}[J_m,J_n]&=&-m\,\d_{m+n}\nn
{}[L_m,J_n]&=&-n\,J_{m+n}~.
\eea
and the same algebra is satisfied by the $\bar J_n$ and $\bar L_n$. Barred and unbarred quantities commute with each other. In the case at hand, $c=1$.\\

We recognise the first line as the celebrated Virasoro algebra, as expected from the fact that the classical theory has a conformal symmetry group. In the second line, one may recognise the level $k=1$, abelian Kac-Moody algebra. The third line is obtained from the operator product expansion between $T$ and $J$ in \eq{JJope}, and it makes the total algebra into the semi-direct product of the abelian Kac-Moody algebra and the Virasoro algebra. The algebra \eq{ALAk1} is called the {\it enveloping Virasoro algebra} (with $c=1$ and $k=1$). The general enveloping algebra of the affine Lie algebra is given in \eq{ALA} in the Appendix.\footnote{Notice that \eq{ALAk1} satisfies the property that the level $k$ can be changed by a rescaling of $J$. Thus, in the simple case we are dealing with here, in which the affine Lie algebra is based on the commutative Lie group U(1), the level has no real meaning. This is not important for us, since we will not use it: rather, our analysis in Section \ref{bfdual} will be based on the fact that we are here dealing with a special case of the general enveloping algebra of the affine Lie algebra.}

This, i.e.~Eq.~\eq{ALAk1}, is the central result from the physics literature which we have been seeking, and will use in Section \ref{bfdual}. For, together with \eq{normalT} and the mode expansions \eq{Laurent}, the tensor product of the holomorphic enveloping algebra of the affine Lie algebra \eq{ALAk1} and its anti-holomorphic copy contains all of the information about the quantum version of the model. This is because the states now live in the vector space on which this algebra acts (a Hilbert space), and the quantities and the dynamics are constructed from the operators satisfying the algebra. We will show this in Section \ref{bfdual}.

At this point, we notice that the generators $L_\pm$ and $L_0$ span an $\mbox{SL}(2,\mathbb{R})$ subalgebra. Together with the generators  $\bar L_\pm$ and $\bar L_0$, which satisfy the same algebra, they form the conformal algebra of $\mbox{SL}(2,\mathbb{C})$, which is the symmetry group of the vacuum of this model. 

\subsection{The free, massless Dirac fermion}\label{fdirac}

In this subsection, we consider our second model: of a free, massless Dirac fermion in two dimensions. Our goal is to rederive the infinite-dimensional algebra \eq{ALAk1} in this model. We will follow similar steps as in Section \ref{fmB}: we will derive the classical symmetries, quantise the model, and obtain operator product expansions. (The discussion of how this accommodates to the sense of `model', in Sections \ref{models;was22B} and \ref{notation;was22C}, is postponed to Section \ref{twois}.)

The massless Dirac fermion is a two-component, complex spinor which can be decomposed as $\Psi=:(\psi,\ti\psi)$, where $\psi$ and $\ti\psi$ are chiral (Weyl) fermions, called left- and right-chiral, respectively. The action is:
\bea\label{Dirac}
S_{\tn{Dirac}}={1\over4\pi}\int\dd^2x\,\bar\Psi\,\slashed\pa\Psi={1\over4\pi}\int\dd^2z\left(\psi^\dagger\,\bar\pa\psi+\ti\psi^\dagger\,\pa\ti\psi\right),
\eea
where in the middle expression we used the real form of the action, and in the last expression we used complex coordinates. The equations of motion imply that $\psi$ and $\ti\psi$ are, respectively, holomorphic and anti-holomorphic (respectively meromorphic, in the quantum version of the model). 

It is customary to further decompose the Dirac fermions into real, i.e.~Majorana fermions, as follows: $\Psi={1\over\sqrt{2}}\left(\Psi_1+i\,\Psi_2\right)$. In terms of the chiral (Weyl) fermions, we get $\psi={1\over\sqrt{2}}\left(\psi_1+i\psi_2\right)$, where $\psi_{1,2}$ are Weyl-Majorana fermions, and so the action takes the form of the sum of two copies of a single Weyl-Majorana fermion, which is conventionally called $\chi$ (where $\chi=\psi_{1,2}$). The action for a single Weyl-Majorana fermion is:
\bea
S_{\tn{WM}}={1\over8\pi}\int\dd^2z\left(\chi\,\bar\pa\chi+\ti\chi\,\pa\ti\chi\right),
\eea
and again $\chi,\ti\chi$ are (if extended from the real line to the entire complex plane) 
meromorphic and anti-meromorphic, respectively. 

Like in the case of the free, massless boson studied in Section \ref{fmB}, this action is invariant under two sets of symmetries: \\
(i)~~{\bf Conformal transformations:} $z\rightarrow f(z)$, $\bar z\rightarrow\bar f(\bar z)$, the same transformations on the complex plane that we found in the bosonic model, Section \ref{fmB}.\\
(ii) {\bf Left-holomorphic-chiral and right-anti-holomorphic-chiral transformations:} which act on the Weyl-Majorana fermions as follows:
\bea\label{chiralt}
\psi\rightarrow\psi'&=&e^{i\,\a(z)}\,\psi\nn
\ti\psi\rightarrow\ti\psi'&=&e^{i\,\ti\a(\bar z)}\,\ti\psi~.
\eea

Quantisation now proceeds similarly to what we did for the free, massless boson. One can do canonical quantisation (see Frishman and Sonnenschein (2010:~Section 2.12)) or again use circle integration and radial ordering to calculate operator product expansions, and get from them the commutation relations for the generators. \\

First, let us notice that the definition of primary fields, Eq.~\eq{prim} in the Appendix, applies equally well to fermions as it does to bosons. Also, the stress-energy tensor for fermions can be introduced, analogously to Eq.~\eq{freeT} (as we do below). 

The Weyl-Majorana fermion is a primary field, as one finds from its operator product expansion with the stress-energy tensor \eq{ope}. Like before, we have conserved currents (i.e.~annihilated by $\bar\pa$ for the meromorphic, and by $\pa$ for the anti-meromorphic current) associated with the sets of symmetries (i) and (ii) above:
\bea\label{fJT}
J(z)&=&:\psi^\dagger\psi:~,~~~~~\bar J=:\ti\psi^\dagger\,\ti\psi:\nn
T(z)&=&-{1\over2}:J(z)\,J(z):=-{1\over2}:\left(\psi^\dagger\,\pa\psi-\pa\psi^\dagger\,\psi\right):~,
\eea
with a similar expression for $\bar T$ in terms of $\ti\psi$. $\chi$ has conformal dimension $({1\over2},0)$, and $\ti\chi$ has conformal dimension $(0,{1\over2})$. The central charge \eq{TTope} is $c=\half$ (and so, a Dirac fermion, which is the sum of two Majorana-Weyl fermions, has twice this amount, i.e.~$c=1$). 

Notice that despite the half-integral values of the conformal dimensions, the currents $J(z), \bar J(\bar z)$ in \eq{fJT} actually have the same conformal dimensions as the currents \eq{normalT} in the bosonic model. This is because they are now {\it quadratic} in the fermions. 

As it turns out, the operator product expansion of $J$ with itself is identical to that in \eq{JJope}. Consequently, because $T$ in \eq{fJT} satisfies the Sugawara construction (cf.~after Eq.~\eq{freeT}), the product expansions of $T$ with itself and of $T$ and $J$ are also identical to those in \eq{TTope} and in \eq{JJope}. Now since also here, $J$ and $T$ are meromorphic operators, we can expand them in coefficients $J_n$ and $L_n$, as in \eq{Laurent}. The resulting algebra is thus the very same as in the bosonic case, \eq{ALAk1}, i.e.~the semi-direct product of the Virasoro algebra with $c=1$ with the abelian affine algebra at level $k=1$ (and its anti-meromorphic copy)! Therefore, also the state-space will be the same, since it is constructed as the Hilbert space on which the algebra acts. 

The implication of this basic fact---the agreement of the two models on their algebra of currents---will be explored in the next Section. We will see that it naturally leads to the existence of a duality, and thus to the formulation of a theory comprising the two models.

\section{Boson-Fermion Duality Illustrates the Schema}\label{bfdual}

In this Section, we show how Section \ref{bfd}'s bosonic and fermionic models (models in our sense!) illustrate the schema set out in Sections \ref{schema} and \ref{symdual}, leading up to the definition of duality in Section \ref{DlyDef}. In Section \ref{ddict}, we state the basic `dictionary' of the duality, mapping fields, currents and stress-energy tensors. Then Section \ref{twois} builds on this, to show that the two models are isomorphic, in exactly the sense of our schema: i.e.~as regards the whole trio of states, quantities and dynamics. Then in Section \ref{mat}, we return to Section \ref{RandI}'s theme: of defining a theory by abstraction from its models. We first note some special features of our case-study, in particular that it has just two isomorphic models; and then we define a theory, a common core, from the two models. In Section \ref{abstraction}, we discuss other ways one might define a theory from these models, i.e.~so as to have them be representations of it. Finally, we briefly discuss generalizations of our case-study: to include massive particles, and to include non-abelian degrees of freedom (Section \ref{mgd}).

\subsection{The duality dictionary}\label{ddict}

In Section \ref{bfd}, we derived the algebraic structures of the bosonic and the fermionic models. In particular, we saw that both models satisfy the enveloping algebra of the affine algebra (Eq.~\eq{ALAk1} and Eq.~\eq{ALA} in the Appendix) with $c=1$ and $k=1$, for the quantum currents corresponding to the Noether symmetries.

At this point, Frishman and Sonnenschein (2010:~Section 6.1) conclude that the two theories are equivalent. They write: \\
\indent`Due to the uniqueness of the irreducible unitary $k=1$ representation of the affine Lie algebras, and the fact that the infinite-dimensional algebraic structure fully determines the theories, we conclude that {\it in two space-time dimensions the theories of massless free scalar field and Dirac field are equivalent.} The equivalence implies that every operator of one model should have a partner in the other model, in such a way that the operator product expansions of these operators should be identical.' (p.~133:~Italics in the original).

While we basically agree with this statement, one of its clauses---that the infinite-dimensional algebraic structure fully determines the theories (in our jargon form Section \ref{TandM}: the models)---has not been proven, and requires some explanation and justification. Doing that is our plan for this subsection and the next. In this subsection, we give the basic dictionary that Frishman and Sonnenschein refer to in the quoted passage. (Then in Section \ref{states}-\ref{dyns}, we will argue that this infinite-dimensional algebraic structure indeed is sufficient to fully specify what we have called a model.)

The duality dictionary is given by the correspondence of the bosonic affine current algebra currents with the corresponding fermionic currents, and between the stress-energy tensors, as follows (cf.~Frishman and Sonnenschein (2010:~Section 6.1))\footnote{The dictionary thus relates \eq{Jcurrents} and \eq{normalT} to \eq{fJT}. We here add the subscripts `B' and `F' for `bosonic' and `fermionic', respectively.}:
\bea\label{dd}
J_{\tn{B}}(z)=\pa\f(z)~&\leftrightarrow&~J_{\tn{F}}(z)=~:\psi^\dagger(z)\,\psi(z):\\
T_{\tn{B}}(z)=-{1\over2}:\pa\f(z)\,\pa\f(z):~&\leftrightarrow&~T_{\tn F}(z)=-{1\over2}:\left(\psi^\dagger(z)\,\pa\psi(z)-\pa\psi^\dagger(z)\,\psi(z)\right):~,\nonumber
\eea
and similarly for the anti-meromorphic currents. We have already seen that both sides satisfy the same operator product expansion, and therefore they satisfy the same algebra. 

Notice that only fermion bilinears appear in \eq{dd}. This is what we expect, since a single boson ($h=1$) should correspond to a pair of fermions ($h=\half$ and $\bar h=\half$). But it is not a priori clear which bosonic field a single fermion should correspond to. One would here expect to take some kind of `square root' of the boson. But, surprisingly, the dictionary turns out to extend to a single fermion field as follows:
\bea\label{expf}
:e^{i\,\f(z)}:~~&\leftrightarrow&~~\psi(z)~,~~~~:e^{-i\f(z)}:~~\leftrightarrow~~\psi^\dagger(z)\nn
:e^{-i\bar\f(\bar z)}:~~&\leftrightarrow&~~\ti\psi(\bar z)~,~~~~:e^{i\,\bar\f(\bar z)}:~~~\leftrightarrow~~\ti\psi^\dagger(\bar z)~,
\eea
and again the operator product expansions agree. At first, this is a very surprising result, reminiscent of the construction of a coherent state. However, the operator product expansion shows that the conformal dimension of $:e^{i\a\,\f(z)}:$ is $h=\a^2/2$, so the above dictionary indeed reproduces $h=\half$ and $\bar h=\half$ for the left- and right-chiral Dirac fermions, respectively. This is indeed a purely quantum result, with no straightforward classical analogue.

One of the most surprising features of \eq{expf} is that $\f$ satisfies canonical commutation relations, while $\psi$ satisfies canonical {\it anti-}commutation relations. How can this be? To calculate the anticommutator of two $\psi$'s, one uses the formula $e^Ae^B=e^{[A,B]}\,e^Be^A$, which holds when $[A,B]$ is a c-number. Using this formula to evaluate the commutator of two exponentials, and using the canonical commutation relations for the boson and its conjugate canonical momentum, one finds that, indeed, the left- and right-chiral fermions anti-commute! 

\subsection{Two isomorphic model triples}\label{twois}

In Section \ref{ddict}, we gave the `dictionary' between the bosonic and fermionic models. This is a bijection between the basic operators of the theory (the fields). But there is, of course, more to duality than this. We need to show that the two models are {\it isomorphic,} as triples.
Recall our second sense of `model', in Section \ref{models;was22B}: as a representation of a theory. So a model is a triple (what we called the `model triple'), together with some specific structure. And the model triple was not a `pure copy' of the theory, but a representation of it using the specific structure. The next three subsections deal with the triples of states, quantities, and dynamics. In each of the subsections we show the existence of an isomorphism between the states, quantities, and dyamics of the two models (in the third case, an equivariance relation). This will justify that the boson-fermion equivalence is indeed a duality, in the sense of our schema. The final subsection considers the interpretation of this duality.

\subsubsection{States}\label{states}

We begin by showing that the state spaces of the two models are isomorphic. This will form the first item in our model triple. 
The state spaces of the bosonic and the fermionic models were introduced at the ends of Sections \ref{fmB} and \ref{fdirac}, as the Hilbert spaces obtained from the representations of the enveloping algebra of the affine Lie algebra \eq{ALAk1} (which, for short, we shall also call the `enveloping algebra'). We give some more details here.

The enveloping algebra is realized in the bosonic and the fermionic cases in terms of different fields: the bosonic $J$-currents \eq{Jcurrents} and the fermionic $J$-currents \eq{fJT} are defined in terms of different fields. Consequently, also the stress-energy tensors, given in the two cases by the Sugawara construction, differ, and so do the mode operators $J_n$ and $L_n$ which enter the enveloping algebra of the affine Lie algebra. In other words, we have two representations of the same algebra. Let us call these ${\cal A}_{\tn B}$ for the bosonic representation and ${\cal A}_{\tn F}$ for the bosonic representation.

The state spaces are, as mentioned in Sections \ref{fmB} and \ref{fdirac}, the Hilbert spaces obtained from the representations of the enveloping algebra, Eq.~\eq{ALAk1}: or, better, its generalisation, Eq.~\eq{ALA} in the Appendix, to general underlying Lie group and general value of $k$. The vacuum state is obtained by requiring $J^a_n\,|0\ket=0$, for $n\geq0$, where $a$ labels the generators of the Lie algebra. Here, the generators $J^a_n$ are the ladder operators of the algbera. This condition can also be understood physically as the regularity condition $J(z)\,|0\ket=0$ at $z=0$ for the affine currents. 

The non-trivial representations are constructed from the irreducible representations of the algebra, which are uniquely characterised by the highest-weight states (analogous to states of maximal $J$ for SU(2)), obtained by application of a primary field (cf.~Eqs.~\eq{ope} and \eq{prim}) to the vacuum. If we denote such a state by $|l,\bar l\ket$, where $l,\bar l$ are the representations, the remaining states in the representation (so-called `descendants') are obtained by appropriately applying generators, and take the generic form: \\
$L_{-m_1}\cdots L_{-m_M}\,\bar L_{-\bar m_1}\cdots\bar L_{-\bar m_{\bar M}}\,J^{a_1}_{-n_1}\cdots J^{a_N}_{-n_N}\,\bar J^{a_1}_{-\bar n_1}\cdots \bar J^{\bar a_{\bar N}}_{-\bar n_{\bar N}}|l,\bar l\ket$, for some integers $M,\bar M,N,\bar N$. For instance, on the fermionic model, expanding the fermion into modes:\footnote{Here, $\nu=\half$ for periodic boundary conditions, and $\nu=0$ for anti-periodic.} $\psi(z)=\sum_{r\in{\mathbb{Z}}+\nu}{\psi_r\over z^{r+\half}}$, the $\psi_{-n}$ operators are creation operators, and the $\psi_n$ are annihilation operators. The fermionic vacuum is defined as $\psi_n\,|0\ket=0$ ($n>0$), and the states in the Hilbert space are of the type $\psi_{-n_1}\cdots\psi_{-n_k}\,|0\ket$. 

We will use the fact that {\it the irreducible unitary representations of the enveloping algebra, thus constructed, are unique up to unitary equivalence.} (See e.g.~Frishman and Sonnenschein (2010, p.~133).\footnote{For more details, see Di Francesco et al.~(1997:~Chapter 14) or Kac (1990).}

So, for each of the models, we construct a representation of the algebra on a Hilbert space, ${\cal H}_{\tn B}$ for the bosons and ${\cal H}_{\tn F}$ for the fermions; and these representations are unitarily equivalent. Let us denote the unitary operator in question $U$. 

\subsubsection{Quantities}\label{physq}

Next, we show that also the quantities of the two models are isomorphic, thus providing the second item of the model triple. What are the relevant physical quantities? We already mentioned in Section \ref{DlyIsom} that we take the physical quantities to be all the renormalizable, self-adjoint operators constructed from a more basic set of quantities (which in a moment we will identify with the currents) and respecting the appropriate symmetries. 

Let us look again at the two representations, ${\cal A}_{\tn B}$ and ${\cal A}_{\tn F}$, of the enveloping algebra of the affine Lie algebra, which we discussed in Section \ref{states}. We can say more about these representations: for we already have the `dictionary', Eq.~\eq{dd}-\eq{expf}, between the fields. At this point we know that this dictionary is a bijection, but we wish to find a condition for it to be an isomorphism. First: notice that this bijection induces a similar bijection between the respective mode operators, i.e.~it induces a bijective map between the algebras, $d:{\cal A}_{\tn B}\rightarrow{\cal A}_{\tn F}$: ($d$ for `duality')
\bea\label{isoJL}
d(J_{n,{\tn B}})&=&J_{n,{\tn F}}\nn
d(L_{n,{\tn B}})&=&L_{n,{\tn F}}~.
\eea
This is of course just the statement that we have two equivalent representations of the enveloping algebra. 

From the previous section, we know that the representation spaces ${\cal H}_{\tn B}$ and ${\cal H}_{\tn F}$ of these algebras are constructed from the irreducible representations of highest weight, which are unique up to unitary equivalence. Since the construction of the representations of the states is the same for the two models (i.e.~in terms of the algebra generators mapped by Eq.~\eq{isoJL}), the  algebra generators themselves must be compatible with the bijection which maps the highest weight representations on the two sides. So, for consistency we must require:
\bea\label{iso}
d(\cdots)=U^\dagger\,(\cdots)~U~,
\eea
where $U$ is the {\it same} map used to map the states of the highest-weight representations in the previous subsection. If the maps $U$ and $d$ were not related {\em a la} \eq{iso}, the structure of the theory would not be preserved by the duality (since  an operator acting on a state would not be mapped to the the corresponding operator acting on the corresponding state). In short: once we have fixed the states to be mapped by $U$, as in Section \ref{states}, then the {\it same} transformation must map the quantities, as in \eq{iso}.\\

But Eq.~\eq{iso} is precisely the condition for the map to be an {\it isomorphism}, in addition to a bijection. And it now follows from the above that the modes of the currents of the two models are mapped as:
\bea\label{JLn}
U^\dagger \,J_{n,{\tn B}}~U&=&J_{n,{\tn F}}\nn
U^\dagger \,L_{n,{\tn B}}~U&=&L_{n,{\tn F}}~.
\eea
It also follows that the same relations hold for the currents \eq{dd} themselves:
\bea\label{Udd}
U^\dagger\, J_{\tn B}(z)~U&=&J_{\tn F}(z)\nn
U^\dagger\,T_{\tn B}(z)~U&=&T_{\tn F}(z)~.
\eea
The same map also maps the fermionic operators:
\bea\label{Uepsi}
U^\dagger:e^{i\,\f(z)}:U&=&\psi(z)~,~~~~~U^\dagger:e^{-i\,\f(z)}:U=\psi^\dagger(z)\nn
U^\dagger:e^{-i\,\bar\f(\bar z)}:U&=&\ti\psi(\bar z)~,~~~~~U^\dagger:e^{i\,\bar\f(\bar z)}:U=\ti\psi^\dagger(\bar z)~.
\eea
This is the formalization of \eq{expf}, which was justified by checking its properties (including the correct statistics, but also all the correct correlation functions). Given \eq{expf}, the factors of $U$ follow from the fact that  $\psi(z)$ creates or annihilates a {\it fermionic} state in ${\cal H}_{\tn F}$, and this state is obtained from the corresponding bosonic state in ${\cal H}_{\tn B}$ using the same map $U$. The operator creating or annihilating this state must therefore transform bilinearly in $U$. One can also check that the map between the affine currents, the first equation in \eq{JLn}, follows from \eq{Uepsi}. 

We have discussed the maps between the operators of the enveloping algebra of the affine Lie algebra, and the corresponding fields. We now discuss the {\it physical} quantities, ${\cal Q}_{\tn B}$ and ${\cal Q}_{\tn F}$, of self-adjoint operators respecting the relevant symmetries.
In the fermionic model, all the self-adjoint operators are quadratic in the fermions: and taking into account the chiral symmetry algebra, they must necessarily be powers of the fermionic currents \eq{Udd} (and their anti-chiral counterparts), appropriately normally-ordered. In fact, arbitrary analytic functionals of the currents are allowed. There can be a mixing between the meromorphic and anti-meromorphic sectors in the analytic functionals, but only such that the chiral symmetry algebra is preserved. Thus, the enveloping algebra indeed contains all of the information about the physical quantities: ${\cal Q}_{\tn F}$ consists of arbitrary analytic functionals in the fermionic currents \eq{Udd} (and their anti-meromorphic counterparts), appropriately normally-ordered. The normal ordering automatically ensures that these correlation functions are well-defined.

One can in principle enlarge the set ${\cal Q}_{\tn F}$ to also contain the correlation functions of operators which violate the chiral symmetry algebra, evaluated on the same states, i.e.~without changing the Lagrangian of the model. However, one is then changing the symmetries of the model triple $f=\bra{\cal H}_{\tn F},{\cal Q}_{\tn F},{\cal D}_{\tn F}\ket$, and hence one is defining a {\it new model triple} (and, consequently, if there is a duality for this larger class of models, one is defining a new theory). We will return to this possibility when we discuss sine-Gordon/massive Thirring duality in Section \ref{TSG}.

In the bosonic model, there is a similar structure. Now it is not the mixing of {\it chiralities} of the fields which the symmetries forbid, but the appearance of the {\it underived} field $\f(z)$ in ${\cal Q}_{\tn B}$. Namely, the translation symmetry $\Phi\rightarrow\Phi+a$ (or, more generally, the affine current symmetry algebra \eq{aca}), forbids the appearance of operators which depend on $\f(z)$ or $\bar\f(\bar z)$ {\it directly}, i.e.~as opposed to depending on them through their derivatives. When the model is written in terms of the meromorphic and anti-meromorphic parts of $\Phi$, translation symmetry is preserved iff all operators depend on the {\it derivatives} of the scalar field. This is precisely how the scalar field appears in the currents \eq{Udd}. Thus, once again, the affine current symmetry algebra of the model tells us that all the operators which are physical quantities of ${\cal Q}_{\tn B}$, must be analytic functions of the two currents (and their anti-meromorphic counterparts). ${\cal Q}_{\tn B}$ consists of all possible analytic functionals of those currents. As for the renormalizability constraint on the physical quantities: as we mentioned before, normal ordering automatically takes care of getting the correct renormalized quantities. 

To end this subsection, let us get a possible worry out of the way. Namely, one might be concerned that \eq{Uepsi} seems to be introducing complex operators, such as $:e^{i\,\f(z)}:$ , into the set of physical quantities of the bosonic model, which is supposed to be entirely real. But this is not quite right. For \eq{Uepsi} are in fact {\it not} physical operators, on our conception of the term. It is indeed true that $:e^{i\,\f(z)}:$ is not self-adjoint in the bosonic model, but then neither is $\psi(z)$ self-adjoint in the fermionic model. Neither of these two operators are thus to be taken as physical on either of the two sides, even if we can build physical operators, such as the currents \eq{dd}, by taking powers of them. Thus, none of the operators in \eq{Uepsi} belong to either ${\cal Q}_{\tn B}$ or ${\cal Q}_{\tn F}$. Rather we should think of such operators as {\it states} in the Hilbert space, using the state-operator correspondence. Indeed, recall, from Section \ref{states}, that the states in the fermionic Hilbert space are of the type: $\psi_{-n_1}\cdots\psi_{-n_k}\,|0\ket$. So this explains how $\psi(z)$ can be a physical operator---namely, it creates and annihilates physical states in ${\cal H}_{\tn F}$---while it is not self-adjoint and does not belong to the set ${\cal Q}_{\tn F}$ of physical quantities. 

Perhaps the most surprising aspect of the above is the fact that, in two dimensions, {\it the Hilbert space of the bosonic model contains  fermionic states}, e.g.~states with conformal weight $(\half,0)$. In fact, because the conformal weight of the operator $:e^{i\,\a\,\f(z)}:$ is $h=\a^2/2$, the Hilbert space contains a 1-parameter family of states with a continuous range of Euclidean spin values.\footnote{Our use of the word `Euclidean spin' here follows the jargon in the physics literature, for the eigenvalue under Euclidean rotations, as we mentioned in Section \ref{fmB}. It is questionable whether such jargon is justified by the physical interpretation in 1+1 dimensions. But we will not need to dwell on this point, since our main aim in this Section is formal.} This feature of the quantised models is indeed surprising---and illustrates our theme of surprise announced in (2) of Section \ref{orient}.

\subsubsection{Dynamics}\label{dyns}

Finally, we discuss the equivariance of the dynamics of the two model triples. `Dynamics' can be understood in different ways in different theories, and even in one formulation of a single theory. Think, for example, of the difference in the dynamics if it is formulated in the Heisenberg or in the Schr\"odinger pictures of a theory. 

First of all, we have formulated our model triples in the Heisenberg picture: operators are generally time-dependent and the states are time-independent. Let us consider the bosonic model first. We are working in Euclidean spacetime, but when analytically continued to Minkowski spacetime the operator $H_{\tn B}:=L_{0,{\tn B}}+\bar L_{0,\tn B}$ (which generates dilatations on the plane) is the generator of time translations, and is to be identified with the Hamiltonian. It is indeed the 00-component of the stress-energy tensor. The same is true in the fermionic model: the 00-component of the stress-energy tensor, which is the generator of time translations, is the operator $H_{\tn F}:=L_{0,{\tn F}}+\bar L_{0,\tn F}$. These two Hamiltonians are mapped to each other by the map $U$ in \eq{JLn} (and the anti-meromorphic version of that equation). Thus the dynamics is correctly preserved by the duality map: more precisely, it is equivariant with the unitary transformation.

We see that the requirement that the dynamics is correctly mapped between the two theories does not give any additional piece of the duality map, but simply follows from the other two: since all the quantities were already dual. This seems to be a general fact in dualities between quantum field theories, though we do not necessarily expect it to be the case for any duality. Namely, in a quantum field theory model, once we require that all states and physical quantities between the two theories map correctly, i.e.~according to the {\it isomorphism} \eq{iso}, then all correlation functions of the model automatically map correctly as well. But the correlation functions of a quantum field theory model exhaust all the dynamical information about the model, so that knowing all the correlation functions we can, in principle, reconstruct the dynamics. 

Thus, an alternative way to formulate duality is in terms of the correlation functions. This is, in fact, how duality is usually formulated in the string theory literature. However, we find our own conception of duality more illuminating, because it allows us to map the states and the operators individually and directly, without having to unscramble this information from the full set of correlation functions. But either way, the main point is that the dynamics is correctly equivariant with the duality map.

\subsubsection{Interpretation}\label{interpn524}

With our boson-fermion duality now in hand, we return to Section \ref{DlyInterpn}'s theme that a duality should respect the interpretation maps introduced in Section \ref{interp}. Our initial point in Section \ref{DlyInterpn} was that since duality relates two model triples, and interpretation maps apply to model triples (strictly speaking: their components, such as the set of quantities), interpretation simply proceeds independently on the two sides of the duality. The range of these interpretation maps could  be: either two distinct but isomorphic `sectors of reality', or the very same sector of reality---in which case there is a triangular rather than square diagram, as in Figures \ref{interpdlydiag1} and \ref{interpdlydiag2}. We also said that the choice between these cases was a matter of an `internal' vs.~an `external' interpretation.

External interpretations for the models are straightforward to read off from the model triples: they are the bosonic, respectively fermionic, interpretations which the two models come with in the first place (i.e.~in this case, their original historical interpretations). There is a natural map $I_{\tn{Int}}$ which assigns intensions: for example, it maps a bare or abstract bosonic state to the meaning `boson on a line with such-and-such properties'. And the map $I_{\tn{Ext}}$ assigns extensions. For example, it maps the abstract expectation value of a bosonic field to a measurable property of a specific boson that is fixed  as the reference by the context of use. For example, the property might be the boson's amplitude or probability of being at a specific place at a specific time: we say `amplitude or probability', since we understand a value or expectation value, written schematically as $\langle Q , s \rangle$, to also represent all the matrix elements $\langle s_1 |{\hat Q} | s_2 \rangle$ (cf.~Section \ref{DlyIsom}-(1)). In the fermionic model, there are similar interpretation maps for the fermionic model triple: where the codomains of the maps are now not the properties of bosons, but the properties of (say) electrons! Thus clearly, on the external interpretation the codomains of the maps do not need to agree. In philosophical jargon: the set of possible worlds where the bosons and the fermions appear need not be the same sets of worlds.

The internal interpretation abstracts from the specific structure of the two models, so that the two model triples receive the same interpretation. This interpretation is therefore best worked out at the level of the theory. Since the duality is exact, each model contains both bosonic and fermionic states, as we discussed in Section \ref{physq}. This means that the world (or set of worlds) which is the codomain of our interpretation, should contain both bosons and fermions. The interpretation maps, applied to the models, must commute with the duality map $d$ (or, equivalently, $U$) in Eqs.~\eq{isoJL}-\eq{iso}. The interpretations will map $J(z)$ and $T(z)$ to a current and a stress-energy tensor, respectively; (as intensions or as extensions, as appropriate: in the ensuing discussion, we shall not make this distinction, since everything we say applies equally well to both kinds of maps). The internal interpretation also maps the abstract expectation value of a bosonic field (whether written as $\f(z)$ or as a fermion bilinear) to the amplitude or probability of some bosonic event. Likewise, the abstract expectation value (matrix element) of a fermionic field between two suitable given states  will be mapped to the transition amplitude between two fermionic states. 

\subsection{Defining a theory from the two model triples}\label{mat}

We now return to Section \ref{RandI}'s theme: of defining a theory by abstraction from its model triples. We first discuss how our duality's having just two isomorphic model triples makes this enterprise vulnerable, in two ways (Section \ref{531}). Then we undertake to define a theory, a common core, from the two models; and discuss their specific structures (Section \ref{532}). Thus the general issues of Section \ref{531}---which return us to the Hilbertian themes at the start of Section \ref{intro}---will lead in to the specific details of Section \ref{532}.

\subsubsection{Two vulnerabilities}\label{531}

At first sight, there might seem to be no issues about the definition of a theory (in our sense, from Sections \ref{introschema} and \ref{bare;was22A}) from a pair of  model triples that are isomorphic: or indeed, from any set of two or more isomorphic model triples. Can we not simply define the theory as the structure of which the isomorphic model triples are isomorphic `concrete' copies? More precisely: here we should clarify the phrase `as the structure of which', in order to respect Section \ref{models;was22B}'s point that a model (in our sense!) usually realizes a theory by being a representation of it, and representation allows mere homomorphism, rather than isomorphism. So the thought is: can we not simply define the theory as the structure, $S$ say, of which the (two or more) isomorphic model triples are representations---as it happens, isomorphic ones? Talk of $S$ thus defined as a theory will then engender admitting all its representations as models in Section \ref{models;was22B}'s sense.

We agree that this relaxed attitude is tenable. After all, `theory' is a term of art: so one is at liberty to define it as one sees fit. But in both pure mathematics and theoretical physics, the abstraction (`extraction') of a general pattern or structure from a given set of examples is a matter of judgement, to be made in the light of one's aims and intuitions, including the aim of representing the world as accurately as possible. 
This means that there are two related worries one might have: that is, two limitations to which the above strategy for defining a theory is vulnerable. 

(i): Suppose that---as in our bosonization case-study---the given examples are isomorphic: where `isomorphic' is itself a term of art, made precise by some mathematical structure we see exactly copied in the examples. Then one worries that there might be non-isomorphic examples which one has not thought of (has not `been given') that are equally good examples of the general pattern one is trying to write down---that equally fit one's aims and intuitions. 

(ii): Suppose the given examples are not isomorphic; (again, in the sense we have in mind, especially initially). Still one worries that they might not be sufficiently varied, so that the pattern one writes down after considering them is too restrictive. That is: the pattern encodes aspect(s) that, given one's aims and intuitions, are really accidental commonalities of the examples. Recall (from the end of Section \ref{RandI}) the example of colour as an artefact that could  beset Frege's definition of direction as an equivalence class of mutually parallel lines. 

The general answer to these worries lies in the point, argued in Section \ref{logweak}, that for models which purport to describe the physical world, the distinction between what is in the triple and what is specific structure cannot be blurred.\footnote{As an example, consider the bosonic and fermionic models, but now weakened by the stipulation that the Virasoro algebra belong to the model triple, while the affine Kac-Moody algebra (and the third line in Eq.~\eq{ALAk1}) belongs to the specific structure. As we argued in Section \ref{logweak}, this stipulation {\it changes the physical content} of the models, and so it is not innocuous. The models thus obtained contain different numbers of (uninterpreted) physical degrees of freedom, and so cannot describe the bosons or the fermions of Section \ref{bfd}. This is because the boson and the fermion CFTs (even before they are physically interpreted) treat the Kac-Moody degrees of freedom not as `accidental commonalities', in the sense of Section \ref{RandI}: but as physical, and related to the Virasoro generators by the Sugawara construction. (For example, if we drop the chiral symmetry on the fermionic model, we lose the reason to restrict to chiral quantities only: cf.~Section \ref{physq}; and likewise for the boson's affine current symmetry algebra.) Thus the boson and fermion models are {\it not} dual, if based on just the Virasoro algebra. We thank Josh Hunt for bringing up this example.\label{josh}} So these worries can be set aside: for a give target system, typically only a single triple will provide a complete description---up to isomorphism, that is. 

So much by way of rehearsing the general issues about defining a theory from models: specifically, from model triples. For the purposes of this paper, what matters is how these issues play out in our case-study, bosonization. Section \ref{532} will give details about this. But to summarise:--- The theory we will construct in Section \ref{532} is the simplest one that can be constructed from our bosonic and fermionic model triples, using the isomorphism at hand from Section \ref{twois}. But in line with this Subsection's comments, we make no claim that is the only (or even best) way to `extract a pattern' from the model triples. Indeed, our final two Subsections, Sections \ref{abstraction} and \ref{mgd},  {\em will} consider other such ways: in particular, by taking into account other models.

\subsubsection{Defining the theory}\label{532}

In this subsection, we collect our results from Section \ref{twois} about the two model triples, in order to define our {\it theory}. We have so far defined two {\it models:---}

(a) A bosonic model triple, $b:=\bra{\cal S}_{\tn B},{\cal Q}_{\tn B},{\cal D}_{\tn B}\ket$, with: ${\cal S}_{\tn B}={\cal H}_{\tn B}$, the Hilbert space which is constructed from the irreducible highest-weight representations of the enveloping algebra of the affine Lie algebra, represented on the bosons. `Represented on the bosons' here means that the generators of the affine Lie algebra \eq{ALAk1}, which proceed from the Laurent expansion \eq{Laurent}, are constructed from the {\it bosonic} fields \eq{Jcurrents} and \eq{freeT} (and their anti-holomorphic counterparts). Thus, ${\cal Q}_{\tn B}$ is the set of normally-ordered analytic functions in the bosonic currents $(J(z)_{\tn B},\bar J(\bar z)_{\tn B}, T(z)_{\tn B}, \bar T(\bar z)_{\tn B})$; and ${\cal D}_{\tn B}=L_{0,\tn B}+\bar L_{0,\tn B}$ the dilatation operator of the bosonic model. 

(b) A fermionic model triple, $f:=\bra{\cal S}_{\tn F},{\cal Q}_{\tn F},{\cal D}_{\tn F}\ket$, with: ${\cal S}_{\tn F}$ the same Hilbert space, represented on the fermions, ${\cal H}_{\tn F}$; ${\cal Q}_{\tn F}$ the set of normally-ordered analytic functions of the currents $(J(z)_{\tn F},\bar J(\bar z)_{\tn F}, T(z)_{\tn F}, \bar T(\bar z)_{\tn F})$; and ${\cal D}_{\tn F}=L_{0,\tn F}+\bar L_{0,\tn F}$ the dilatation operator of the fermionic model.

What structures make up the `specific structure' $\bar M$, of each of the models $B:=\bra b,\bar B\ket$ and $F:=\bra f,\bar F\ket$, in the sense of our notation in Section \ref{notation;was22C}, and especially Eq.~\eq{modelq}? On the bosonic side, the specific structure $\bar B$ clearly contains the field $\f(z)$ (and functionals of it), together with the symmetry algebra \eq{aca} acting on it. This symmetry algebra will, however, manifest itself in the model triple through the affine currents and their algebra. Also the defining relations of the field (equation of motion, etc.) are specific to $\bar B$. 

On the fermionic side, it is the field $\psi(z)$, with its chirality symmetry (and a different set of defining relations, equation of motion, etc.), which are parts of the specific structure $\bar F$. Though $\psi(z)$ defines a state in the Hilbert space, as discussed in Section \ref{physq}, thinking of this state as created by $\psi(z)$, i.e.~a fermion with certain meromorphic and chirality properties, it is part of the specific structure. All the Hilbert space knows about this fermion is that there is a state of conformal weight $(\half,0)$.\\

We are now ready to discuss the {\it theory} which we can construct from these two models: by discussing the common core of the two models, i.e.~the model triples (or roots), to which the theory is isomorphic: $b\cong f\cong T$. It is a theory based on four currents $(J(z),\bar J(\bar z),T(z),\bar T(\bar z))$ of conformal dimensions $(1,0)$, $(0,1)$, $(2,0)$, $(0,2)$, satisfying the enveloping algebra with $c=1$ and $k=1$. Its states are the unitary representations of this algebra. The dynamics is given by singling out the Hamiltonian $H=L_0+\bar L_0$. \\

{\it Sets of symmetries of the theory:} The theory has two built-in sets of symmetries: (i) a conformal group and (what we shall call), generated by the stress-energy tensor, (ii) an affine current symmetry algebra, generated by the $J$-currents. 

(i) The conformal symmetry group, Eq.~\eq{ct}, is represented in the same way in the two models---since this is a symmetry group of the background spacetime. This is also related to the fact that the Hamiltonian, and hence the dilations, are related by equivariance between the two theories. Again: $T(z)$,  the stress-energy tensor of the theory, is represented in both the  models.

(ii) The affine current symmetry algebra is represented very differently in the two models! Namely, as affine current algebra transformations (Eq.~\eq{aca} in Section \ref{fmB}(ii)) on the bosonic model, but as left- and right-chiral symmetry algebra (Eq.~\eq{chiralt} in Section \ref{fdirac}(ii)) on the fermionic model. This symmetry algebra restricts the kinds of physical quantities in the theory, as we have explicitly discussed in Section \ref{physq}, in the same way. Yet the theory does not ``see'' any of the features from which the affine current algebra arises: which are very different in the two models, viz.~the transformations of the fields Eq.~\eq{aca} vs.~\eq{chiralt}. 

We have discussed this duality in some detail because it is a good model to the more general, and technically involved, boson-fermion dualities in two-dimensional conformal field theory: to which we turn in Section \ref{mgd}. 

\subsection{Further abstraction}\label{abstraction}

The discussion in Sections \ref{physq} and \ref{mat} illustrates our theme of the limitations of abstraction. We constructed our {\it theory} from two isomorphic model triples: the model triple of the theory was built from the representations of the enveloping algebra of the affine Lie algebra \eq{ALAk1}, which give a unique set of {\it states} (discussed in Section \ref{states}), and a basic set of {\it quantities}: the generators of the algebra themselves. The {\it dynamics} was a choice of  a Hamiltonian among the quantities (Section \ref{dyns}). As we saw in Section \ref{physq}, the {\it full} set of quantities ${\cal Q}$ of the theory contains more than just the basic set: so that arbitrary analytic functionals of the currents ($J(z),\bar J(\bar z),T(z),\bar T(\bar z))$, and their derivatives, are allowed. Compare the discussion of the symmetry algebra (ii) in Section \ref{mat}. 

But we also know, from Sections \ref{RandI} and \ref{531}, that there are no general rules, fixed once-and-for-all, for defining theories. So one asks: to what extent is our procedure above unique? It is of course unique if what one wishes to describe is a boson or a fermion, as given systems with known degrees of freedom. But the procedure of abstraction suggests three natural ways in which our theory might be modified. The first way would make for a more {\it restrictive} theory; the other two entail further abstraction, thus allowing for a more {\it general} one. \\

(a) The {\it conformal symmetry group} (Section \ref{532}-(i)) was used to form the enveloping algebra. Hence it is realised by the states and the quantities, in the sense that the states and quantities form representations of this symmetry group. But one can construct a new theory in which the class of operators is {\it reduced}: namely, by placing restrictions on the conformal transformation properties of the quantities. This leads to a theory with a smaller set of quantities (and subsequently to bosonic and fermionic models with smaller sets of quantities). Alternatively, one can reduce the space of states by placing similar restrictions on them.

(b) The {\it affine current symmetry algebra} (Section \ref{532}-(ii)) limited the set of quantities to those that can be constructed from the currents $(J(z),\bar J(\bar z),T(z),\bar T(\bar z))$ (as mentioned at the start of this subsection). But one might decide that this is not a symmetry algebra one wishes to keep, for the physical system of interest; (for example, in the presence of mass terms, this symmetry algbera will be dropped). One then allows a larger class of, or even all, self-adjoint, renormalisable operators constructed from \eq{expf}, not just the ones that preserve this symmetry algebra. In this way, one clearly gets a richer theory (with a larger set of operators and states), of which Section \ref{bfd}'s two models are still representations.\footnote{Notice that the ambiguity here is in the best definition of the {\it theory}, not of the {\it duality}: cf.~footnote \ref{josh}.} And as we stressed in Section \ref{logweak}, it is a matter of physical judgment, which kinds of operators one needs to admit in order to describe the physics at hand. In particular: if one wants to add a mass, one is forced to generalise the theory in this way. The boson-fermion duality continues to hold, thanks to the existence of the maps Eq.~\eq{expf} and \eq{JLn}. But we get a {\it more general class of theories}, which will not necessarily be each other's duals: the generality of the class depends on which additional set of operators one takes on board with the quantities. This will be illustrated explicitly in Section \ref{mgd}. 

(c) Though the two models share the {\it spacetime coordinates} $z,\bar z$, these coordinates do not enter the basic considerations that led to building the states, quantities, and dynamics of the theory, in Sections \ref{states}-\ref{dyns}. Indeed, the basic object of interest, in constructing the triple, is the algebra of the mode operators $L_n$ and $J_n$ (and powers of them): and these are spacetime-independent. Furthermore, these modes contain essentially the same information as the spacetime-dependent currents $(J(z),\bar J(\bar z),T(z),\bar T(\bar z))$, i.e.~the latter can be reconstructed from the former through the Laurent expansions \eq{Laurent}, which take identical form in the fermionic case. So, one might decide that $z,\bar z$ are just book-keeping devices with no essential information about the theory. On this view, one can construct a theory just based on the algebra \eq{ALAk1} and its representations, without its spatio-temporal realization. The bosonic and fermionic model triples then still form (spatio-temporal) representations of this algebra: but one can envisage the existence of other representations, which are not spatio-temporally realised. This would presumably give rise to non-isomorphic models, in the sense of Sections \ref{RandI} and \ref{531}. While doing away with spacetime may seem a radical suggestion, it is not so uncommon: think e.g.~of spin chains as possible models.\\

Point (a) is a straightforward modification of our theory, but also of the models. So it should not be seen as illustrating the limitations of abstraction, in the sense of Sections \ref{RandI}  and \ref{531}. Rather, it is  a {\it method to obtain more restrictive theories,} by consistently strengthening the symmetry requirements of the models.

But points (b) and (c) do illustrate our remarks, in Sections \ref{RandI}  and \ref{531}, about the need for models to be `sufficiently varied'. By taking, in (b) and (c), some of the symmetry shared by the models to be accidental, one gets a larger class of models, which is likely to include non-isomorphic ones. 

In the next subsection, we will give such examples of dualities between isomorphic models which are more general: either because they have less symmetry, or because they have more fields, with additional symmetries. 

\subsection{General boson-fermion dualities}\label{mgd}

In this Subsection, we briefly discuss two generalizations of our basic duality. The two generalizations are important for our discussion since they fulfill, even better than our basic duality does, Section \ref{introbosonzn}'s two desiderata  for the choice of examples of dualities. Recall that these desiderata were: on the one hand, (i) an example should be mathematically precise  and represent established physics; and on the other hand, (ii) an example should be scientifically important (as described more fully in Section \ref{orient}). 

As we will see, these desiderata will be amply fulfilled by this Subsection's two generalizations. The first generalization especially illustrates scientific importance. The second illustrates, not so much mathematical precision {\em per se}: but rather, {\it mathematical richness and generality}, thus showing that the boson-fermion duality described in this Section is not an isolated `coincidence' that holds for free, massless models, but part of a very rich class of mathematically interesting (as well as rigorous!) isomorphic models. Thus we will here define a rich class of theories, each of which is an equivalence class of exactly two isomorphic model triples.

We will discuss the two kinds of generalization in turn, in the next two subsections. Section \ref{TSG} discusses the duality between the massive Thirring model and the sine-Gordon model. Section \ref{NAD} considers non-abelian versions of boson-fermion duality. In both cases, we must be brief and must suppress technical details.

\subsubsection{Duality between the Thirring model and the sine-Gordon model}\label{TSG}

The basic boson-fermion duality can be extended to include mass terms for the fermions, and interaction terms for both fermions and bosons. This generalisation is important, because it shows that the duality is not a special property that only occurs in the free, massless case, in which the action is conformally invariant. Massive, and interactive, theories are also subject to duality. So this strengthens the scientific importance of duality: it brings duality into the `real world'. In fact, the Thirring model-sine-Gordon duality is quite important in condensed matter systems. See Giamarchi (2003), Altland and Simons (2010). 

The massless Thirring model generalises the free, massless Dirac fermion by the addition of a quartic interaction term for the fermions, with coupling constant $g$. This quartic interaction is built from the $J$-currents, and so preserves the chiral symmetry algebra described in (ii), Section \ref{fdirac}. The model can be solved exactly, and the quantum theory is well-defined only for $g>-\pi$ (Coleman (1975:~p.~2094)).\footnote{Looking at the relation between the couplings \eq{couplings} (Section \ref{introbosonzn}), this will correspond to the value $\b^2<\infty$ of the bosonic coupling.}

The fermionic mass term in the massive Thirring model explicitly breaks the chiral symmetry algebra described in (ii), Section \ref{fdirac}. This is because the mass term which is added to the action of the Dirac fermion, Eq.~\eq{Dirac}, mixes the left- and right-chiral (Weyl) fermions. It takes the following form:
\bea\label{mass}
\D S_{\tn{mass}}=m\left(\ti\psi^\dagger\,\psi+\psi^\dagger\,\ti\psi\right)~.
\eea
Remember that the action \eq{Dirac} did not mix the left-chiral Weyl fermion $\psi,\psi^\dagger$ with the right-chiral Weyl fermion $\ti\psi,\ti\psi^\dagger$. The above mass term explicitly mixes the two, and so breaks the chiral symmetry algebra. 

The mass term \eq{mass} can be translated into a bosonic term using a straightforward generalisation of the dictionary \eq{expf}.  The generalisation is straightforward in that it takes the same form, but now depends on a bosonic coupling $\b$, which is related to the fermionic coupling through \eq{mass}. The equation of motion of the sine-Gordon model is:
\bea\label{sine}
\Box\f+{\m^2\over\b}:\sin(\b\f):=0~,
\eea
and so the fermionic coupling is related, through \eq{couplings}, to the bosonic coupling $\b$. Notice that, to linear order in $\b$, the above reduces to a standard bosonic mass term, with mass $\m$. We are here, however, interested in the exact model. 

There are divergences that need to be renormalised. The bosonic term corresponding to \eq{mass} is:
\bea
\D S_{\tn{sine-G}}={\m^2\over\b^2}\,:\cos\b\f:~,
\eea
where $\m$ is a scale which originates in the normal ordering procedure, and appears in the boson-fermion dictionary as an overall multiplicative constant.\footnote{The scale $\m$ is already present in the massless theory. But it does not play any important role, since it is just an overall renormalisation constant.}

The algebra underlying the model triple of this model is still the enveloping algebra of the affine Lie algebra with $c=1$, but the level now depends on the coupling: $k={\b^2\over4\pi}={1\over1+{g/\pi}}$. We see that in the limit of zero fermionic coupling, we reproduce the algebra at level 1. 

Our discussion, in Section \ref{twois}, of the isomorphic model triples, thus generalises to the massive Thirring and sine-Gordon models, with appropriate modifications. In both cases, the algebra is the enveloping algebra, now with a coupling-dependent level. Therefore, the discussion of the {\it states} is analogous to the one in Section \ref{states}: the Hilbert space consists of the highest-weight representations of the enveloping algebra, and their descendants. The {\it quantities} are constructed from a wider class of operators, compared to our discussion in Section \ref{physq}. Namely, the quantities now include non-chiral operators (respectively, operators which break affine current algebra transformations: see Section \ref{fmB}-(ii) for bosons, Section \ref{fdirac} for fermions) built from the fields: such as \eq{sine} for the bosons, and \eq{mass} for the fermions. Finally, the dynamics is still given by the Hamiltonian, which is the zero component of the Virasoro generator $L_0$, in the bosonic or fermionic representation. Thus, we get a theory by abstraction from these two isomorphic triples, as outlined in Section \ref{mat}. 

As we remarked before, this generalised theory explicitly illustrates our comment (b) in Section \ref{abstraction}, about the contingent nature of the chiral symmetry algebra. By allowing the theory to break the chiral symmetry algebra, we get a wider class of theories (which depend on the coupling and the mass): a class of which the basic free, massless case is just a special case. 

\subsubsection{Non-abelian boson-fermion dualities}\label{NAD}

As we mentioned in this Subsection's preamble, the free, massless bosons and fermions can be generalised in another direction (Witten (1984)): to include non-abelian degrees of freedom. We will not here provide any technical details, but we will simply list some of the important examples of dualities studied in the literature; all of which are conformal field theories, except for (c) and (e):\\

(a)~$N$ free Majorana (real) fermions, in an $N$-dimensional vector representation of $\mbox{O}(N)$. They are dual to the bosonic, Wess-Zumino-Witten (WZW) model with an $\mbox{O}(N)$ symmetry group. The Wess-Zumino-Witten model is a model whose action is built from a bosonic group element (rather than an algebra element), in this case an $N\times N$ matrix of $\mbox{O}(N)$. The bosonic and fermionic models are both invariant under the affine Lie algebra transformations of $\mbox{O}_{\tn L}(N)\times\mbox{O}_{\tn R}(N)$ (for left- and right-action, respectively) at level $k=1$. In both cases, the central charge is $c=N/2$. 

(b)~$N$ free, massless Dirac (complex) fermions are dual to a bosonic WZW model with group $\mbox{U}(N)$. The two models satisfy the affine Lie algebra of $\mbox{SU}_{\tn L}(N)\times\mbox{SU}_{\tn R}(N)\times\mbox{U}(1)$, with central charge $c=N$ and $k=1$. 

(c)~Mass terms can be added to the Dirac fermions in case (b), with modifications in the dictionary and in the bosonic theory which are similar to the ones discussed in the previous subsection. 

(d)~The Majorana and Dirac fermion models can be endowed with colour and flavour charges. Majorana fermions with $N_{\tn F}$ flavours and $N_{\tn C}$ colours, thus transforming under the group $[\mbox{O}(N_{\tn F})\times\mbox{O}(N_{\tn C})]_{\tn L}\times[\mbox{O}(N_{\tn F})\times\mbox{O}(N_{\tn C})]_{\tn R}$, are dual to the Wess-Zumino-Witten action with two bosonic fields, taking values in $\mbox{O}(N_{\tn F})$ and $\mbox{O}(N_{\tn C})$. In the same way, $N_{\tn F}\times N_{\tn C}$ Dirac fermions can be expressed as the sum of two Wess-Zumino-Witten actions, and a third term for an aadditional field. The three bosonic fields take values in the group manifolds $\mbox{SU}(N_{\tn F})$, $\mbox{SU}(N_{\tn C})$, and $\mbox{U}(1)$. Again, one finds two copies of the affine Lie algebra, now with levels different from one, viz.~$k=N_{\tn F}$ and $k=N_{\tn F}$, respectively. This duality can also be generalised to other gauge groups. 

(e)~Mass terms can be added to the theories (d) with flavour and colour charges, with appropriate modifications in the dictionary and in the bosonic theory, as before. \\

All of the above models end up having model triple structures which are constructed as representations of the enveloping algebra of the affine Lie algebra \eq{ALA} in the Appendix, for various values of the level, $k$, and the central charge $c$, and for different Lie groups. Thus their Hilbert spaces are constructed from the highest-weight representations, which as we already saw in Section \ref{states} are unique up to unitary transformations. Clearly, all of these theories can be given a set of states, as discussed in Section \ref{states}, of which the bosonic and fermionic models form two representations. Also, all of these theories are based on the same class of algebras, with generators $J_n$ and $L_n$ as in \eq{ALAk1} for the bosonic case, and likewise for the fermionic case in Section \ref{fdirac} and their anti-holomorphic counterparts  (see Eq.~\eq{ALA} in the Appendix). Thus, each of these theories can also be given a set of quantities, in the way discussed in Section \ref{physq}. For the non-chiral theories (c) and (e), this set of quantities is enlarged by the addition of non-chiral quantities, as we already discussed in detail in the abelian case in Section \ref{TSG}. Finally, the dynamics of these models is as discussed in Section \ref{dyns}. \\

In conclusion, this large class of examples, based on a general enveloping algebra of an affine Lie algebra, exemplifies our schema for duality in Sections \ref{schema} and \ref{symdual}, just as the basic case did in Section \ref{twois}. Namely: a duality is an isomorphism between models. More specifically: it is an isomorphism  between model triples; since models also have their own specific structures. And the theory obtained for each of the dualities accords with what we said in Section \ref{RandI}: equivalence classes of isomorphic model triples give rise to a theory which is itself a triple, in which the models' specific structures have been abstracted away. And finally: our comments about non-isomorphic models (in Sections \ref{RandI}, \ref{531}, and \ref{abstraction}) are illustrated by the examples (c)-(e). For these models have less symmetry: the theory which then results is more general. 

\section*{Envoi}
\addcontentsline{toc}{section}{Envoi}

In the physics and philosophy of physics literature, a duality in physics is agreed to be a matter of two theories being in some sense `the same'. In this paper, we have answered the question how this can be made precise, and illustrated our answer with a case-study: bosonization.

We have proposed that a duality is best understood formally, i.e.~in terms of uninterpreted theories: hence our term, `schema'. Namely: there is a bare theory---the common core of the two given theories---which has various models, among which are the two given theories. The duality then consists in the fact that these two models are isomorphic as regards the structure and notions given in the bare theory. (Thus each of the two models also has specific structure of its own, which is unmatched by the other; and the bare theory also has, in general, other non-isomorphic models.) Often, this isomorphism is a surprising fact, since the two given theories are presented in very different terms.

We spelt out this schema in detail, in Sections 2 to 3. Among the themes we emphasized are: (i) the distinction between theories and models, (ii) the role of interpretation, (iii) the relations between a duality and the symmetries of the two given theories, and (iv) the presence of non-isomorphic models in physics.

Then in Sections 4 and 5, we illustrated the schema with bosonization. This is a matter of a quantum field theory of bosons being in some sense `the same' as a quantum field theory of fermions. Nowadays, many such boson-fermion pairings are known. Our discussion emphasized the simplest, and earliest, case, which is known to hold exactly: a duality between a free, massless bosonic quantum field theory, and a free, massless fermionic theory, both in two spacetime dimensions. But we ended with a brief overview of other examples: involving, in particular, interacting and massive theories. (And there are extensions to higher dimensions: see e.g.~Kopietz (2008); as well as an experimental interest in these systems as realising e.g.~one-dimensional spin chains: Giamarchi (2003:~Chapter 2), Altland and Simons (2010:~Sections 4.3 and 9.4.4).)

Our schema, and this illustration of it, of course leaves plenty of work still to be done. As to physics, one should seek other illustrations of the schema: maybe some of these will prompt revision, or at least augmentation, of the schema. As to philosophy, one should ask what light this schema casts on philosophical debates about the interpretation of physical theories, and about such theories' equivalence. But we postpone these topics to another occasion. 

\section*{Acknowledgements}

We thank Joseph Kouneiher, not least for his patience! We also thank an anonymous referee for comments, the participants at the Symmetries and Asymmetries in Physics conference in Hannover, and especially Josh Hunt for comments. SDH thanks several audiences: the British Society for the Philosophy of Science 2016 annual conference, the Oxford philosophy of physics group, LSE's Sigma Club, the Munich Center for Mathematical Philosophy, and DICE2016. SDH's work was supported by the Tarner scholarship in Philosophy of Science and History of Ideas, held at Trinity College, Cambridge.

\section*{Appendix. Some Elements of Conformal Field Theory}
\addcontentsline{toc}{section}{Appendix. Some Elements of Conformal Field Theory}

In Section \ref{fmB}, we used the notion of a primary field. A primary field of conformal weight $(h,\bar h)$ is defined to transform, under a conformal transformation \eq{ct},  as follows:
\bea\label{prim}
\Phi(z,\bar z)\rightarrow\left({\pa f\over\pa z}\right)^h\left({\pa\bar f\over\pa\bar z}\right)^{\bar h}\,\Phi(f(z),\bar f(\bar z))~.
\eea
This is in analogy with the transformation law for covariant tensors in ordinary QFTs: it takes the transformation property of the field under conformal transformations as defining for the class of primary fields. The physical significance of primary fields is discussed around Eq.~\eq{ope}. 

In our analysis in Sections \ref{bfd} abd \ref{bfdual}, an essential role was played by the enveloping Virasoro algebra \eq{ALAk1}, with $c=1$ and $k=1$. This algebra is a special case of the following general enveloping algebra of an affine Lie algebra:
\bea\label{ALA}
{}[L_n,L_m]&=&(n-m)\,L_{n+m}+{c\over12}\,n\,(n^2-1)\,\d_{n+m}\nn
{}[L_n,J_m^a]&=&-m\,J^a_{n+m}\nn
{}[J^a_n,J^b_m]&=&i\,f^{ab}_c\,J^c_{n+m}+k\,n\,\d_{ab}\,\d_{n+m}~.
\eea
Here, $c$ is the central charge and $k$ is the level, and $f^{ab}{}_c$ are the structure constants of the underlying Lie algebra of the affine Lie algebra. Notice that the above algebra contains, in the first line, the ordinary Virasoro algebra. And the last line is the affine Lie algebra. The middle line gives the commutation relation between generators of the two algebras.\footnote{For more on affine Lie algebras, see Di Francesco et al.~(1997:~Chapter 14) or Kac (1990).}



\section*{References}
\addcontentsline{toc}{section}{References}

\noindent
Ammon, M. and Erdmenger, J. (2015), {\em Gauge/Gravity Duality: Foundations and Applications}, Cambridge: Cambridge University Press.\\
\\
Atiyah, M.~F.~(2007) ``Duality in Mathematics and Physics'', lecture delivered at the Institut de Matem\`atica de la Universitat de Barcelona: {\tt http://www.imub.ub.es}.\\
\\
Atland, A. and Simons, B.~(2010). {\em Condensed Matter Field Theory}. Cambridge: Cambridge University Press.\\
\\
Barrett, T.W.~and Halvorson, H.~(2016). ``Glymour and Quine on Theoretical Equivalence''. {\it Journal of Philosophical Logic}, 45(5), pp.~467-483.\\
 http://philsci-archive.pitt.edu/id/eprint/11341\\
\\
Barrett, T.W.~and Halvorson, H.~(2016a). ``Morita Equivalence''. {\it The Review of Symbolic Logic}, 9(3), pp.~556-582 \\
http://philsci-archive.pitt.edu/id/eprint/11511\\
\\
Carnap, R. (1947) {\em Meaning and Necessity}, Chicago: University of Chicago Press\\
\\
Castellani, E.~(2017). ``Duality and `particle' democracy''. {\it Studies in History and Philosophy of Modern Physics}, 59, pp.~100-108.\\
\\
Caulton, A.~(2015). `` The role of symmetry in the interpretation of physical theories''. {\it Studies in History and Philosophy of Modern Physics}, 52, pp.~153-162.\\
\\
Coleman, S.~(1975). ``Quantum sine-Gordon equation as the massive Thirring model''. {\it Physical Review D} 11 (8), pp.~2088-2097.\\
\\
Corfield, D.~(2017). ``Duality as a Category-Theoretic Concept''. {\it Studies in History and Philosophy of Modern Physics}, 59, pp.~55-61.\\
\\
Corry, L. (1999),  `Hilbert and Physics 1900-1915', in Gray J., ed. (1999) {\em The Symbolic Universe: geometry and physics 1890-1930}, Oxford University Press: pp.~145-188. \\
\\
Corry, L. (2004), {\em David Hilbert and the Axiomatisation of Physics}, Springer-Science.\\
\\
Corry, L. (2006), `On the origins of Hilbert's sixth problem: physics and the empiricist approach to axiomatization', {\em Proceedings of the ICM 2006: Madrid Spain}, pp.~1697-1718.\\
\\
Corry, L. (2018),  `Mie's Electromagnetic Theory of Matter and the Background to Hilbert's 
Unified Foundations of Physics', this volume.\\
\\
De Haro, S.~(2016). ``Spacetime and Physical Equivalence''. In {\it Space and Time after Quantum Gravity}, Huggett, N. and W\"uthrich, C.~(Eds.), to appear.\\
\\
De Haro, S.~(2016a). ``Duality and Physical Equivalence''. \\
 http://philsci-archive.pitt.edu/id/eprint/12279 (the title of this preprint has changed). \\
\\
De Haro, S. (2017). ``Dualities and emergent gravity: Gauge/gravity duality''.
{\em Studies in History and Philosophy of Modern Physics}, 59, 2017, pp.~109-125.\\ 
\\
De Haro, S. (2017a). ``Invisibility of Diffeomorphisms''. {\it Foundations of Physics}, 47 (11), 2017, pp.~1464–1497.\\
\\
De Haro, S., Mayerson, D., Butterfield, J.N. (2016). ``Conceptual Aspects of Gauge/Gravity Duality'', {\it Foundations of Physics}, 46 (11), pp.1381-1425. doi:~10.1007/s10701-016-0037-4.\\
\\
De Haro, S., Teh, N., Butterfield, J.N.~(2017). ``Comparing Dualities and Gauge Symmetries''. {\em Studies in History and Philosophy of Modern Physics}, 59, pp.~68-80.\\
\\
Dell'Antonio, G.F., Frishman, Y., Zwanzinger, D.~(1972). ``Thirring Model in Terms of Currents: Solution and Light-Cone Expansions''. {\it Physical Review D} 6(4), pp.~988-1007.\\
\\
Dewar, N.~(2016). ``Sophistication about symmetries'', forthcoming in {\em British Journal for the Philosophy of Science}; available at: http://philsci-archive.pitt.edu/id/eprint/12469\\
\\
Dieks, D., Dongen, J. van, Haro, S. de~(2015), ``Emergence in Holographic Scenarios for Gravity''. 
{\it Studies in History and Philosophy of Modern Physics} 52 (B), pp.~203-216. doi:~10.1016/j.shpsb.2015.07.007.\\
\\
Di Francesco, P., Mathieu, P., S\'en\'echal, D.~(1997). {\em Conformal Field Theory}. New York: Springer-Verlag.\\
\\
Fraser, D.~(2017). ``Formal and physical equivalence in two cases in contemporary quantum physics''. {\it Studies in History and Philosophy of Modern Physics}, 59, pp.~30-43. doi:~10.1016/j.shpsb.2015.07.005.\\
\\
Frege, G. (1884/1950) {\em The Foundations of Arithmetic}, (translated by J. L. Austin), Oxford: Blackwell \\
\\
Frege, G. (1892), `\"{U}ber Sinn und Bedeutung', {\em Zeitschrift f\"{u}r Philosophie und philosophische Kritik}, pp.~25-50; translated as `On Sense and reference', in P.T. Geach and M. Black eds. (1960), {\em Translations from the Philosophical Writings of Gottlob Frege}, Oxford: Blackwell.\\
\\ 
Frishman, Y.~and Sonnenschein, J.~(2010). {\em Non-Perturbative Field Theory}, Cambridge: Cambridge University Press. \\
\\
Giamarchi, T.~(2003), {\em Quantum Physics in One Dimension}, Oxford: Oxford University Press.\\
\\
Ginsparg, P.~(1990). ``Applied Conformal Field Theory'', in {\it Fields, Strings, Critical Phenomena: Proceedings}. E.~Brezin and J.~Zinn-Justin (Eds.). Amsterdam: North-Holland. hep-th/9108028. \\
\\
Gogolin, A.O., Nersesyan, A.A. and Tsvelik, A.M. (2004). {\em Bosonization and Strongly Correlated Systems}, Cambridge: Cambridge University Press.\\
\\
Grattan-Guinness, I. (2000), `A sideways look at Hilbert's twenty-three problems of 1900', {\em Notices of the American Mathematical Society} {\bf 47}, pp/ 752-757.\\
\\
Gray J. (2000), {\em The Hilbert Challenge}, Oxford University Press.\\
\\
Gray, J. (2008), {\em Plato's Ghost: the modernist transformation of mathematics}, Princeton University Press.\\
\\
Gray, J. (2012), {\em Henri Poincare, a scientific biography}, Oxford University Press.\\
\\
Huggett, N. (2017), ``Target space $\neq$ space''. {\em Studies in the History and Philosophy of Modern Physics}, 59, pp.~81-88. doi:10.1016/j.shpsb.2015.08.007.\\
\\
Kac, V.~(1990), {\em Infinite-Dimensional Lie Algebras}, Cambridge: Cambridge University Press.\\
\\
Kennedy, H. (1972), `The origins of modern axiomatics: from Pasch to Peano', {\em The American Mathematical Monthly}, {\bf 79} pp.~133-136.\\
\\
Kopietz, P.~(2008), ``Bosonization of Interacting Fermions in Arbitrary Dimensions'. Lecture Notes in Physics Monographs, Vol. 48. Springer. \\
\\
Kouneiher, J.~(2018), `QFT, dualities, integrable systems, and supersymmetry: where we stand today', this volume.\\
\\
Lewis, D. (1970), `General semantics', {\em Synthese} {\bf 22}, pp.~18-67; reprinted in his {\em Philosophical Papers: volume 1}  (1983), Oxford University Press.\\
\\
Lewis, D. (1970a), `How to define theoretical terms', {\em Journal of Philosophy} {\bf 67}: 427-446; reprinted in his {\em Philosophical Papers: volume 1}  (1983), Oxford University Press. \\
\\
Lewis, D. (1980), `Index, context and content', in S. Kanger and S. Ohman (eds.), {\em Philosophy and Grammar}, D. Reidel, pp.~79-100.\\
\\
L\"ust, D.~and Theisen, S.~(1989). {\it Lectures on String Theory}. Lecture Notes in Physics, 346. Berlin: Springer-Verlag. \\
\\
Mandelstam, S.~(1975). ``Soliton Operators for the Quantized sine-Gordon Equation''. {\it Physical Review D} 11 (10), pp.~3026-3030.\\
\\
Matsubara, K.~(2013). ``Realism, underdetermination and string theory dualitites''. {\it Synthese}, 190: 471-489.\\
\\
Putnam, H. (1962) `The Analytic and the Synthetic', in H. Feigl and G. Maxwell
(eds.) {\em Minnesota Studies in the Philosophy of Science} vol. III, Minneapolis: University of Minnesota Press, pp.~358-397.\\
\\
Rickles, D.~(2013). ``Duality and the emergence of spacetime''. {\it Studies in History and Philosophy of Modern Physics}, 44, pp.~312-320.\\
\\
Rickles, D.~(2017). ``Dual theories: `same but different' or different but same'?'' {\em Studies in the History and Philosophy of Modern Physics}, 59, pp.~62-67. doi:~10.1016/j.shpsb.2015.09.005.\\
\\
Smolin, L. (2018), `What are we missing in our search for quantum gravity', this volume.\\
\\
Stachel, J. (2018),  `Einstein and Hilbert', this volume.\\
\\
St\"{o}ltzner, M. (2000), `Opportunistic Axiomatics � Von Neumann on the Methodology of Mathematical Physics', in {\em John Von Neumann and the Foundations of Quantum Physics}, M. Redei and M. St\"{o}ltzner (eds.): Springer: Volume 8 of the series Vienna Circle Institute Yearbook (2001) pp.~35-62.\\
\\
St\"{o}ltzner, M. (2001), `How Metaphysical is ``Deepening the Foundation''? � Hahn and Frank on Hilbert's Axiomatic Method', in {\em History of Philosophy of Science}, M. Heidelberger and F. Stadler (eds.): Springer: Volume 9 of the series Vienna Circle Institute Yearbook (2001) pp.~245-262.\\
\\
Teh, N.J.~(2013). ``Holography and emergence.'' {\it Studies in History and Philosophy of Modern Physics}, 44(3), pp.~300-311.\\
\\
Teh, N.~(2016). ``Gravity and Gauge''. {\it British Journal for the Philosophy of Science}, 67(2), pp.~497-530.\\ 
\\
von Neumann, J. (1932), {\em Mathematical Foundations of Quantum Mechanics}:  English translation published in 1955 by Princeton University Press. \\
\\
Weatherall, J.O.~(2015). ``Categories and the Foundations of Classical Field Theories''. Forthcoming in {\it Categories for the Working Philosopher}, E.~Landry (Ed). Oxford: Oxford University Press.\\
\\
Witten, E.~(1984). ``Nonabelian bosonization in two dimensions''. {\it Communications in Mathematical Physics}, 92, pp.~455.\\
\\
Witten, E.~(2018), ``What Every Physicist Should Know About String Theory'', this volume.


\end{document}